\documentclass[preprint,aps,10pt,showkeys,nofootinbib]{revtex4-1}

\usepackage{amsmath,amssymb,amsfonts,amsthm}
\usepackage{amsbsy} 
\usepackage{epsfig}
\usepackage{latexsym}
\usepackage{dbnsymb}
\input amssym.def
\input amssym.tex

\usepackage{hyperref}

\newcommand{\oarX}[1]{\href{http://arxiv.org/abs/#1}{{\ttfamily #1}}}
\newcommand{\arX}[1]{\href{http://arxiv.org/abs/#1}{{\ttfamily arXiv:#1}}}

\def\barr{\begin{array}}
\def\earr{\end{array}}
\def\half{\frac{1}{2}}
\def\ben{\begin{equation}}
\def\een{\end{equation}}
\def\bs{\begin{subequations}}
\def\es{\end{subequations}}
\def\bena{\begin{eqnarray}}
\def\eena{\end{eqnarray}}

\def\vol{\rm Vol}
\def\const{\rm constant}
\def\bR{\mathbb{R}}
\def\bC{\mathbb{C}}

\def\SO{{\rm SO}}
\def\O{{\rm O}}
\def\GL{{\rm GL}}
\def\SU{{\rm SU}}
\def\SL{{\rm SL}}
\def\GG{\mathfrak{G}}
\def\Sp{{\rm Spin}}

\def\im{{\rm i}}

\def\M{\mathcal{M}}
\def\be{\begin{equation}}
\def\ee{\end{equation}}
\newcommand{\ie}{\textit{i.e.}~}
\newcommand{\eg}{\textit{e.g.}~}
\def\bes{\begin{eqnarray}}
\def\ees{\end{eqnarray}}

\newcommand{\dd}{\mathrm{d}}

\newcommand{\bra}[1]{\left\langle #1 \right|}
\newcommand{\ket}[1]{\left| #1 \right\rangle}
\newcommand{\hphi}{\hat{\varphi}}
\newcommand{\hphid}{\hat{\varphi}^{\dagger}}
\newcommand{\perm}{\text{permut.}}
\def\DD{{\mathrm{D}}}

\newcommand{\sig}[5]{{\sigma^{(i_{#1}^{+} i_{#1}^{-})(j_{#2}^{+},j_{#2}^{-})(j_{#3}^{+},j_{#3}^{-})(j_{#4}^{+},j_{#4}^{-})(j_{#5}^{+},j_{#5}^{-})}
_{(q_{#2}^{+},q_{#2}^{-})(q_{#3}^{+},q_{#3}^{-})(q_{#4}^{+},q_{#4}^{-})(q_{#5}^{+},q_{#5}^{-})}
}}

\newcommand{\inter}[3]{\iota^{J_{#1}j_{#2}^{+}j_{#2}^{-}j_{#3}^{+}j_{#3}^{-}}_{q_{#2}^{+}q_{#2}^{-}q_{#3}^{+}q_{#3}^{-}}}

\begin{document}

\title{\boldmath Homogeneous cosmologies as group field theory condensates}

\preprint{AEI-2013-259}

\author{Steffen Gielen}
\email{sgielen@perimeterinstitute.ca}
\affiliation{Perimeter Institute for Theoretical Physics, 31 Caroline St. N., Waterloo, Ontario N2L 2Y5, Canada}
\author{Daniele Oriti}
\email{doriti@aei.mpg.de}
\author{Lorenzo Sindoni}
\email{sindoni@aei.mpg.de}
\affiliation{Max Planck Institute for Gravitational Physics (Albert Einstein Institute), Am M\"uhlenberg 1, 14476 Golm, Germany, EU}

\begin{abstract}
We give a general procedure, in the group field theory (GFT) formalism for quantum gravity, for constructing states that describe macroscopic, spatially homogeneous universes. These states are close to coherent (condensate) states used in the description of Bose--Einstein condensates. The condition on such states to be (approximate) solutions to the quantum equations of motion of GFT is used to extract an effective dynamics for homogeneous cosmologies directly from the underlying quantum theory. The resulting description in general gives nonlinear and nonlocal equations for the `condensate wavefunction' which are analogous to the Gross--Pitaevskii equation in Bose--Einstein condensates. We show the general form of the effective equations for current quantum gravity models, as well as some concrete examples. We identify conditions under which the dynamics becomes linear, admitting an interpretation as a quantum-cosmological Wheeler--DeWitt equation, and give its semiclassical (WKB) approximation in the case of a kinetic term that includes a Laplace--Beltrami operator. For isotropic states, this approximation reproduces the classical Friedmann equation in vacuum with positive spatial curvature. We show how the formalism can be consistently extended from Riemannian signature to Lorentzian signature models, and discuss the addition of matter fields, obtaining the correct coupling of a massless scalar in the Friedmann equation from the most natural extension of the GFT action. We also outline the procedure for extending our condensate states to include cosmological perturbations. Our results form the basis of a general programme for extracting effective cosmological dynamics directly from a microscopic non-perturbative theory of quantum gravity.
\end{abstract}

\date{May 27, 2014}

\keywords{Quantum gravity, loop quantum gravity, spin foam models, group field theory, quantum cosmology, analogue gravity}

\maketitle
\section{Introduction}
One of the central challenges faced by any proposed theory of quantum gravity is the derivation, from a fundamental theory describing the degrees of freedom presumably relevant at the Planck scale, of effective physics at scales large enough to be relevant for observation and for the connection to other areas of physics. Any such effective large-scale description, describing physical regimes where quantum-gravity effects are not directly relevant, must be consistent with the predictions of General Relativity, with the standard model of particle physics and with cosmological observations such as those done recently in WMAP \cite{wmap} and Planck \cite{planck}. It should also suggest new phenomena or new explanations for existing observations. This challenge, of course, is independent from the similarly fundamental challenge of showing that the proposed theory is in itself mathematically consistent.

In background-independent approaches to quantum gravity this task is particularly challenging: in such theories, the most natural notion of a fundamental vacuum state is one that describes no spacetime at all. Macroscopic, approximately smooth geometries corresponding to  physically interesting solutions of General Relativity are thought of as states with a very large number of quantum-geometric excitations \cite{ThiemannCS,OPS,PS,Corichi,collective}, at least if one allows for fluctuations 
beyond those with very large wavelengths (which are necessary to be able to speak of a smooth geometry at all). 

In understanding the predictions of the theory for such geometries, one faces two basic issues. 

The first issue is the definition and interpretation of appropriate states within the structures of the given theory, such that one can associate (at least in an approximate sense) a spacetime metric description to them. This involves at least two kinds of approximations, a priori independent from one another: one has to map the fundamental degrees of freedom, often thought of as discrete spacetime structures, to a continuum field on a differentiable manifold, and one has to describe states that are sufficiently semiclassical to be associated to a classical geometric configuration, again at least in an approximate sense. 

The second issue is the dynamical description of these states in the quantum theory, at some effective continuum level. In order to be compatible with what we know, this should reproduce, at least in some low-energy regime, \ie up to possible corrections at short distances, the dynamics of General Relativity. This issue is at least as important as the first, lest one would be constrained to a purely kinematical description of spacetime within quantum gravity.

In this paper, we describe in some detail the different steps of a generic procedure that can address both of these issues in the group field theory (GFT) formalism for quantum gravity, at least as far as spatially homogeneous geometries are concerned, as shown already in \cite{prl}. We work in a Fock space picture in which the quantum GFT fields create and annihilate elementary building blocks of space (interpreted as $(d-1)$-simplices in $d$ spacetime dimensions) with a finite number of degrees of freedom encoding discrete geometric data. These states have an equivalent description in terms of the spin network states of loop quantum gravity \cite{LQG}, and their dynamics is encoded, at the perturbative GFT level, in spin foam amplitudes and simplicial gravity path integrals as in the covariant formulation of the same theory \cite{SF}. We use the physical intuition from Bose--Einstein condensation, where one faces a similar problem of relating the microscopic quantum dynamics of atoms to an effective large-scale description of the condensate as a quantum fluid. The picture we propose is indeed that of a {\em condensate} of elementary GFT quanta which make up a continuum, approximately smooth and spatially homogeneous spacetime. This picture can be put on firm footing using the geometric interpretation of the GFT Fock space in terms of elementary parallel transports (giving a discretised spin connection) or simple bivectors (giving a discretised metric) which derives from interpreting the same data in loop quantum gravity or spin foam models. We are able to give a general {\em reconstruction procedure} which maps a given configuration of $N$ such building blocks to an approximate continuum geometry given in terms of a metric on a differentiable manifold. While a finite number of such building blocks can only contain finite information about the reconstructed metric, in the quantum theory we can consider arbitrarily high numbers of quantum-gravitational quanta of space, to improve arbitrarily the same approximation. We give a criterion for the reconstructed metric to be spatially homogeneous, the most relevant case for cosmology.

In the quantum physics of Bose--Einstein condensates, the simplest states one considers are coherent states that are eigenstates of the field operator. This property allows the derivation of an effective dynamics for the condensate wavefunction directly from the underlying microscopic dynamics of the atoms. This dynamics, given by the Gross--Pitaevskii equation, has a direct hydrodynamic interpretation, and provides the kind of effective macroscopic physics that we seek to derive for the case of gravity. The states we consider are such coherent states, but have to satisfy the additional property of being consistent with the gauge invariance of GFT under local Lorentz transformations (or rotations in Riemannian signature). A central conceptual issue in drawing this analogy is the adaptation of the very notion of a hydrodynamic interpretation to the background-independent context:  we cannot expect to obtain a description in terms of a `fluid' on spacetime; instead a field capturing some effective degrees of freedom of quantum geometry is defined on the configuration space for gravity, \ie {\em superspace}, the space of geometries \cite{supersp}. A natural possibility is to view the collective wavefunction appearing in the definition of a condensate state as a wavefunction {\em \`a la} Wheeler--DeWitt quantum cosmology. This interpretation is consistent with the geometric content of the states we consider, but the collective wavefunction will satisfy in general a non-linear and non-local (on minisuperspace) extension of the usual equations of quantum cosmology. This general feature calls for a rethinking of the relation between quantum cosmology and full quantum gravity, and possibly of the interpretation and use of quantum cosmology itself. However, it is not totally unheard of; in fact, a nonlinear extension of quantum cosmology has been suggested, for example, in the loop quantum cosmology context in \cite{nonlincosm}.

We proceed as follows. In Sec.~\ref{gftintro}, we give an introduction to the group field theory (GFT) formalism, emphasising a Fock construction of the kinematical Hilbert space and its interpretation in terms of discrete geometries. As we show in detail, this construction is a reformulation of the basic structure of loop quantum gravity and spin foam models, introducing a second-quantised language that will allow us to directly define condensate states describing macroscopic universes. In Sec.~\ref{approxgeo}, we outline the general procedure for associating to a given discrete geometry, of the type described by states in the GFT Fock space, a reconstructed metric geometry on a manifold that can be interpreted as a spatial hypersurface in canonical gravity. We focus on the case of spatial homogeneity, in which our procedure requires no additional input beyond a choice of 3-dimensional Lie group $\GG$ acting on the space manifold with respect to which the reconstructed metric can be homogeneous. This essentially classical discussion is then used in Sec.~\ref{condensec} to motivate the definition of condensate states in group field theory. We consider two types of condensates, both possessing the right type of (pre-)geometric data: `single-particle' condensates, which are particularly simple to construct, and `dipole' condensates which are automatically gauge-invariant (with respect to local (Lorentz) rotations). We discuss properties of these states, comparing them to coherent and squeezed states used in quantum optics, looking at correlation functions and their interpretation as exact vacuum states of GFT. In Sec.~\ref{effcosm}, we look at the dynamics of these condensate states. In order to be exact solutions to the equations of motion, an infinite number of expectation values must vanish for these states. We focus on two of them which we express as conditions on the collective wavefunction defining the condensate. This defines an effective dynamical equation for such collective, cosmological wavefunctions. We write down the general form of this equation, which holds for any of the current models of 4$d$ quantum gravity in the GFT or spin foam formulation, with special care in ensuring that simplicity constraints are imposed. Then, we specify some conditions in which this becomes a linear differential equation. We give its semiclassical (WKB) approximation for a Laplacian kinetic operator and show that it reduces to the classical Friedmann equation in the isotropic case. We then extend the formalism to Lorentzian signature. We find that the semiclassical analysis done for Riemannian signature can be done analogously and results in equations that are `analytic continuations' of the previous ones, \ie they contain the sign changes corresponding to a change in the metric signature.

The last part of the paper, Sec.~\ref{beyond}, deals with extensions of this formalism beyond the simple case of homogeneous universes without matter:  adding matter fields and perturbations. In a simple example where a massless scalar field is incorporated into the GFT field as an additional argument (corresponding to an additional dimension in superspace), a natural choice of kinetic term on the extended configuration space leads directly to the right coupling of a massless scalar field to gravity, again in the isotropic and WKB approximation. We present ideas for introducing inhomogeneities (which require extending the class of states we have been considering) by adding fluctuations over the exact condensate. We discuss the possibility of identifying arguments of the perturbation field with coordinates in the background geometry defined by a GFT condensate, which could be used to develop a systematic cosmological perturbation theory.

For the convenience of the reader, we discuss divergences arising for Lorentzian models that are associated with the infinite volume of the gauge group, the classical dynamics of Bianchi IX universes, as well as some facts about the geometry of the homogeneous space ${\rm SL}(2,\bC)/\SU(2)$ in the appendix.

\

To summarise, in this paper we give a general procedure for extracting cosmology from quantum gravity, that can be applied to any GFT or spin foam model incorporating data interpretable as a discrete metric or connection. Our examples show that it can give the correct semiclassical limit corresponding to a classical theory of gravity, which is very promising, but clearly more work is needed to apply this procedure to various models discussed in the literature. This paper is a first step in a programme for deriving effective dynamics for an emergent spacetime geometry from a theory of pre-geometric degrees of freedom.

\section{Group field theory} 
\label{gftintro} 

Group field theories (GFT) are field theories over a group manifold $G^d$, not interpreted as spacetime, for models of $d$-dimensional gravity (where we will be only interested in $d=4$), where $G$ is the Lorentz group, its Riemannian counterpart, or some appropriate subgroup (usually $\SU(2)$). In this, they can be viewed both as a generalisation of matrix models, and as an enrichment of tensor models through the addition of group-theoretic data interpreted as pre-geometric or geometric degrees of freedom. The basic variable in GFT is a complex field $\varphi(g_1,\ldots,g_d)$ and the dynamics is encoded in an action, $S[\varphi,\bar\varphi] = \mathcal{K}[\varphi,\bar\varphi] + \sum_i \lambda_i \mathcal{V}_i\left[\varphi,\bar\varphi\right]$, for a kinetic (quadratic) term $\mathcal{K}$ and interaction (higher order) polynomials $\mathcal{V}_i$ weighted by appropriate coupling constants.

\

Let us first define the basic structures of GFT, and then motivate them from loop quantum gravity and spin foam models. The classical field $\varphi$ is a function $G^d\rightarrow\bC$ usually endowed with the invariance 
\ben
\varphi(g_1,\ldots,g_d) = \varphi(g_1 h,\ldots,g_d h)\quad\forall h\in G\,,
\label{firstgauge}
\een
corresponding to local gauge transformations (Lorentz transformations for $G=\SL(2,\bC)$) in gravity.

The quantum field theory can be defined in operator language by imposing the basic (non-relativistic) commutation relations which are consistent with (\ref{firstgauge}),
\be
\left[ \hat{\varphi}(g_I),\hat{\varphi}^\dagger(g'_I) \right]\,=\, \mathbb{I}_G(g_I, g_I') \,,\qquad \left[ \hat{\varphi}(g_I),\hat{\varphi}(g'_I) \right] = \left[ \hat{\varphi}^\dagger(g_I),\hat{\varphi}^\dagger(g'_I) \right] \,=\, 0\,,
\label{commrel}
\ee
where $\mathbb{I}_G(g_I, g_I')$ is the identity operator on the space of gauge-invariant fields. For compact group $G$, this can be defined by $\mathbb{I}_G(g_I, g_I')=\int_G \dd h \prod_{I=1}^d\delta(g_I h (g_I')^{-1})$; for non-compact $G$ one has to be more careful to avoid divergences, as we will discuss in detail in Appendix~\ref{diverge}. 

\

One can now proceed to expand the field in a basis of functions on $L^2(G^d/G)$ (indexed by some set of labels $\vec{\chi}$) and promote the expansion coefficients to creation and annihilation operators,
\be
\hat{\varphi}(g_1,\ldots,g_d)\equiv\hat{\varphi}(g_I)=\sum_{\vec{\chi}}\;\hat{c}_{\vec{\chi}}\; \psi_{\vec{\chi}}(g_I)\,,\qquad
\hat{\varphi}^\dagger(g_1,\ldots,g_d)\equiv\hat{\varphi}^\dagger(g_I)=\sum_{\vec{\chi}}\; \hat{c}^\dagger_{\vec{\chi}} \;\psi^*_{\vec{\chi}}(g_I) \,,
\ee 
which will satisfy 
\be
\left[ \hat{c}_{\vec{\chi}} , \hat{c}_{\vec{\chi}'}^\dagger \right] = \delta_{\vec{\chi},\vec{\chi}'}\,,\quad\left[ \hat{c}_{\vec{\chi}} , \hat{c}_{\vec{\chi}'} \right] = \left[ \hat{c}_{\vec{\chi}}^\dagger , \hat{c}_{\vec{\chi}'}^\dagger \right] = 0
\ee
for an appropriate normalisation of the basis functions $\psi_{\vec{\chi}}(g_I)$. These operators can be used to define a Fock space starting from a vacuum state $|0\rangle$ annihilated by all $\hat{c}_{\vec{\chi}}$; they will act as
\ben
\hat{c}_{\vec{\chi}} | n_{\vec{\chi}}\rangle = \sqrt{n_{\vec{\chi}}} | n_{\vec{\chi}} - 1\rangle\,,\quad\hat{c}^\dagger_{\vec{\chi}} | n_{\vec{\chi}}\rangle = \sqrt{n_{\vec{\chi}} + 1} | n_{\vec{\chi}} + 1\rangle\,.
\een
The field operator $\hat\varphi^{\dagger}(g_I)$ itself creates a ``particle'' with data $\{g_I\}$, or more precisely the equivalence class $[\{g_I\}]=\{\{g_I h\},\, h\in G\}$, when acting on $|0\rangle$. This particle is interpreted as an elementary building block of simplicial geometry, a $(d-1)$-simplex with the group elements $g_I$ corresponding to elementary parallel transports of a (gravitational) $G$-connection along the links dual to the $d$ faces (\ie $(d-2)$-subsimplices). Local gauge transformations act on the vertex where these links meet as simultaneous right multiplication of all $g_I$ by a common element $h$ of $G$, which is the motivation for requiring (\ref{firstgauge}) to make the theory gauge-invariant.

At least for finite-dimensional $G$, there is a well-defined notion of non-commutative Fourier transform which takes functions on the group to functions on its Lie algebra $\mathfrak{g}$ \cite{carlosdanielematti}. This can be used to define the equivalent ``momentum space'' representation of the same theory. The non-commutativity of multiplication of the group (and of the Lie algebra) is reflected in the introduction of a $\star$-product in this dual representation. The dual variables $B_I$, which are elements of $\mathfrak{g}$, are interpreted as bivectors associated to the faces, corresponding to $\int e\wedge\ldots\wedge e$ for a $d$-bein $e$ in the case of gravitational models. A basic issue of the spin foam and GFT approach, as far as the construction of gravitational models is concerned, is to restrict the generic Lie algebra variables $B_I$ to those that can be written in this form, at the quantum level. We will come back to that later.

\

What we have described so far is a self-contained formulation of a quantum field theory, and in fact contains everything that will be required for the constructions in the rest of the paper. However, to understand better the motivation of the GFT approach, let us explain more closely the relation of group field theories \cite{GFT} to loop quantum gravity (LQG) \cite{LQG} and specifically its spin foam corner \cite{SF}, which has in fact provided the impetus for the development of group field theories. More details on this are given in a separate publication \cite{danielefock} but we illuminate the most important points here.

\

There is a very close relation between group field theories and spin foam models, which are a proposal for defining a discrete and algebraic sum-over-histories formulation for quantum gravity, based on the variables and states used in loop quantum gravity. In fact, spin foam models and group field theories are in one-to-one correspondence \cite{carlomike}. To any spin foam model assigning an amplitude to a given cellular complex (a possible \lq history\rq ~of spin networks), there exists a group field theory, specified by a choice of field and action, that reproduces the same amplitude for the GFT Feynman diagram dual to this cellular complex. Conversely, any quantum GFT partition function also defines a spin foam model by specifying uniquely the Feynman amplitudes associated to the cellular complexes appearing in its perturbative expansion. In a formula,

\be
Z\,=\,\int\mathcal{D}\varphi\,\mathcal{D}\bar\varphi\,e^{- S[\varphi,\bar\varphi]}\,=\, \sum_\sigma \frac{\prod_i(\lambda_i)^{N_i(\sigma)}}{{\rm Aut}(\sigma)}\, \mathcal{A}_\sigma\,,
\ee
where $\sigma$ are cellular complexes dual to the Feynman diagrams of the GFT model. Their combinatorial structure depends on the model; writing the action in terms of kinetic (free) part and interactions as $S=S_{{\rm k}}+\sum_i \mathcal{V}_i$, the possible interaction vertices are determined by the combinatorial pattern of pairings of field arguments in the different terms $\mathcal{V}_i$. $N_i(\sigma)$ is then the number of vertices of type $i$ in the Feynman diagram dual to $\sigma$, and ${\rm Aut}(\sigma)$ is the order of the automorphisms of $\sigma$. $\mathcal{A}_\sigma$ is the Feynman amplitude that the GFT model assigns to $\sigma$ and, generically, can be represented as a spin foam model (or, equivalently, as a non-commutative discrete gravity path integral \cite{danielearistide}).

Thus we see that group field theories not only encode the same dynamics of quantum geometry as spin foam models, but that they do more than that. Unless a fundamental theory of quantum gravity possesses a finite number of degrees of freedom, a spin foam formulation of it cannot be based on a single cellular complex. A complete definition should involve an infinite class of cellular complexes, in the same way in which the Hilbert space of the canonical theory is defined over an infinite class of spin network graphs (appearing as boundary states), and in particular a prescription for organising the amplitudes associated to all such complexes. Group field theory provides one such prescription\footnote{Another prescription could be some refinement procedure in the spirit of lattice gauge theory, with associated coarse graining methods used to extract effective continuum physics. For this direction of investigation, see \cite{bianca}.} as it generates a sum over complexes, weighted by spin foam amplitudes and the coupling constants $\lambda_i$, as a Feynman diagram expansion and thus with canonically assigned combinatorial weights. Thus one can say that group field theory is actually a {\it completion} of the spin foam approach. 

\

Spin foam models and loop quantum gravity are usually presented as covariant and canonical formulations of the same quantum theory, but their exact correspondence is not yet fully understood. A straightforward second quantisation of spin networks (both their kinematics and dynamics), and thus of loop quantum gravity, leads however directly to the GFT formalism, and this GFT/LQG correspondence defines in turn a correspondence between the canonical LQG formulation and covariant spin foam dynamics. One advantage of the GFT reformulation is that it provides the right tools to study the physics of many LQG degrees of freedom, to bypass the need to deal explicitly with complicated spin networks and spin foams, and to derive effective descriptions for collective variables and features of the non-perturbative sector of the theory. All these are reasons for using quantum field theory reformulations of many-body quantum physics in condensed matter theory and particle physics, so it should come as no surprise that we encounter the same advantages in quantum gravity. Our paper exemplifies this use of the GFT formalism: we will indeed bypass the spin foam formulation of the dynamics, provide both a definition of interesting, albeit very simple, non-perturbative quantum states of the theory, interpreted as cosmological quantum spacetimes, and extract an effective cosmological dynamics for them, using the second quantised features of the GFT formalism. 

We now give some more details on this second quantised formalism, and on the link between LQG and GFT, and thus the direct LQG relevance of our results. For a more extensive treatment, see \cite{danielefock}. 

\

In first-quantised language, one has a Hilbert space $\tilde{\mathcal{H}}_d$ of states associated to $V$ $d$-valent graph vertices (which includes particular states associated to both open and closed graphs,
of the type defining the Hilbert space of LQG). Each such vertex is a node with $d$ outgoing open links, and can be thought as dual to a polyhedron with $d$ faces\footnote{In this paper, we restrict attention to the simplicial case, in which $d$ equals the spacetime dimension, and each GFT quantum (or spin network vertex) is dual to a ($d-1$)-simplex, \ie a tetrahedron in $d=4$.}. $V$-particle states are given by wavefunctions describing $V$ vertices or their dual polyhedra, of the type
\be
\phi(g_i^j)=\phi(g_1^1,g_2^1,\ldots,g_d^{1};\ldots;g_1^V,g_2^V,\ldots,g_d^V) \,,
\label{waveff}
\ee
where each open link outgoing from each vertex is associated a group element of the group $G$ ($G=\SU(2)$, $\Sp(4)$, or $\SL(2,\mathbb{C})$ in quantum gravity GFT models, and $G=\SU(2)$ in standard LQG), with gauge invariance at vertices in $V$: $\phi(g_i^j)=\phi(g_i^j \beta_j)$ for $V$ elements $\beta_j$ of $G$. The set of such functions (restricting to square-integrable ones) can be turned into the Hilbert space $L^2(G^{d\cdot V}/G^V)$ by defining the inner product via the Haar measure on the group, or some right/left-invariant measure in the non-compact case.

The Hilbert space for these functions,  $\tilde{\mathcal{H}}_d$, includes as a special class of states the usual LQG states associated to closed $d$-valent graphs $\Gamma$. \footnote{These are the {\em cylindrical functions} of LQG, where there is a notion of {\em cylindrical consistency}: a state on a graph $\Gamma$ is identified with states on $\tilde\Gamma\supset\Gamma$ that do not depend on the edges $\tilde\Gamma\backslash\Gamma$. In the GFT setting used here, cylindrical consistency is not imposed.} 
There is a relation $E(\Gamma)\subset (\{1,\ldots,V\}\times\{1,\ldots,d\})^2$ (satisfying $[(i\,a)\,(i\,a)]\not\in E(\Gamma)$) which specifies the connectivity of such a graph: if $[(i\,a)\,(j\,b)]\in E(\Gamma)$, there is a directed edge connecting the $a$-th link at the $i$-th node to the $b$-th link at the $j$-th node, with source $i$ and target $j$. LQG wavefunctions are then of the form
$\Psi_\Gamma(\{ G^{ab}_{ij}\})$, where the group elements $G_{ij}^{ab}\in G$ are assigned to each link $e:=[(i\,a)\,(j\,b)]\in E(\Gamma)$ of the
graph. These are labelled by two pairs of indices: the first pair identifies the pair of vertices $(ij)$ connected, while the second pair identifies the outgoing edges $(ab)$ of each vertex
glued together to form the link. 
We
assume the gauge invariance $\Psi_{\Gamma}(\{G^{ab}_{ij}\}) = \Psi_\Gamma(
\{\beta_i
G^{ab}_{ij}\beta_j^{-1}\})$, for any $V$ group elements $\beta_i$ associated to the vertices.

Given a closed $d$-valent graph $\Gamma$ with $V$ vertices (specified by $E(\Gamma)$), a wavefunction $\Psi_\Gamma$ associated to $\Gamma$ can be obtained by group-averaging of any wavefunction $\phi_{\Gamma}$ of the form (\ref{waveff}) associated to the vertices of $\Gamma$,
\be
\Psi_\Gamma(\{G^{ab}_{ij}\})\,=\,
\prod_{e\in E(\Gamma)}\int_{G}\dd\alpha_{ij}^{ab}\,
\phi_\Gamma(\{\alpha_{ij}^{ab} g_{i}^{a} ; \alpha_{ij}^{ab}  g_{j}^{b}\})
\,=\,\Psi_\Gamma(\{ (g^{a}_{i})^{-1}g_{j}^{b} \})\,, \label{gluingGroup} \ee
in such a way
that each edge in $\Gamma$ is associated with two group
elements $g_i^a,g_j^b\in G$. 
The integrals over the $\alpha$'s glue open spin network vertices corresponding to the function $\phi$, pairwise along
common links, thus forming the spin network represented by
the closed graph $\Gamma$. The same construction can be phrased in the flux representation and in the spin representation.

\

Let us denote by $\mathcal{H}_v$ the subspace of single-particle (single-vertex) states, \ie elements of $\tilde{\mathcal{H}}_d$ with $V=1$. A general $V$-particle state can be decomposed into products of elements of $\mathcal{H}_v$,
\be
\phi(g_I^i)=\langle g_I^i| \phi \rangle = \left(\prod_{i=1}^V\sum_{\vec{\chi}_i}\right) \phi^{\vec{\chi}_1...\vec{\chi}_V}\,\langle g_I^1|\vec{\chi}_1\rangle \cdots \langle g_I^V|\vec{\chi}_V\rangle\,, 
\ee
where the complete basis of single-vertex wave functions is given either by the spin network wavefunctions for individual spin network vertices,
\be
\vec{\chi} = \left( \vec{J}, \vec{m}, \mathcal{I} \right) \;\;\;\rightarrow\;\;\; \psi_{\vec{\chi}}(g_I) =  \langle g_I|\vec{\chi}\rangle = \left[ \prod_{I=1}^{d} D^{J_I}_{m_I n_I}(g_I)\right]\,C^{J_1\ldots J_d,\mathcal{I}}_{n_1\ldots n_d} \,,
\ee
where $\mathcal{I}$ label a basis in the intertwiner space between the given group representations, 
or by a product of non-commutative plane waves $e_{g_I}$ constrained by the (non-commutative) closure condition for the fluxes,
\be
\vec{\chi} = \left(B_I\in\mathfrak{g}\;|\; \sum_I B_I =0 \right) \;\;\; \rightarrow \;\;\; \psi_{\vec{\chi}}(g_I) =  \langle g_I|\vec{\chi}\rangle = \left[ \prod_{I=1}^{d} e_{g_I}(B_I)  \right] \, \star \, \delta_\star\left( \sum_I B_I\right) \,.
\ee

Being quantum field theories, GFTs describe this Hilbert space in second quantised language. Assuming bosonic statistics for the spin network vertices, and thus the symmetry 
\be
\phi\left( g_I^1, g_I^2, \dots, g_I^i,\dots, g_I^j,\dots,g_I^V\right) = \phi\left(g_I^1, g_I^2, \dots, g_I^j,\dots,g_I^i,\dots,g_I^V\right)\,, 
\label{bosonic}
\ee
the corresponding Fock space is 
\be
\mathcal{F}(\mathcal{H}_v) = \bigoplus_{V=0}^\infty\left( \mathcal{H}_v^{(1)} \otimes \mathcal{H}_v^{(2)} \otimes \cdots \otimes \mathcal{H}_v^{(V)}\right)
\ee
where only symmetric elements of $\mathcal{H}_v \otimes \mathcal{H}_v \otimes \cdots \otimes \mathcal{H}_v$ are included. The inner product on this Fock space descends, for each summand in the direct sum, from the one on $L^2(G^{d\cdot V}/G^V)$, and is equivalent (at least for $G=\SU(2)$) to the LQG inner product of states on a fixed graph.

One moves to a labelling of quantum states by their \lq\lq occupation numbers\rq\rq, \ie to a new basis of the Hilbert space of a given (finite) number $V$ of spin network vertices, defined by
\ben
| n_1, \ldots, n_a, \ldots\rangle = \sqrt{\frac{n_1! \ldots n_\infty !}{V!}} \sum_{\{ \vec{\chi}_i | n_a \}} \bigotimes_{i=1}^V|\vec{\chi}_i\rangle
\een
where the label $a$ runs over a basis of the single-particle Hilbert space $\mathcal{H}_v$, and one only sums over those configurations compatible with the labels $\{n_a\}$, \ie many-particle states where $n_a$ particles are in the state $a$. For elements of $\mathcal{F}(\mathcal{H}_v)$, we can then rewrite
\bena
\phi_{\tilde{\Gamma}}\left( g_I^i \right) &=& \sum_{\{\vec{\chi}_i\}} \phi_{\tilde{\Gamma}}^{\vec{\chi}_1 \ldots \vec{\chi}_V}\; \prod_{i=1}^V \langle g_I^i|\vec{\chi}_i\rangle = \sum_{\{ n_a\}} \bigotimes_{i=1}^V\langle g_I^i| \left( \sum_{\{\vec{\chi}_i\} | \{n_a\}} \phi_{\tilde{\Gamma}}^{\vec{\chi}_1 \ldots \vec{\chi}_V}\; \bigotimes_{j=1}^V|\vec{\chi}_j\rangle\right)\nonumber
\\&=& \sum_{\{ n_a\}} \left(\sqrt{\frac{V!}{n_1! \ldots n_\infty !}} \sum_{\{\vec{\chi}_i\} | \{n_a\}} \phi_{\tilde{\Gamma}}^{\vec{\chi}_1 \ldots \vec{\chi}_V}\right) \left(\bigotimes_{i=1}^V\langle g_I^i|\right)|n_1,\ldots,n_a,\ldots\rangle\nonumber
 \\ &=: &\sum_{\{ n_a\}} \tilde{C}\left(n_1,\ldots, n_a,\ldots\right) \psi_{\{n_a\}}\left( g_I^i \right) 
\label{stateoccupnumbers} 
\eena
where we denote the coefficients of the new basis elements in the group representation by  $\psi_{\{n_a\}}\left( \vec{g}_i \right)= \langle g_I^i | n_1, \ldots, n_a, \ldots\rangle$, and $\tilde{C}\left(n_1,\ldots, n_a,\ldots\right)$ are the coefficients of $\phi$ in the occupation number basis.
The states of the new basis, and thus all the states of the Fock space of the theory, can be obtained in terms of the creation/annihilation operators defined by
\be
\left[ \hat{c}_{\vec{\chi}} , \hat{c}_{\vec{\chi}'}^\dagger \right] = \delta_{\vec{\chi},\vec{\chi}'}\,,\quad\left[ \hat{c}_{\vec{\chi}} , \hat{c}_{\vec{\chi}'} \right] = \left[ \hat{c}_{\vec{\chi}}^\dagger , \hat{c}_{\vec{\chi}'}^\dagger \right] = 0 \,;\quad
\hat{c}_{\vec{\chi}} | n_{\vec{\chi}}\rangle = \sqrt{n_{\vec{\chi}}} | n_{\vec{\chi}} - 1\rangle\,,\quad\hat{c}^\dagger_{\vec{\chi}} | n_{\vec{\chi}}\rangle = \sqrt{n_{\vec{\chi}} + 1} | n_{\vec{\chi}} + 1\rangle \,.
\ee
It is clear from this algebra that these fundamental operators {\it create} and {\it annihilate} LQG network vertices. One can construct arbitrary spin networks of the type we are considering by acting multiple times on the special state given by the {\it Fock vacuum} $| 0 \rangle$, which is interpreted as the \lq\lq no-space\rq\rq\, (or \lq\lq emptiest\rq\rq) state in which no degree of freedom of quantum geometry is present. It is defined by the property that it is annihilated by all annihilation operators,  $\hat{c}_{\vec{\chi}} | 0 \rangle = 0 \;\; \forall \vec{\chi}$, and so all of its occupation numbers are zero, $|0\rangle = | 0, 0, \ldots, 0\rangle$.

From the linear superposition of creation and annihilation operators, it is then standard to define the bosonic {\it field operators}
\be
\hat{\varphi}(g_1,..,g_d)\equiv\hat{\varphi}(g_I)=\sum_{\vec{\chi}}\;\hat{c}_{\vec{\chi}}\; \psi_{\vec{\chi}}(g_I)\,,\qquad
\hat{\varphi}^\dagger(g_1,..,g_d)\equiv\hat{\varphi}^\dagger(g_I)=\sum_{\vec{\chi}}\; \hat{c}^\dagger_{\vec{\chi}} \;\psi^*_{\vec{\chi}}(g_I) \,,
\ee 
satisfying the commutation relations (here we are using a suitable normalisation of the basis functions $\psi_{\vec{\chi}}(g_I)$)
\be
\left[ \hat{\varphi}(g_I),\hat{\varphi}^\dagger(g'_I) \right]\,=\, \mathbb{I}_G(g_I, g_I') \,,\qquad \left[ \hat{\varphi}(g_I),\hat{\varphi}(g'_I) \right] = \left[ \hat{\varphi}^\dagger(g_I),\hat{\varphi}^\dagger(g'_I) \right] \,=\, 0\,,
\ee
where $\mathbb{I}_G(g_I, g_I')$ is the identity operator on the space of gauge invariant fields. Hence we recover and motivate the definition (\ref{commrel}) from the perspective of spin networks in LQG.

These are the fundamental GFT field operators, expanded in modes either via Peter--Weyl decomposition for $\vec{\chi} = (\vec{J},\vec{m},\mathcal{I})$ or via the non-commutative Fourier transform for $\vec{\chi} = (\vec{B}\in\mathfrak{g} | \sum_{I=1}^d \vec{B}_I = 0)$. In the second case the formula has to be understood as involving a $\star$-multiplication between field modes (creation/annihilation operators) and single-vertex wave functions. 

\

The kinematical operators of GFTs are also obtained naturally from the canonical LQG kinematics (or, equivalently, from quantum simplicial geometry). 

Given a general \lq $(n+m)$-body operator\rq\, $\widehat{\mathcal{O}_{n+m}}$, that is an operator acting on spin network states formed by $n$ spin network vertices and resulting in states with $m$ spin network vertices, one can define its matrix elements
\ben
\langle \vec{\chi}_1, \ldots,\vec{\chi}_m  | \widehat{\mathcal{O}_{n+m}} | \vec{\chi}'_1 , \ldots, \vec{\chi}'_n \rangle = \mathcal{O}_{n+m}\left( \vec{\chi}_1,\ldots,\vec{\chi}_m, \vec{\chi}'_1,\ldots,\vec{\chi}'_n\right)
\een
and a corresponding operator on the GFT Fock space,
\ben
\widehat{\mathcal{O}_{n+m}}\left[\hat{\varphi},\hat{\varphi}^\dagger\right] = \int (\dd g)^{d\cdot m} \;(\dd h)^{d\cdot n}\; \mathcal{O}_{n+m}\left(g_I^1,\ldots,g_I^m,h_I^1,\ldots,h_I^n\right)\prod_{i=1}^m\hat\varphi^{\dagger}(g_I^i)\prod_{j=1}^n\hat\varphi(h_I^j)\,.
\een

The quantum dynamics of spin networks can be encoded in a `projection'\, operator onto physical states (other possibilities can be considered as well), which encodes the action of some Hamiltonian constraint operator. We take the condition on physical states to be of the form
\be
\widehat{P} \, | \Psi \rangle_{{\rm phys}} = | \Psi \rangle_{{\rm phys}}\,.
\ee
The operator $\widehat{P}$ will in general decompose into 2-body, 3-body, $\ldots$, $(n+m)$-body operators, \ie into operators whose action involves 2, 3, $\ldots$, $(n+m)$ spin network vertices, and possibly an infinite number of components, weighted by suitable coupling constants,
\be
\widehat{P} |\Psi \rangle_{{\rm phys}} = \left[\lambda_2\widehat{P}_2 + \lambda_3\widehat{P}_3 + \ldots\right] | \Psi \rangle_{{\rm phys}} = | \Psi \rangle_{{\rm phys}} \,.
\ee 
The second quantised counterpart of the quantum dynamics in the Fock space is then given by an operator $\widehat{F}$ on the GFT Fock space that corresponds to the operator $\mathbb{I}-\widehat{P}$ acting on first quantised spin-network states. It is defined by
\bena
\widehat{F} &\equiv&\sum_{\vec{\chi} }\hat{c}_{\vec{\chi}}^\dagger \hat{c}_{\vec{\chi}}\;-\;\sum_{n,m}^\infty \lambda_{n+m} \sum_{\{ \vec{\chi},\vec{\chi}'\} }\hat{c}_{\vec{\chi}_1}^\dagger \ldots \hat{c}_{\vec{\chi}_m}^\dagger P_{n+m}\left( \vec{\chi}_1,\ldots,\vec{\chi}_m, \vec{\chi}'_1,\ldots,\vec{\chi}'_n\right) \hat{c}_{\vec{\chi}_1'} \ldots \hat{c}_{\vec{\chi}_n'}
\\
&=&\int (\dd g)^d\; \hat\varphi^\dagger(g_I)\hat\varphi(g_I)\;-\;\sum_{n,m}^\infty \lambda_{n+m} \left[\int  (\dd g)^{d\cdot m}\; (\dd h)^{d\cdot n}\;P_{n+m}\left(g_I^i,h_I^j\right)\prod_{i=1}^m\hat\varphi^{\dagger}(g_I^i)\prod_{j=1}^n\hat\varphi(h_I^j)\right] \nonumber
\eena
and acts on states in $\mathcal{F}(\mathcal{H}_v) = \oplus_{V=0}^\infty\left( \mathcal{H}_v^{(1)} \otimes \mathcal{H}_v^{(2)} \otimes \cdots \otimes \mathcal{H}_v^{(V)}\right)$.

These are the GFT dynamical operator equations, which can be also encoded in the Schwinger-Dyson equations for $n$-point functions. We have chosen here the normal ordering for creation and annihilation operators, as is customary in many-body quantum physics. Different operator orderings would give quantum corrections to be absorbed in the interaction kernels of the theory. 

The identification of the corresponding GFT action, starting from the above quantum dynamics expressed in canonical (second quantised) form, requires some assumption. We consider a {\it grandcanonical} ensemble 

\be
Z_g = \sum_{s} \langle s | \hat{\rho}_g | s\rangle = \sum_{s} \langle s | e^{-\, \left(\widehat{F} \, -\, \mu \widehat{N}\right)}  | s\rangle
\label{granc}
\ee

where the sign of the \lq chemical potential\rq\, $\mu$ selects quantum states with many or few spin network vertices as  dominant. In order to turn this expression into the GFT path integral, we introduce then a second quantised basis of eigenstates of the  GFT quantum field operator, $| \varphi \rangle  = \exp\left(\sum_{\vec{\chi}} \varphi_{\vec{\chi}} \hat{c}_{\vec{\chi}}^\dagger\right) |0 \rangle = \exp\left(\int (\dd g)^d\; \varphi(g_I) \hat{\varphi}^\dagger(g_I)\right) |0 \rangle$, satisfying
\ben
\hat{c}_{\vec{\chi}} | \varphi \rangle = \varphi_{\vec{\chi}} \, | \varphi \rangle\,,  \qquad \langle \varphi | \hat{c}_{\vec{\chi}}^\dagger  = \overline{\varphi_{\vec{\chi}}} \, \langle \varphi | \,,
\een
or equivalently
\ben
\hat{\varphi}(g_I) | \varphi \rangle = \varphi(g_I) \, | \varphi \rangle\,,  \qquad \langle \varphi | \hat{\varphi}^\dagger(g_I)  = \overline{\varphi(g_I)} \, \langle \varphi |\,,
\een
and a completeness relation as is usual for such coherent states,
\ben
\mathbb{I} = \int \mathcal{D}\varphi \;\mathcal{D}\overline{\varphi}\; e^{-|\varphi|^2}\, | \varphi \rangle \langle \varphi |\,, \qquad |\varphi|^2 \equiv \int (\dd g)^d \;\overline{\varphi}(g_I)\,\varphi(g_I) \, =\, \sum_{\vec{\chi}}\, \overline{\varphi_{\vec{\chi}} }\,\varphi_{\vec{\chi}} \,.
\een

 The functions $\varphi$ and $\overline{\varphi}$ are indeed the classical GFT fields, and the measure over them is the (formal) GFT path integral measure. Inserting the corresponding resolution of the identity in the formula for the quantum partition function, one obtains
 
 \be
Z_g = \sum_{s} \langle s | e^{-\, \left(\widehat{F} \, -\, \mu \widehat{N}\right)}  | s\rangle \, = \,   \int \mathcal{D}\varphi \;\mathcal{D}\overline{\varphi}\; e^{- S_{{\rm eff}}[\varphi,\overline{\varphi}]}
\label{stat}
\ee
in terms of an effective action 
\be
e^{-  S_{{\rm eff}}[\varphi,\overline{\varphi}]}\,\equiv\,e^{- |\varphi|^2}\, \langle \varphi |\, e^{-\, \left(\widehat{F} \, -\, \mu \widehat{N}\right)} \,  | \varphi \rangle \qquad . 
\ee
The effective action $S_{{\rm eff}}$ is the quantum corrected version of the {\it classical GFT action}
\be
S_0\left[\varphi,\overline{\varphi}\right]=\, \frac{\langle \varphi | \widehat{F} | \varphi \rangle}{\langle \varphi | \varphi \rangle} \,;
\ee 
the chemical potential becomes a mass term in the effective action, rescaling the term coming from the identity operator (which became the number operator in $\widehat{F}$) by $m^2 = 1 - \mu$.

The same type of quantum corrections would have appeared from a different operator ordering in the very definition of $\widehat{F}$. Because of this, and because both sides of the two possible definitions of the quantum partition functions would have anyway to be properly defined, with a careful handling of such quantum corrections, one can just define the quantum GFT theory starting from the above classical action. To give proper meaning to the corresponding path integral and to the partition function, one has then to go through the usual renormalisation and constructive procedures of quantum field theory.

The corresponding classical GFT action is then of the form
\bena
S\left[ \varphi, \bar\varphi\right]\; &=&\; \int (\dd g)^d\; \bar\varphi(g_1,\ldots,g_d)\,\varphi(g_1,\ldots,g_d) \\&& -\sum_{n,m}^\infty \lambda_{n+m} \left[\int  (\dd g)^{d\cdot m} (\dd h)^{d\cdot n}\; V_{n+m}\left(g_I^i,h_I^j\right)\prod_{i=1}^m \bar\varphi(g_I^i)\prod_{j=1}^n\varphi(h_I^j)\right]\,,\nonumber  \\ V_{n+m}\left( g_I^i,h_I^j\right)& =& P_{n+m}\left(g_I^i,h_I^j\right)\,. \nonumber
\eena
The GFT interaction kernels (or spin foam vertex amplitudes) are thus nothing else than the matrix elements of the canonical projection operator in the basis of (products of) single-vertex states.

In this section we define the quantum GFT through the path integral. Later on, we will use an operator formalism in which we impose the operator version of the Euler--Lagrange equations $\delta S/\delta\varphi=\delta S/\delta\bar\varphi=0$ on physical states. The relation between the two is given by the Schwinger--Dyson equations, which give expectation values of general functionals of the field from the path integral. We will come back to those in the discussion of the effective GFT dynamics in Sec.~\ref{effcosm}.

\

The GFT formalism as outlined above is quite general, and lends itself to different constructions. In the following, we will consider models (the most studied ones) aiming at describing 4$d$ quantum gravity. These have been defined via a strategy inspired by the formulation of gravity as a constrained topological BF theory, also known as the Plebanski formulation. We will give more details at a later stage, when needed for explicit manipulations. Now we only recapitulate some features of this strategy, only some of which have a corresponding justification in the canonical LQG theory. The formulation of gravity as a constrained BF theory suggests: 1) to choose for the group $G$ the local gauge group of gravity in 4$d$, $G= \SL(2,\mathbb{C})$ in Lorentzian signature and $G=\Sp(4)$ in the Riemannian case; 2) to start from a GFT model describing topological BF theory, such as the Ooguri model, and thus quantising only flat connections, and 3) to impose suitable conditions on the Lie algebra-valued variables conjugate to such a flat connection (the $B$ field of BF theory) which enforce the geometric nature of them, \ie force them to be a function of a tetrad field. In the continuum, classical theory, the result of such constraining is the Palatini formulation of gravity. The corresponding GFT and spin foam construction takes place at the discrete level, more precisely in a simplicial context. The basic objects of the theory are combinatorial tetrahedra, labelled by the discrete counterpart of BF fields: group elements or (conjugate) Lie algebra elements associated to the faces of the tetrahedra (or links of the dual spin network vertices). These are the same variables used above to label states and amplitudes in the GFT formalism. The interaction of these tetrahedra, encoded in the GFT vertex, is given in terms of a 4-simplex having five of these tetrahedra in the boundary. This corresponds to a specific choice of \lq locality principle\rq\, in GFT. One then imposes restrictions (\lq\lq simplicity constraints\rq\rq) on such algebraic data, intended as discrete counterpart of the Plebanski geometricity conditions in the continuum. How to do this imposition correctly is a main focus of activity in the spin foam/GFT literature, and different models have been proposed. The general point is, however, that after such conditions have been imposed, the Lie algebra elements labelling the faces of the tetrahedron can be interpreted as derived from a discrete tetrad field associated to the 4-simplex, with edge vectors associated to tetrahedron edges. In models incorporating the Immirzi parameter, and aiming at a quantisation of the Plebanski--Holst formulation of gravity, the same constraints imply that one can move from covariant $\SL(2,\mathbb{C})$ (or $\Sp(4)$) variables to $\SU(2)$ ones, with the detailed embedding of $\SU(2)$ data in the full group encoded in the dynamics of the theory, \ie in the details of the kinetic and/or interaction terms of the GFT action. 

\

Our strategy for the rest of the paper will be to use the Fock space construction of GFT to define states that we can interpret geometrically as spatially homogeneous macroscopic geometries, and that are comparably easy to manipulate in the quantum theory. For the geometric interpretation, one can focus on either metric or connection variables, corresponding to the Lie algebra or group representation of GFT, as we have described. 

GFT Fock states with $N$ particles that could be interpreted purely in terms of bivector data, with completely undetermined parallel transports, can be constructed out of basis states of the form
\ben
|B_{I(m)}\rangle := \frac{1}{N!}\prod_{m=1}^N \hat{\tilde{\varphi}}^{\dagger}(B_{I(m)})|0\rangle\,,
\label{state}
\een
where $\hat{\tilde{\varphi}}(B_I)$ is the Fourier transform of the field operator $\hat\varphi(g_I)$. For such a state, the Lie algebra variables $B$ specify the bivectors to be attached to each face, while there is no information about the group elements to be attached to the dual links. Unlike in the analogous case of scalar field theory on Minkowski space, and for compact $G$ where a non-commutative Fourier transform can be defined, the states (\ref{state}) are normalisable.

A slightly more general construction allows for the $N$ bosons to have general wavefunctions,
\ben
|\Psi_{m}\rangle :=  
\int (\dd B)^{4N}\,\prod_{m=1}^N\,
\Psi_m(B_{1(m)},\ldots,B_{d(m)})\star|B_{I(m)}\rangle\,.
\label{statewf}
\een
Such a state corresponds to a set of $N$ tetrahedra with geometric data attached to its faces and dual links according to the properties of the wavefunctions $\{\Psi_{m}\}_{m=1}^{N}$, keeping in mind that these are indistinguishable particles, so that there is no sense in which individual tetrahedra can be associated with particular wavefunctions $\Psi_i$.  For states of the form (\ref{statewf}), requiring normalisability means that pairwise star-products of the functions $\Psi_m$ have to be integrable,
\ben
\int (\dd B)^4 \;\overline{\Psi_i(B_I)}\star\Psi_j(B_I)<\infty\,.
\een
Of course, such states still form a rather small subset of general $N$-particle states in which there would be a general function $\Psi(B_{1(1)},\ldots,B_{4(N)})$ on $\mathfrak{so}(4)^{4N}$ which does not factorise as a product.

If we use a smooth state like \eqref{statewf},
and consider, for example, coherent states such as the ones introduced in the LQG literature 
\cite{ThiemannCS,OPS,PS}, the assignment of geometric data can be visualised in terms
of the position of the peak in the Lie algebra \emph{and} in the group at the same time,
in a controlled way (\ie with a given finite spread around the peak). In this way, we can attach 
simultaneously intrinsic and extrinsic geometric data to each tetrahedron. 

We next turn to a classical discussion of the type of discrete geometries corresponding to GFT states, consisting of $N$ building blocks (simplices) with certain pre-geometric data, to identify the ones that are useful for cosmology.

\section{Approximate geometries and homogeneity} 
\label{approxgeo}

As said, the Fock vacuum of group field theory represents the no-space state, containing no geometry at all. In order to model a macroscopic geometry we need an excited state, \ie a state obtained from superpositions of states, possibly with a high occupation number. 
Among the various possibilities, we need to construct a class of states that is naturally adapted to
the concept of homogeneity. To understand what this entails, in this section we first focus on classical discrete geometries, characterised by the data appearing in the GFT Fock space (parallel transports and bivectors), like (\ref{state}). Our goal is to select those discrete geometries that correspond to macroscopic, spatially homogeneous metric geometries. Here we focus on metric rather than connection variables, but this not essential for the criterion of homogeneity which one could similarly give for the gravitational connection instead of the metric. 

The gauge invariance of the GFT field results, in the Lie algebra formulation,
in a multiplication by a (noncommutative) Dirac delta of the sum of the Lie algebra elements. Because of this (closure) constraint $\sum_I B_I = 0$, \eqref{state} is parametrized by $3N$ 
linearly independent
bivectors $\{B_{i(m)}\}$ $(i=1,2,3,\;m=1,\ldots,N)$. 
Furthermore, we imagine we can impose the simplicity constraints and ignore the discrete ambiguities in their solution; we then take the independent bivectors to be of the form $B_i^{AB}={\epsilon_i}^{jk}e_j^Ae_k^B$ for three $\bR^4$ vectors $e_i^A$.

On the space of bivectors, or alternatively the space of $e^A_{i(m)}$, there is an action of $\SO(4)
^N$ (or $\SO(3,1)^N$ in the Lorentzian case),
\ben
B_{i(m)}\mapsto \left(h_{(m)}\right)^{-1} B_{i(m)} h_{(m)}\,,\quad e_{i(m)}\mapsto e_{i(m)} h_{(m)}\, .
\label{lortraf}
\een
This group of transformations is a gauge symmetry of gravity, corresponding to local frame 
rotations.  

The resulting gauge-invariant configuration space for each tetrahedron is six-dimensional and may 
be parametrised by the quantities
\ben
g_{ij(m)} = e_{i(m)}^A\,e_{Aj(m)}
\label{metric}
\een
which can be interpreted, in light of the discrete geometric meaning of the variables $e$ and $B$, as defining discrete metric coefficients. We will shortly justify further this interpretation.
These metric coefficients can be expressed directly in terms of the bivectors alone:
\ben
g_{ij}=\frac{1}{8\,{\rm tr }(B_1 B_2 B_3)}{\epsilon_i}^{kl}{\epsilon_j}^{mn}\tilde{B}_{km}\tilde{B}_{ln}\,,\quad \tilde{B}_{ij}:=B^{AB}_{i}B_{j\,AB}\,.
\label{tetmetric}
\een
This formula bears some resemblance to the well known Urbantke metric, a spacetime metric constructed out of a triple of spacetime two-forms, but the two are not obviously related. 

The coefficients $g_{ij}$ are precisely the gauge-invariant data that we are interested in in the construction
of the states. Since the $g_{ij}$ have been constructed from the bivectors $B$ which are invariant under coordinate transformations (being integrals of a 2-form over the faces), they must be scalars under diffeomorphisms, and hence cannot be interpreted as metric coefficients in a coordinate basis. Instead, as we will see below, they are naturally interpreted as giving the metric in a given fixed frame that transforms covariantly under diffeomorphisms. Indeed, the $g_{ij}$ are the gauge-invariant content of states for a single tetrahedron; the remaining information can be understood as specifying a choice of local frame, removed by the condition of invariance under (\ref{lortraf}).

The next step is to relate classical discrete quantities given by $\{g_{ij(m)}\}$, for $1\le m\le N$, with continuum geometries. The problem
of reconstructing continuum geometries from discrete data has been discussed several times
in the past and remains, in its full generality, largely open. In particular, for LQG, which uses the variables that we are manipulating here, see
\cite{ThiemannCS,OPS,PS,Corichi,collective}. 

Such discrete geometries can be seen as a sampling of a continuous geometry at $N$ different points. Furthermore, given the interpretation of the geometric data, this geometry is only a spatial slice of a four-dimensional manifold (information about the extrinsic curvature, and hence the embedding of the slice, is given by the conjugate connection variables).
Clearly, without further instructions, it is impossible to associate to a finite set of numbers (\ref{metric}) a unique smooth
continuous geometry.

We are interested here in those spatial geometries that correspond to 
three-dimensional homogeneous spaces. We must be able to characterise a certain discrete geometry given by $\{g_{ij(m)}\}$ in such a way that it is clear whether it is \emph{compatible} with a homogeneous
spatial geometry or not. In order to do this we need to construct an embedding of the discrete geometry,  such that a comparison with a homogeneous continuum geometry is possible. 

Let us recall, first, that homogeneity of a Riemannian geometry on $\M$ is defined with respect to a Lie group of isometries $\GG$, acting transitively on $\M$. 
In fact, homogeneous manifolds can be classified in terms of their isometry group $\GG$ \cite{cosmobook}.
For our purposes, we assume that $\M \simeq \GG/X$, where $X\subset \GG$ is a discrete subgroup of a three-dimensional Lie group $\GG$. 

We proceed as follows. Each of the tetrahedra (corresponding to a GFT quantum) can be embedded
into a three-dimensional topological manifold $\M$, our spatial slice. 
The embedding of
each tetrahedron requires a sufficiently smooth function, mapping each point of the 
tetrahedron (boundary
and bulk) to our manifold.
Given the structure of the tetrahedron, the embedding can be completely determined once we
specify the point $x_{m}\in \M$ at which one of the vertices is embedded, and the three tangent vectors $\mathbf{v}_{i(m)} \in T_{x_{m}}\M$
emanating from it, corresponding to the three edges incident at the given vertex. 
The exponential
map naturally defined by the Maurer--Cartan connection on $\GG$, pulled back on $\M$, allows us to
embed the whole tetrahedron, since any point in it can be represented in terms of a linear combination of the three independent vectors $\mathbf{v}$ which is then exponentiated\footnote{We stress that we do not have to make reference to any notion of sprinkling of points in a given manifold by using a Poisson process, as it is customarily done in these contexts, \eg for discrete geometries with LQG-type data \cite{collective}, or in the causal set approach \cite{causal}. Defining a Poisson process requires a choice of measure, associating a volume to a given region, and while the group action of $\GG$ on $\M$ provides a natural measure (fix a volume form at one point $x\in\M$ and define it everywhere else by the pull-back of the group action) that could be used, we regard this measure as a fiducial background structure: only the dynamical geometric variables derived from (\ref{tetmetric}), such as the determinant of $g_{ij}$, should be used to make statements about densities and volumes. We hence consider arbitrary embeddings, while ensuring that our statements about spatial homogeneity do not depend on this arbitrary choice. Once this is guaranteed, everything we say will also hold if one chooses to restrict to sprinklings which are a subset of all possible embeddings.}.

With the construction described above, we are able to interpret data of the form $\{g_{ij(m)}\}$ for $1\le m\le N$ in terms of a discrete sampling of continuum geometric data defined on $\M$, in correspondence with the embedded tetrahedra
\ben
\tetrahedron_m\,\mapsto\,\left\{x_m \in \M,\;\left\{{\bf v}_{1(m)},{\bf v}_{2(m)},{\bf v}_{3(m)}\right\}\subset T_{x_m}\M\right\}\,.
\een

The exact translation of the data associated to each tetrahedron (bivectors, tetrads, or parallel transports) to continuum fields on the manifold $\M$ requires however some extra care. We want to interpret all the above pre-geometric quantities  as resulting from continuum geometric fields integrated over domains of finite size. Furthermore, being gauge variant, such integrations require the use of appropriate parallel transports with the continuum connection (see for example the construction in \cite{collective} and references therein).

For example, in order to reconstruct an approximate tetrad, and hence the metric, as it is induced on $\M$ by
the embedding of the tetrahedra, we need to assume that the associated \emph{reconstructed}
curvature is small over the size of the same tetrahedra, so that we can approximate their interior as flat and regard the needed parallel transports as acting trivially. More precisely, the linear size of each tetrahedron has to be much smaller than
different possible curvature radii inferred using the reconstructed metric and connection on $\M$. This is a condition on our procedure to be \emph{self-consistent} in its geometric interpretation. 

This approximation
is the first source of error in our discussion which has to be taken into account, especially
in the construction of an effective dynamics. Indeed, an effective dynamics predicting a regime of very high curvature compared to the scale of the embedded tetrahedra would make the above geometric interpretation less reliable, as the approximation on which it hinges breaks down. In this case 
 the corresponding state could not be trusted to have such simple geometric interpretation, and will have to be replaced with a better
guess, or be reanalysed in terms of a more subtle geometric reconstruction procedure and interpretation (see also the related work on loop quantum cosmology \cite{LQC}). 

Now assuming that the approximation of near-flatness holds, 
we interpret the $\bR^{4}$ vectors $e^A_{i(m)}$ associated to a tetrahedron as physical tetrad vectors integrated along the edges specified by ${\bf v}_{i(m)}$, a natural choice for which is  
a basis of left-invariant vector fields on $\GG$:
$
{\bf v}_{i(m)} = {\bf e}_i(x_m),
$
where $\{{\bf e}_i\}$ are the vector fields on $\M$ obtained by push-forward of a basis of left-invariant vector fields on $\GG$. 
Fixing a $\GG$-invariant inner product in the Lie algebra ${\bf g}$ of $\mathfrak{G}$, such a basis is unique up to a global action of $\O(3)$. 

If we did not make such a choice but left the vectors ${\bf v}_{i(m)}$ unspecified, even two embeddings in which all tetrahedra are embedded at the same points $x_m$, but with different tangent vectors ${\bf v}_{i(m)}$, would lead to physically distinct reconstructed metrics, as a local $\GL(3)$ transformation on $g_{ij}$ in general cannot be undone by a diffeomorphism. We could have chosen a different set of vector fields and assume that all tetrahedra are embedded with such tangent vectors, of course, but the existence of a group action on space $\M$ provides us with a natural, canonical way of fixing ${\bf v}_{i(m)}$, avoiding such issues: the tetrahedra are always oriented along the local frame given by the left-invariant vector fields. 

Within the approximation of near-flatness, we can approximate the integral of the 1-form $e^A$ representing the physical tetrad over an edge by its value at the point $x_m$. This implies we assume the edges to be of unit coordinate length, which is a statement about the coordinate system we are expressing the metric in. Some choice of this sort is always required when passing from diffeomorphism-invariant to diffeomorphism-variant quantities. We then have the following relation between the vectors $e^A_{i(m)}$ and the physical tetrad:
\ben
e^A_{i(m)}=e^A(x_m)({\bf e}_i(x_m))\,.
\een
For the gauge-invariant quantities $g_{ij}$, this implies that
\ben
g_{ij(m)}=g(x_m)({\bf e}_i(x_m),{\bf e}_j(x_m))\,,
\label{physmet}
\een
and thus $g_{ij(m)}$ are the metric components in the frame of the left-invariant vector fields $\{{\bf e}_i\}$. 

Clearly, if the spatial geometry was homogeneous, these coefficients would be constant in space. We are interested in the converse question: given the coefficients $g_{ij(m)}$, is the underlying metric geometry compatible with spatial homogeneity? At this stage, any positive answer to this question can only hold in the approximate sense where one looks at only $N$ points in the manifold. This is indeed our second main source of approximation. The larger $N$ is, the more confident we can be of the association between our discrete data and a continuum geometry.

Within this approximation, the criterion for which a
state like \eqref{state} or \eqref{statewf} can be interpreted as a discrete $N$-point sampling
of a homogeneous geometry $\M$ is clear. With the embedding discussed above,
making use of the action of the group $\GG$ on $\M$,
we can say that the state is {\em compatible with spatial homogeneity} if
\ben
g_{ij(m)} = \bar{g}_{ij}, \qquad \forall m=1,\ldots N.
\label{homocrit}
\een

Again we should stress that we focussed the discussion only on the intrinsic geometry 
(the three-dimensional metric) but a perfectly analogous discussion holds for the connection (which includes, in the Ashtekar formulation, the extrinsic curvature), or in fact for any other field, such as matter degrees of freedom added to the GFT configuration space as additional data characterising the elementary tetrahedra (see Sec.~\ref{beyond})\footnote{Note also that we had only to specify a homogeneity criterion for the gauge-invariant quantities $g_{ij}$ because we had previously fixed a unique reference frame for all our tetrahedra as part of our embedding and reconstruction procedure. Had we not done this, an additional criterion for homogeneity would have been to require exactly that the local frame in which the metric quantities were expressed was the same for all tetrahedra.}. For a scalar field, the criterion of homogeneity would simply be that its value be the same for all tetrahedra; fields with tensor indices would be interpreted as given in the frame of left-invariant vector fields, so that a criterion analogous to (\ref{homocrit}) results. In the next section, we will also lift this homogeneity criterion to the quantum setting, as a condition of wavefunctions for quantum states, and thus to probability distributions over the space of discrete geometric data. 

\

As we have said, the procedure is subject to some self-consistency conditions. First, one has to
ensure that the flatness condition (small curvature with respect to the size of the tetrahedra)
is satisfied. Second, while in principle the group $\GG$ is unspecified and its choice provides an additional input, the effective dynamics coming from a particular GFT model can provide conditions on the possible consistent choices for $\GG$: we shall see later on that, for a special choice of GFT action and quantum state representing an isotropic homogeneous geometry, the semiclassical regime of the effective dynamics corresponds to a positively curved 3-geometry, which suggests that $\GG=\SU(2)$ for consistency.

In general the reconstructed continuum geometries can be arbitrary anisotropic  homogeneous metrics, corresponding to all possible Bianchi types. It is a dynamical question whether an approximately isotropic geometry emerges from the GFT dynamics, just like in classical general relativity.

\

Let us summarise the conditions on GFT states to have an interpretation as describing macroscopic homogeneous spatial geometries. The first is the criterion of {\em homogeneity}, which at the classical level corresponds to the condition (\ref{homocrit}), \ie the microscopic geometric data must be the same for all tetrahedra. This is our primary motivation for considering {\em condensate} states, characterised by just a single macroscopic `wavefunction' for many elementary building blocks. The second condition is that of {\em near-flatness} of the elementary tetrahedra: the components of the curvature, given by appropriate gauge-covariant combinations of elementary parallel transports, must be small, \ie close to the identity in the gauge group ($\SO(4)$ in what we have considered so far -- in the GFT formalism one usually considers the universal covering group, $\Sp(4)$ in this case). In the quantum theory this must be phrased in terms of expectation values: the condensate wavefunction must be peaked around small values of elementary curvatures (recall that these represent geometric quantities integrated over the size of the tetrahedra). The third condition is that the sampling size $N$ should be reasonably large, and the larger the better the approximation of the continuum. The fourth condition is {\em semiclassicality}: at least in some regime, in order to be able to speak about a classical universe emerging from a given quantum state, the state must have semiclassical properties. The standard choice would be to use coherent states \cite{ThiemannCS,OPS,Corichi,collective}, as done for example in \cite{meLorenzo}. In the context of this paper, we will extract semiclassical physics from the quantum dynamics by means of a WKB approximation; the condition is then to be in a regime where this approximation is valid. This is a standard procedure in quantum cosmology to discuss spacetime histories emerging from a quantum state, see \eg \cite{hhh}.\footnote{Notice that, along the way from the microscopic description of the quantum spacetime degrees of freedom to the effective macroscopic cosmological ones, the semiclassical approximation can be taken at various points. The specific stage at which one takes it will in general affect the result. In our procedure, since the notion of quantum condensate is needed to obtain a cosmological approximation, the semiclassical approximation cannot be taken until the very end. In fact, we will obtain quantum cosmology-like effective equations, and it is on these final equations (better, on the corresponding cosmological wavefunction, here arising as the collective variable for the GFT condensate) that we will use the semiclassical approximation.} The problem of emergence of classical physics from a quantum theory is of course a major one \cite{emergbook}, which we do not really tackle in its generality in this paper.


\section{GFT condensates as continuum homogeneous geometries}
\label{condensec}

In this section we will describe continuum homogeneous geometries as GFT states, lifting to the quantum level the classical considerations of the previous section. We work first in the Riemannian context where the gauge group is $\Sp(4)$. The immediate quantum version of the homogeneity criterion (\ref{homocrit}) is easy to identify. Working in the Lie algebra representation, for example, thus using $B$ variables to label our states (assumed to satisfy simplicity conditions, so to be interpretable in geometric terms), as in (\ref{statewf}), one can consider special states of the product type
\ben
|\Psi_N \rangle := \frac{1}{N!}\left( 
\int (\dd B)^4
\Psi(B_{1},\ldots,B_{4})\star
 \hat{\tilde{\varphi}}^{\dagger}(B_{1},\ldots,B_{4})\right)^N|0\rangle\,,
\een 
containing $N$ quanta each associated to the same wavefunction. These can be given an interpretation in terms of discrete 
geometries that naturally approximate homogeneous (and possibly anisotropic) spatial slices. The classical homogeneity criterion (\ref{homocrit}) of identical geometric data for all tetrahedra (in the appropriate local frame) becomes, at the quantum level, the requirement of {\it identical distribution over the space of geometric data} for all tetrahedra.
Notice that this requirement does not refer to the shape of the wavefunction $\Psi(B_{1},\ldots, B_{4})$, or to whether it represents a semiclassical state (\eg a heat kernel) or not (\eg a Dirac-delta-like distribution in the Lie algebra variables): at the end of Sec.~\ref{approxgeo}, we distinguished the independent conditions of {\em homogeneity} and {\em semiclassicality}. We will discuss specific conditions on the choice of wavefunction $\Psi$ within the more general class of condensate states in Sec.~\ref{selfc}. 
 
\
 
Again, there are two sources of approximations.
First, the flatness condition, \ie the smallness of the ratio between the size of the tetrahedra
and the radius of curvature. Second, the value of $N$, giving the size of the sampling. We ignore the condition of semiclassicality for this discussion, focussing on the GFT quantum dynamics first.

Having a Fock space structure for our states gives a straightforward way to go beyond a formalism in which the number of quanta is fixed, and consider
superpositions of states with different particle numbers.
In addition, the structure of the Fock space allows us also to take states with an infinite number of particles, \eg
\ben
|\Psi\rangle := \sum_{N=0}^{\infty}\frac{c_N}{N!}\left( 
\int (\dd B)^4
\Psi(B_{1},\ldots,B_{4})\star
 \hat{\tilde{\varphi}}^{\dagger}(B_{1},\ldots,B_{4})\right)^N|0\rangle\,.
\label{statehom}
\een 

\

To summarise, we assume that the underlying continuum geometry describes a compact region in space, which can be all of space if $\M$ is compact, as will be the case later on (where $\M$ is a three-sphere). One can associate a length scale $L$ to this region, \eg by setting $L=V^{1/3}$ where $V$ is the total volume with respect to some arbitrary fixed measure (such as the left-invariant measure on $\M$ mentioned before). Then, probing the geometry of this region by $N$ tetrahedra at $N$ different points can be understood as a restriction to wavelengths longer than $L/N^{1/3}$ in the reconstructed geometry. Sending $N$ to infinity, we take this approximation scale associated to the discrete sampling to zero. In this way,
a second-quantised state will allow us to approximate continuous discrete geometries, with the
only source of error encoded in the flatness condition. Since this condition refers only to the 
discrepancy between the continuum geometric quantities and the discrete ones, we can
say that sending $N$ to infinity will allow us to recover homogeneity to arbitrary accuracy.

\

We will call these states {\it GFT condensates}, since they correspond to particular second-quantised states having macroscopic occupation numbers for given modes, controlled
by the wavefunction $\Psi$, thus following the standard terminology used in condensed matter theory \cite{BEC}. 

\

The above definition of condensate states can be generalised. Indeed, there are many ways to construct states that, as many-body states, can be
interpreted as a condensate of a given building block. Besides the obvious freedom to 
choose the coefficients $c_N$ in \eqref{statehom},
one can imagine to consider more general states that include correlations between particles.
One possibility is to consider states of the form
\ben
|\Psi\rangle := \sum_{N=0}^{\infty}\frac{c_N}{N!}\left( 
\int (\dd B)^4 (\dd B')^4
\Psi(B_{1},\ldots,B_{4},B'_{1},\ldots,B'_{4})\star
 \hat{\tilde{\varphi}}^{\dagger}(B_I)
  \hat{\tilde{\varphi}}^{\dagger}(B'_I)
 \right)^N|0\rangle \, .
\label{statebog}
\een 
Such states are used, for instance, in the discussion of the Bogoliubov approximation
of the dynamics of Bose--Einstein condensates \cite{BEC}.

In principle, nothing prevents us from considering states that are built out of larger
elementary building blocks, \ie states that are encoding correlations among a larger
and larger number of quanta, or even combinations of them, and possibly depending on a slightly larger set of pre-geometric data. These may not be exactly homogeneous geometries according to our simple criterion (\ref{homocrit}), but could be more physically appropriate to describe approximately homogeneous (and thus more realistic) geometries or correspond to the outcome of a more refined reconstruction procedure. The information about
the best state that encodes the appropriate physical properties that our system, a macroscopic homogeneous universe, possesses, should come from elsewhere. For
instance, a better control on the properties of the GFT phase transition leading to condensation of the GFT quanta might give
hints. 

\ 

In this paper, we will focus on just two of these possible states. These are the simplest
possible choices that will allow us to work with states that contain the idea of sampling
a continuous homogeneous geometry in the sense specified in the previous section.
Therefore, for our purposes, the only restriction that we are going to impose
is that the state contains exactly the geometric data of a tetrahedron. Consequently,
besides the gauge invariance of GFT fields, it has to include the gauge symmetry
\eqref{lortraf}.

We model the states after coherent states for single-particle modes and for pairs, constraining
the coefficients $c_{N}$ appearing in \eqref{statehom} and \eqref{statebog} to define exponential operators (giving the desired coherence properties), and reducing the freedom in the definition of the state to the choice of a single function. 

The simplest class of states is a \lq single-particle\rq \, condensate,
\ben\label{simple}
|\sigma\rangle := \mathcal{N}(\sigma) \exp\left(\hat\sigma\right)|0\rangle \quad\text{with}\quad \hat\sigma := \int (\dd g)^4\; \sigma(g_1,\ldots,g_4)\hat\varphi^{\dagger}(g_1,\ldots,g_4) \,
\een
where we require $\sigma(k g_1 ,\ldots,k g_4 )=\sigma(g_1,\ldots,g_4)\, \forall k\in \Sp(4)$, and $\mathcal{N}(\sigma)$
is a normalisation factor,
\ben
\mathcal{N}(\sigma):=\exp\left(-\half\int(\dd g)^4\;|\sigma(g_1,\ldots,g_4)|^2\right)\,.
\een
At least for the `single-particle' condensate, the requirement of normalisability, \ie of obtaining an element of the GFT Fock space, is equivalent to a finite expectation value for the number operator. Hence, while involving a superposition with states of arbitrarily high particle number, the condensate always has a finite average number of GFT quanta. It is only in the first sense that the limit of infinite particle number $N\rightarrow\infty$ is taken.

Despite the fact that, strictly speaking, these states do not correspond to a system with infinite number of particles, for sufficiently large particle number, acting on them with creation and annihilation operators (that is, adding and removing a relatively small number of quanta) does not change their shape in an essential way. In this sense they capture part of the thermodynamic limit, while still allowing the use of the original Fock space.
This is the reason why coherent states are extensively used to describe weakly interacting Bose--Einstein condensates, where the number of atoms will be finite for any particular condensate state
(as dictated by the fact that the system has a large but finite size, typically of $10^3$ bosons), although the formal thermodynamic limit $N\rightarrow\infty$ is often useful as an approximation.

Since (\ref{firstgauge}) imposes automatically that
$\sigma( g_1 k',\ldots,g_4 k')=\sigma(g_1,\ldots,g_4)\,\forall k'\in \Sp(4)$, we have two restrictions
telling us that the function $\sigma$ effectively depends on less arguments.
Defining the Fourier transform of $\sigma$,
\ben
\tilde{\sigma}(B_1,\ldots,B_4) = \int (\dd g)^4 \prod_{I=1}^4 e_{g_{I}}(B_{I}) \sigma(g_1,\ldots,g_4),
\een
the invariances that we impose imply that first $\tilde{\sigma}(B_1,\ldots,B_4) = e_{k}\left(\sum_I B_I\right)\star \tilde{\sigma}(B_1,\ldots,B_4)$ so that integrating over $k$ we obtain
\ben
\tilde{\sigma}(B_1,\ldots,B_4) = \delta_{\star}\left(\sum_{I} B_{I}\right)\star\tilde{\sigma}(B_1,\ldots,B_4)
\een
and the closure constraint is satisfied; from $\sigma(g_1,\ldots,g_4)=\sigma(k^{-1}g_1k,\ldots,k^{-1}g_4k)$ we obtain
\bena
\tilde{\sigma}(B_1,\ldots,B_4) & = & \int (\dd g)^4 \prod_{I=1}^4 e_{k^{-1}g_{I}k}(B_{I}) \sigma(g_1,\ldots,g_4)\nonumber
\\&=& \int (\dd g)^4 \prod_{I=1}^4 e_{g_{I}}(kB_{I}k^{-1}) \sigma(g_1,\ldots,g_4)\nonumber
\\&=& \tilde{\sigma}(kB_1k^{-1},\ldots,kB_4k^{-1})
\eena
which takes care of (\ref{lortraf}). Here we have used some elementary properties of the plane waves $e_g$ and of the corresponding non-commutative Fourier transform \cite{danielearistide, carlosdanielematti}.

We see that the wavefunction $\sigma$ stores exactly and only the gauge-invariant data needed
to reconstruct the metric from the bivectors, as in \eqref{tetmetric}. This is indeed the simplest choice of quantum states that possesses all the properties we identified as corresponding to continuum quantum homogeneous geometries.

\

The second class of states that we are considering is
\bena
|\xi\rangle &:=& \mathcal{N}(\xi) \exp\left(\hat\xi\right)|0\rangle \quad \text{with} \\  \quad \hat\xi &:=& 
\half\int (\dd g)^4 (\dd h)^4\; \xi(g_1^{-1} h_1,\ldots,g_4^{-1} h_4)\hat\varphi^{\dagger}(g_1,\ldots,g_4)\hat\varphi^{\dagger}(h_1,\ldots,h_4)\, ,
\label{dipolestate}
\eena
where, thanks to (\ref{firstgauge}) and $[\hat\varphi^{\dagger}(g_I),\hat\varphi^{\dagger}(h_I)]=0$, the function $\xi$ \emph{automatically} satisfies $\xi(g_I)=\xi(kg_Ik') \; \forall k,k' \in \Sp(4)$ and $\xi(g_I)=\xi(g_I^{-1})$, and $\mathcal{N}(\xi)$ is a normalisation factor ensuring the state $|\xi\rangle$ has unit norm in the Fock space. Using the fundamental commutation relations and a bit of combinatorics, one can show that 
\ben
\langle 0|\exp(\hat\xi^{\dagger})\exp(\hat\xi)|0\rangle = \exp\left(\sum_{k\ge 1} \frac{1}{2k}\langle |\xi|^{2k} \rangle\right)=1+\half \langle |\xi|^{2} \rangle + \frac{1}{8}\langle |\xi|^{2} \rangle^2 + \frac{1}{4} \langle |\xi|^{4} \rangle +\ldots
\label{norm}
\een
where
\ben
\langle|\xi|^{2k}\rangle := \int (\dd g)^{4k} (\dd h)^{4k} \prod_{p=1}^k\overline{\xi(g_{4p-3}^{-1} h_{4p-3},\ldots,g_{4p}^{-1} h_{4p})}\,\xi(h_{4p-3}^{-1} g_{4p+1},\ldots,h_{4p}^{-1} g_{4p+4})
\een
and for $p=k$ in the product it is understood that $g_{4k+i}=g_i$ in the arguments of $\xi$. In order for the state $\exp(\hat\xi)|0\rangle$ to be in the Fock space, these moments of the profile function $\xi$ must go to zero fast enough so that the argument of the exponential in (\ref{norm}) is finite, $\sum\frac{1}{2k}\langle |\xi|^{2k} \rangle<\infty$, which is expected to be a rather strong constraint. If it is satisfied, we can set
\ben
\mathcal{N}(\xi) := \exp\left(-\sum_{k\ge 1} \frac{1}{4k}\langle |\xi|^{2k} \rangle\right)\,.
\label{normcons}
\een

Following recent work  \cite{dipoleLQC}, we call
$\xi$ a \lq dipole\rq ~function on the gauge-invariant configuration space of a single tetrahedron.
This second class of states, while possessing the same gauge invariance and the same geometric 
data as \eqref{simple}, provides two kinds of  improvements:
first, invariance under \eqref{lortraf} is imposed in a natural way, without any further external restriction. 
Second, the state encodes some very simple two-particle correlations, a feature that should be
expected to be necessary in the true vacuum state of the system, due to its highly interacting nature. 

It would be straightforward to define, along the same lines, condensate states whose elementary building blocks are more complicated multi-particle states in the GFT Fock space. For instance, in similar contexts in discrete geometry one often thinks of cubical graphs, where the elementary cell is a ``cube'' that can be thought of as composed of several tetrahedra. Most notably, recent work in loop quantum gravity \cite{alescicianfrani} aims at deriving cosmological dynamics from LQG using cubulations of space. A condensate of cubes would represent the closest analogue to such a construction in our setting, and could be used to compare our approach with the work of \cite{alescicianfrani} in more detail.

According to our previous analysis, the GFT condensate states $|\sigma\rangle$ and $|\xi\rangle$ encode continuous homogeneous (but possibly anisotropic) quantum geometries.
The distribution of geometric data is encoded in the functions $\sigma$ or $\xi$, which can be seen as functions over 
the minisuperspace of homogeneous geometries. Let us stress at this point that there is a
difference with the standard minisuperspace approach. These states are not states of a
symmetry reduced theory. Rather, they are symmetry reduced states of the full theory.
Furthermore, given the sum over samplings, they are independent of a chosen reference
lattice structure, or fixed discretisation of space, and in particular they support arbitrary perturbations over homogeneous geometries.

Despite being general enough to encode all
the Bianchi cosmologies, the above states are simple enough to lead to explicit
calculations and allow the extraction of an effective cosmological dynamics, as we will see in the following. 

\subsection{Self-consistency conditions}
\label{selfc}

Before moving on to the extraction of the effective cosmological dynamics, let us make a bit more precise the approximations mentioned above that are necessary for the geometric interpretation of the above states, according to our simple reconstruction procedure. A detailed analysis of the geometric conditions that can or should be imposed on our quantum states to ensure the correct geometric interpretation is left for future work. This should entail a refined version of the reconstruction procedure we outlined in Sec.~\ref{approxgeo}, possibly reproducing and extending to the full quantum theory (here, in its GFT formulation), the kinematical set-up of loop quantum cosmology \cite{LQC}. The same reconstruction procedure should be generalised beyond the homogeneous case, using previous results on semiclassical quantum geometry states \cite{ThiemannCS,OPS,PS,Corichi,collective}. Also, the geometric conditions so identified should then be incorporated into the quantum theory at the dynamical level, \ie as a choice of quantum ensemble being considered, and thus in the fundamental dynamics of the theory. Here, we limit ourselves to a basic discussion of the form that such geometric condition should take, for the class of states we consider, in a second quantised formulation. 

Following the arguments of Sec.~\ref{approxgeo}, we expect one of the first sources of error for
the effective theory that we are going to derive from these states to consist in the discretisation error associated to the approximation of a continuum manifold in terms of (a large number of) discrete constituents. 
Once the states have been specified, this error can be expressed in terms of
expectation values of certain operators in the given state. For simplicity,
we will consider just the simple condensates \eqref{simple}.

The number of tetrahedra contained in the state can be obtained as the expectation
value of the one-body operator
\begin{equation}
\hat{N}_{\tetrahedron} = \int (\dd g)^4\; \hat{\varphi}^{\dagger}(g_1,\ldots,g_4)
\hat{\varphi}(g_1,\ldots,g_4)\, ,
\end{equation}
which is
\begin{equation}
N_{\tetrahedron} = \langle\sigma|\hat{N}_{\tetrahedron}|\sigma\rangle = \int (\dd g)^4\; |\sigma(g_1,\ldots,g_4)|^2 \, .
\end{equation}

Similar one-body operators can be used to extract other geometric observables. For instance, the \emph{total} volume encoded by the state is
\bena
V_{{\rm tot}} &=& \int (\dd g)^4
(\dd \tilde{g})^4
 \;\bra{\sigma} \hat{\varphi}^{\dagger}(\tilde{g}_1,\ldots,\tilde{g}_4)
{V}(\tilde{g}_1,\ldots,\tilde{g}_4;g_1,\ldots,g_4)
\hat{\varphi}(g_1,\ldots,g_4)
 \ket{\sigma}
 \nonumber\\&=& \int (\dd g)^4
 (\dd \tilde{g})^4
 \;\overline{\sigma}(\tilde{g}_1,\ldots,\tilde{g}_4)
{V}(\tilde{g}_1,\ldots,\tilde{g}_4;g_1,\ldots,g_4)
\sigma(g_1,\ldots,g_4)
\eena
where, for $G=\SU(2)$, ${V}(\tilde{g}_1,\ldots,\tilde{g}_4;g_1,\ldots,g_4)$ is the matrix element of the 
volume operator in LQG between two spin network nodes with geometric
data specified by $\tilde{g}_1,\ldots,\tilde{g}_4$ and $g_{1},\ldots, g_{4}$. Its form can be obtained
from a Peter--Weyl decomposition into the familiar representations in terms of spins. As we have already said, we have to ask that the volume of each tetrahedron
is small compared to the total volume of the spatial slice captured by the state,
\begin{equation}
\frac{V_{\tetrahedron}}{V_{{\rm tot}}} := \frac{1}{N_{\tetrahedron}} \ll 1 \, .
\end{equation}
The volume of a tetrahedron defines the typical scale at which we are probing geometry,
\begin{equation}
L \sim \left(V_{\tetrahedron} \right)^{1/3}\, .
\end{equation}
This scale uses only the geometric information encoded in the quantum state, and is not fixed externally. These very simple considerations are 
consistent with the intuition that the continuum limit can be seen as a thermodynamic
limit for this gas of tetrahedra. At this stage it is not clear what $L$ should be, or how it is related to fundamental (Planckian) units. To understand physically how a particular scale $L$ emerges, one must use the GFT dynamics. One possibility is that a condensation of tetrahedra of the type proposed only occurs within a certain range of values for $L$, which can be derived from the theory, instead of being put into the definition of the states. One can also try to constrain $L$ by changing the type of ensemble one is considering, \eg by adding a thermodynamic variable conjugate to the extensive quantity `volume'. Understanding the precise nature of these constraints is crucial for clarifying the relation between the microscopic GFT dynamics and the resulting effective cosmological scenario and goes together with a more detailed analysis of the kinematical reconstruction procedure to be applied to our quantum states.

\

Another condition that is needed to ensure that the treatment is self-consistent comes
from the fact that the tetrahedra have to be close to flat. Therefore, the curvature of the 
connection that they encode (including the extrinsic curvature) has to be
small on the scale of the tetrahedra themselves. 

Information about the connection can be extracted from expectation values
of suitable operators, for instance
\bena
\chi_{{\rm tot}}[\sigma] &=& \int (\dd g)^4\;
\bra{\sigma} \hat{\varphi}^{\dagger}(g_1,\ldots,g_4)
\chi(g_1,\ldots,g_4)
\hat{\varphi}(g_1,\ldots,g_4)
 \ket{\sigma}\nonumber\\
& = &\int (\dd g)^4\;
|\sigma(g_1,\ldots,g_4)|^2\,
\chi(g_1,\ldots,g_4)\,,
\label{chieq}
\eena
where $\chi$ can be seen as a character of a suitable product of group elements. In the case of $G=\SU(2)$, for instance, taking $\chi$ to be the trace in the $j=\half$ representation, the functions $\chi_{ij}:=\chi(g_ig_4^{-1}g_jg_4^{-1})$ give a complete set of functions on $\SU(2)^4$ that are invariant under $g_I\mapsto g_Ik$ and $g_I\mapsto k'g_I$, which motivates their identification with components of the curvature for a single tetrahedron. In order to measure correlations between different tetrahedra, one has to go beyond using a one-body operator as in (\ref{chieq}). The interpretation of $\chi_{{\rm tot}}[\sigma]$ is that of a sum of curvature components (defined by $\chi$) for all tetrahedra in the condensate. In general, our flatness condition then can be stated as a condition on the deviation of
the curvature expectation values per tetrahedron from their values
at the identity,
\begin{equation}
\frac{\chi_{\tetrahedron}[\sigma] }{\chi(e)} - 1 = \frac{1}{\chi(e)}
\frac{\chi_{{\rm tot}}[\sigma]}{N_{\tetrahedron}}-1
\ll1 \,.
\end{equation}
In terms of the scales defined by the tetrahedron, these conditions are
nothing else than the observation that the curvature scale $L_{c}$ should
be much larger than the typical length scale of the tetrahedron $L$, $L/L_{c} \ll1 $.

In fact, this condition might involve
not only a statement about the expectation values of connection-related operators,
but also conditions on the fluctuations around the mean values. It is clear that if
the fluctuations around the flat connection are too large, the flatness
requirement is not satisfied. One can then expect that the relevant states should be
rather peaked in the connection variables, albeit not necessarily semiclassical
in the Lie algebra variables.

This reasoning can be exported to the case of more general states, with
the appropriate modifications. They will be expressed as restrictions on
certain functionals of the wavefunctions (\eg $\sigma$ or $\xi$) that are used in
the parametrisation of the states.
In turn, these functions are determined dynamically by the equations of motion.
In particular, since these equations are nonlinear, the average number of tetrahedra
$N_{\tetrahedron}$ in the state cannot be tuned by hand.
Therefore, these conditions represent truly nontrivial constraints to be imposed on
the resulting effective dynamics, for its continuum geometric interpretation to be trusted. They contain information about the dynamics of the condensate as a whole that cannot be captured by properties of the individual tetrahedra alone.

\subsection{GFT condensates vs. coherent and squeezed states}
\label{squeezedsec}
Next, it is worth stressing some further properties of the quantum states (\ref{simple}) and (\ref{dipolestate}), especially in their relationship with
states commonly used in quantum optics and in the physics of quantum fluids.

\

The states of the form \eqref{simple} are coherent states, \ie eigenstates of the 
field annihilation operator, as a straightforward calculation shows. Therefore, they
represent a natural class of states for a sort of Hartree--Fock or mean field approximation
in which the GFT field acquires a nontrivial vacuum expectation value,
\ben
\hat{\varphi}(g_{I})| \sigma \rangle ={\sigma}(g_{I})| \sigma \rangle .
\een
As said, this state does not encode multiparticle correlations; it gives rise to
correlation functions that are products and convolutions of a single one-point
correlation function, the mean field.

\

The second class of states, dipole condensates, are coherent states
only in a rough sense.  
In fact, the states $|\xi\rangle$ are more similar to squeezed states \cite{squeeze}.

For a single mode, a squeezed state
is defined as
\ben
|w\rangle = \hat{S}(w)| 0 \rangle,\qquad \hat{S}(w) = \exp\left(\frac{w}{2} \hat{a}^{\dagger}\hat{a}^{\dagger}
-\frac{\overline{w}}{2}\hat{a}\hat{a}
\right)
\een
where $w$ is a complex number, $\hat{a},\hat{a}^{\dagger}$ are ladder operators and the unitary operator $\hat{S}(w)$ is the so-called squeezing operator.
It follows from the definitions that
\ben
\exp\left(\frac{z}{2}\hat{a}^{\dagger}\hat{a}^{\dagger}\right)|0\rangle \propto  |f(z)\rangle\,, \qquad
f(z) = - \frac{z}{|z|} \sinh^{-1}\left( \frac{|z|}{\sqrt{1-|z|^2}}\right) \,.
\label{4.16}
\end{equation}
Indeed, using the properties of the ladder operators, one sees that
\begin{equation}
(\hat{a} - z \hat{a}^{\dagger})\exp\left(\frac{z}{2}\hat{a}^{\dagger}\hat{a}^{\dagger}\right)|0\rangle = 0\,.
\end{equation}
With a simple rescaling we can complete the Bogoliubov transformation and define the ladder operator
\begin{equation}
\hat{b} =\frac{ \hat{a} - z \hat{a}^{\dagger}}{\sqrt{1-|z|^2}},
\label{bogoliubov}
\end{equation}
provided that\footnote{We mention, for completeness, that in the case of $|z|=1$ the two would-be ladder operators commute, while for $|z|^2>1$ the role of annihilation and creation operators is exchanged.} $|z|^2 < 1$.
The state $\exp\left(\frac{z}{2}\hat{a}^{\dagger}\hat{a}^{\dagger}\right)|0\rangle$ is annihilated by it, and hence it is proportional
to the corresponding Fock vacuum.
A general Bogoliubov transformation can be expressed as
\begin{equation}
\hat{b} = S(w) \hat{a} S(w)^{\dagger} = \cosh(|w|) \hat{a} + \frac{w}{|w|} \sinh(|w|) \hat{a}^{\dagger}\,;
\label{genbogoliubov}
\end{equation}
comparing \eqref{bogoliubov} and \eqref{genbogoliubov} we get the desired \eqref{4.16}.

We would like to show that our states \eqref{dipolestate} are squeezed states, \ie to write down squeezing operators that correspond to a Bogoliubov transformation of the ladder operators, presumably requiring appropriate conditions on $\xi$. A proof of such a statement, however,  is not straightforward at all. An alternative path is to use the characterisation of squeezed
states as Fock vacua of Bogoliubov rotated annihilation operators, and to show that these states are
annihilated by an appropriate linear combination of ladder operators $\hat{\varphi},\hat{\varphi}^{\dagger}$, corresponding to another annihilation operator. It is easy to see that
\ben
[\hat{\varphi}(g_{I}),\hat{\xi}] = \int (\dd h)^4\;\xi(g_1^{-1} h_1,\ldots,g_4^{-1} h_4) \hat{\varphi}^{\dagger}(h_1,\ldots,h_4) 
=:\hat{\xi}_{g_{I}}\,,
\een
\ben
[\hat{\varphi}(g_{I}),(\hat{\xi})^{n}] = n  \hat{\xi}_{g_{I}}(\hat{\xi})^{n-1}\,, \qquad
[\hat{\varphi}(g_{I}),\exp(\hat{\xi})] =   \hat{\xi}_{g_{I}}\exp(\hat{\xi})\,.
\label{commu}
\een
Notice that $\hat{\xi}_{g_{I}}$ is \emph{linear} in the creation operator field $\hat{\varphi}^{\dagger}$. As a consequence of this,
\ben
\left(\hat{\varphi}(g_{I})-\hat{\xi}_{g_{I}}\right)  | \xi \rangle = 0\,,
\een
and so the states $| \xi \rangle$ are squeezed states if the operators $\hat{\varphi}(g_{I})-\hat{\xi}_{g_{I}}$ and their Hermitian conjugates satisfy the (suitably gauge-invariant) algebra of creation and annihilation operators, 
\ben
[\hat{\varphi}(g_{I}) - \hat{\xi}_{g_I}, \hat{\varphi}^{\dagger}(h_{I}) - \hat{\xi}^{\dagger}_{h_I}] \propto \delta_{\Sp(4)^3}(g_{I}^{-1}h_I)\equiv\int \dd k\;\delta^4(kg_I^{-1} h_I)\,,
\een
which will in general only be true for specific choices of $\xi$. Evaluating the commutator we find that
\ben
[\hat{\varphi}(g_{I}) - \hat{\xi}_{g_I}, \hat{\varphi}^{\dagger}(h_{I}) - \hat{\xi}^{\dagger}_{h_I}]=
\delta_{\Sp(4)^3}(g_{I}^{-1}h_{I}) + \int (\dd k)^4\;\xi(g_{I}^{-1}k_{I})\overline{\xi(k_{I}^{-1}h_{I})}\,.
\een
Therefore, the states \eqref{dipolestate} can be interpreted as squeezed states only if
the function $\xi$, convoluted with its complex conjugate, is proportional to
a Dirac delta distribution on the group manifold. A trivial case of this is that $\xi$ is itself proportional to a group-averaged delta function, $\xi(g_I)\propto\delta_{\Sp(4)^3}(g_I)$, but more generally $\xi$ has to be the infinite-dimensional analogue of a unitary symmetric matrix for $|\xi\rangle$ to be a squeezed state.

\subsection{Correlation functions}
\label{corrfun}
The particular form of the state chosen as a trial vacuum state of our quantum gravity system implies specific properties of the correlation functions of the group field theory, which are the true encoding of the fundamental quantum dynamics.

For the single-particle condensate (\ref{simple}), the correlation functions are simply factorised in terms of the one-point
correlation function, as we anticipated. This is just the Hartree approximation,
\ben
G^{(n,m)}(g_I^{1},\ldots g_I^{n}; h_I^{1}\ldots h_I^{m})
=
\bra{\sigma} 
\hphid(g_I^{1})\ldots  \hphid( g_I^{n})
\hphi(h_I^{1})\ldots  \hphi(h_I^{m})
\ket{\sigma} = \prod_{i=1}^{n}\overline{\sigma}(g_I^{i}) \prod_{j=1}^{m}\sigma(h_I^{j})\,.
\label{hart}
\een
As said, this particular class of states  ignores correlations among the different
quanta. Furthermore, the result immediately leads to the conclusion that any
equation or condition imposed upon the field operators, when considered in terms
of its expectation value on such a state, would lead to the corresponding equation
for the field $\sigma$, with the straightforward replacement $\hphi \rightarrow \sigma$.

Therefore, the mean field theory encoded in the state \eqref{simple} is just provided by the classical
GFT equations, with an additional symmetry imposed on the classical field configurations.

This is a crucial point, because such equations are obviously provided by the very definition of the fundamental GFT model to be used. Therefore, this simple class of states offers an immediate and straightforward way to obtain an effective cosmological dynamics from any given GFT definition of fundamental quantum gravity dynamics.

\

In the case of dipole condensate states, all the correlation functions are written in terms of the two-point function,
parametrised by $\xi$. In this case, all correlation functions containing an odd number of arguments simply
vanish; the only non-zero correlation functions are of the form
\ben
G^{(n,m)}(g_I^{1},\ldots g_I^{n}; h_I^{1}\ldots h_I^{m})\,,\quad n+m=2k\,.
\een
Closed expressions for the general case are rather complicated, and we just limit ourselves to the cases of the two-point and four-point functions $G^2\equiv G^{(0,2)}, G^4\equiv G^{(0,4)}$ which enter the calculations for the $4d$ GFT models we are most interested in.
Using (\ref{commu}), it is easy to see that
$G^2(g_I,h_I)\equiv\langle\xi|\hat\varphi(g_I)\hat\varphi(h_I)|\xi\rangle$
satisfies
\bena
G^2(g_I,h_I) & = & \int (\dd k)^4 \xi(h_I^{-1}k_I) G^{(1,1)}(g_I,k_I)\nonumber
\\& = &\xi(g_I^{-1}h_I) + \int (\dd k)^4 (\dd k')^4\;\xi(g_I^{-1}k'_I)   \xi(h_I^{-1}k_I)  \overline{G^2(k'_I,k_I)}\,.
\eena
Hence, $G^2$ does not in general coincide with $\xi$, unless this function satisfies the 
condition
\ben
\int (\dd k)^4 (\dd k')^4\;\xi(g_I^{-1}k'_I)   \xi(h_I^{-1}k_I)  \overline{\xi(k_I^{-1}k'_I)} = 0.
\een
This means that, while the function $\xi$ encodes the geometric data that we need, it
does not immediately correspond to the two-point function of GFT in the given state. Instead, we have 

\bena
G^2(g_I,h_I) &\approx&\xi(g_I^{-1}h_I) + \int (\dd k)^4 (\dd k')^4\;\xi(g_I^{-1}k'_I)   \xi(h_I^{-1}k_I)  \overline{\xi(k_I^{-1}k'_I)}\,,
\\\xi(g_I^{-1}h_I) &\approx &G^2(g_I,h_I) - \int (\dd k)^4 (\dd k')^4\;G^2(g_I,k'_I)   G^2(h_I,k_I)  \overline{G^2(k_I,k'_I)}
\eena
where we are neglecting terms built with convolutions of five or more kernels.

The analysis of the four-point correlation function is slightly more involved, but shows the
general pattern of the calculations. By definition,
\ben
G^4(g_I^a,g_I^b,g_I^c,g_I^d)=
\langle
\hphi(g_I^a) 
\hphi(g_I^b)
\hphi(g_I^c)
\hphi(g_I^d)
\rangle
\propto
\bra{0}e^{\DD^{\dagger}}
[\hphi(g_I^a),[\hphi(g_I^b),[\hphi(g_I^c),[\hphi(g_I^d),e^{\DD}]]]]
\ket{0}
\een
where we are using $\DD$ instead of $\hat{\xi}$ to emphasise that the calculation that follows is totally general and is valid (with appropriate modifications) in the case of general $N$-particle coherent states.

It is convenient to introduce the notation $\DD_{g_I^a} = [\hphi(g_I^a) ,\DD]$, $\DD_{g_I^a g_I^b} = [\hphi(g_I^a),[\hphi(g_I^b),\DD]]$ etc. In the case of the dipole where $\DD=\hat\xi$ and $\DD_{g_I^a g_I^b}=\xi((g_I^a)^{-1} g_I^b)$ all higher commutators vanish. Then

\bena
[\hphi(g_I^a),[\hphi(g_I^b),[\hphi(g_I^c),[\hphi(g_I^d),e^{\DD}]]]]
&=&
[\hphi(g_I^a),[\hphi(g_I^b),[\hphi(g_I^c),D_{g_I^d}e^{\DD}]]]
\\&=&
[\hphi(g_I^a),[\hphi(g_I^b),(\DD_{g_I^c g_I^d}+\DD_{g_I^c}\DD_{g_I^d})e^{\DD}]] \nonumber
\\
&=& [\hphi(g_I^a),
(\DD_{g_I^b}\DD_{g_I^c g_I^d}+\perm+\DD_{g_I^b}\DD_{g_I^c}\DD_{g_I^d}+\DD_{g_I^b g_I^c g_I^d})e^{D}
]\nonumber
\\
&=&\left(
\DD_{g_I^a g_I^b}\DD_{g_I^c g_I^d}+\perm
+
\DD_{g_I^a g_I^b}\DD_{g_I^c}\DD_{g_I^d} + \perm\right.\nonumber
\\&&\left.
+\DD_{g_I^a}\DD_{g_I^b}\DD_{g_I^c}\DD_{g_I^d}+\DD_{g_I^a}\DD_{g_I^b g_I^c g_I^b}+\perm+\DD_{g_I^a g_I^b g_I^c g_I^d}
\right)
e^{\DD}\nonumber
\eena
in general. For $\DD=\hat\xi$, the last two contributions vanish, $\DD_{g_I^a}=\hat\xi_{g_I^a}$, $\DD_{g_I^a g_I^b}=\xi((g_I^a)^{-1} g_I^b)$ and the four-point function is the solution to the following equation:
\bena
G^4(g_I^a,g_I^b,g_I^c,g_I^d)&=&\xi((g_I^a)^{-1} g_I^b)\xi((g_I^c)^{-1} g_I^d) + \perm \nonumber
\\&&+ \,\xi((g_I^a)^{-1} g_I^b)\int(\dd h)^4(\dd k)^4\;\xi(h_I^{-1} g_I^c)\xi(k_I^{-1} g_I^d)\overline{G^2(h_I,k_I)}
+ \perm\nonumber
\\&&+\int(\dd h)^4(\dd h')^4(\dd k)^4(\dd k')^4\;\xi(h_I^{-1} g_I^a)\xi(k_I^{-1} g_I^b)\xi((h'_I)^{-1} g_I^c)\xi((k'_I)^{-1} g_I^d)\times\nonumber
\\&&\times \overline{G^4(h_I,k_I,h'_I,k'_I)} \,.
\eena

As in the case of the two-point functions, regarded as functionals of $\xi$  
the four-point correlation functions are given only implicitly and, in absence of
further conditions, do not correspond simply to bilinears in $\xi$.
However, it is also clear that all the correlation functions are given in terms of the two-point function
alone.

To the same order of approximation used above for the two-point function, the four-point function is
\ben
G^4(g_I^a,g_I^b,g_I^c,g_I^d) =
\underbrace{G^2(g_I^a,g_I^b)G^2(g_I^c,g_I^d)+\perm
}_{\text{Gaussian-like}}+O((G^2)^6)\,.
\een

Therefore, at least at leading order in the expansion, the state is quadratic in the sense that we can express the correlation
functions in terms of the two-point function, itself determined by the function $\xi$
in a highly nonlinear way.

\

In absence of an accurate analysis of the critical limit of the recursion relations among
GFT correlators, it is hard to say much more. However, one can still try to estimate the theoretical error of a truncation
of the tower of correlation functions to only a few representatives in terms of
a Ginzburg-like criterion.
Indeed, following standard procedures, one can split the correlation functions in terms
of mean fields and fluctuations,
\bena
\langle \hphi(g_I^a) \rangle &=:& \phi(g_I^a)\,, \\
G^2(g_I^a,g_I^b)&=&\langle (\hphi(g_I^a) - \phi(g_I^a)+\phi(g_I^a))(\hphi(g_I^b) - \phi(g_I^b) + \phi(g_I^b))\rangle \nonumber
\\&=:& \phi(g_I^a)\phi(g_I^b) + G^{2(c)}(g_I^a,g_I^b)\,,
\\
G^3(g_I^a,g_I^b,g_I^c)&=:&G^{3(c)}(g_I^a,g_I^b,g_I^c) + \phi(g_I^a)  G^{2(c)}(g_I^b,g_I^c) 
+ \perm +  \phi(g_I^a) \phi(g_I^b) \phi(g_I^c)\,,
\\
G^4(g_I^a,g_I^b,g_I^c,g_I^d)&=:&\phi(g_I^a)\phi(g_I^b)\phi(g_I^c)\phi(g_I^d) + 
\phi(g_I^a) \phi(g_I^b) G^{2(c)}(g_I^c,g_I^d) + \perm
\nonumber
\\&&+G^{2(c)}(g_I^a,g_I^b)G^{2(c)}(g_I^c,g_I^d) + \perm+ \phi(g_I^a) G^{3(c)}(g_I^b,g_I^c,g_I^d) + \perm\nonumber
\\&&
+ G^{4(c)}(g_I^a,g_I^b,g_I^c,g_I^d)\,,
\eena
and so on.
Then the truncation of the tower of equations, as deduced from the Schwinger--Dyson equation, to a given order leads to a theoretical error in the resulting effective theory 
that can be estimated by the magnitude of the neglected terms.
For instance, in the case of a Hartree--Fock mean field approximation to
the hydrodynamics of Bose--Einstein condensates, the breakdown of the Gross--Pitaevskii equation
is signalled not necessarily by a singularity of the particular solution itself,
but rather by the large value of the fluctuation with respect to the mean field
associated to the particular quantum state considered. 

\

These considerations allow us to at least estimate how reliable the approximation encoded in the use of the simple states
\eqref{simple} and \eqref{dipolestate} is.

It is clear from the analysis of the states \eqref{simple} and \eqref{dipolestate} that
conditions on them to be good approximations to physically relevant states can be rephrased in an equivalent form
in terms of the properties of the correlation functions. While less clear in terms
of the GFT condensate interpretations, correlation functions (and their relations encoded
in the Schwinger--Dyson equations) might be more accessible from the point of view
of the analysis of the perturbative (spin foam) expansion of GFTs.
Consequently, the validity of the ansatz \eqref{simple} or \eqref{dipolestate} might
be directly verified once the relations between correlation functions are investigated
in the critical limit. For instance, in the case of matrix models for $2d$ gravity it has been
shown that correlation function do factorise in the large N limit \cite{Migdal}. This behaviour
would be matched by the simple condensates.

\subsection{Condensate states as exact GFT vacua?}

As we discussed above, GFT condensates of the simple type we defined can at most be approximations to the true vacuum state of the quantum gravity system, even if one believes that something akin to a GFT condensation is what determines such a true vacuum state for our quantum universe. We have also seen that we can estimate the theoretical error made in using such approximation by analysing the $n$-point functions of the theory. 

In some cases, however, one can do even more, and show that specific condensate states (slightly more involved than the ones presented above and used in the following) are {\it exact solutions} to the microscopic quantum dynamics, and thus true vacuum states.  We show here one example.

\

In the case in which the quantum equation of motion involves a trivial kinetic term and a (non-Hermitian) potential term that depends only on the creation operators,
\begin{equation}
\left( \hat{\varphi}(g_1,\ldots, g_{4}) +
\lambda\,\frac{\delta\hat{\mathcal{V}}[\hat{\varphi}^\dagger]}{\delta\hat\varphi^{\dagger}(g_1,\ldots,g_4)}\right)|\psi\rangle=0\,,
\end{equation}
the special state
\begin{equation}
\ket{E} = \mathcal{N}^{-1} \exp\left( - \lambda \hat{V}[\hat{\varphi}^\dagger] \right) \ket{0}
\end{equation}
is an exact solution to the interacting operator field equations.
 
Applying Wick's theorem to the each term in $\langle E\ket{E}=1$, it turns out that $\mathcal{N}$ is the square root
of the GFT partition function\footnote{To see this, it suffices to take the norm of the state $\exp(\lambda \hat{V}[\hat{\varphi}^\dagger] )\ket{0})$ and to insert an identity written as a (formal)
integral over single field coherent states, 
$$
\mathbb{I} = \frac{1}{Z_0} \int \mathcal{D} \sigma\mathcal{D}\overline{\sigma} \ket{\sigma}\bra{\sigma} \exp(-|\sigma|^2)\, ,
$$
where $Z_0$, needed for the normalisation of the integral, is itself a divergent quantity,
being the partition function for a Gaussian ensemble.
}. In the case of Boulatov--Ooguri theories, this state can be seen as a condensate
of five tetrahedra glued to one another to form a 3-sphere topology, in the combinatorial pattern of the boundary of a 4-simplex.

It will be interesting to investigate further the properties of such states, as well the existence of other exact solutions of the GFT dynamics for other models, obtained in a similar fashion.


\section{Effective cosmological dynamics}  
\label{effcosm}

In the previous section we have constructed and discussed a class of states representing homogeneous spatial geometries. At this stage, these states are kinematical. While we have ensured that they are invariant under local frame rotations, and they represent geometric data invariant under spatial diffeomorphisms by construction, they do not yet satisfy any form of dynamical equation that would correspond to the Hamiltonian constraint in geometrodynamics, or to an appropriate generalisation of the Friedmann equation in the cosmological setting.

The dynamics of a given GFT action provides us with precisely such an equation. We start with a general action that we only assume to consist of a quadratic (kinetic) part and an interaction,
\ben
S[\varphi,\bar\varphi]=\int (\dd g)^4 (\dd g')^4\;\bar\varphi(g_1,\ldots,g_4)\mathcal{K}(g_1,\ldots,g_4,g'_1,\ldots,g'_4)\varphi(g'_1,\ldots,g'_4)+\lambda\,\mathcal{V}[\varphi,\bar\varphi]\,,
\een
where $\mathcal{K}$ is in general a differential operator, but can also be a delta distribution in some models which simply identifies the arguments of $\varphi$ and $\bar\varphi$. Assuming the action to be real, there is one independent classical field equation,
\ben
\frac{\delta S[\varphi,\bar\varphi]}{\delta\bar\varphi(g_1,\ldots,g_4)}=\int (\dd g')^4\;\mathcal{K}(g_1,\ldots,g_4,g'_1,\ldots,g'_4)\varphi(g'_1,\ldots,g'_4)+\lambda\,\frac{\delta\mathcal{V}[\varphi,\bar\varphi]}{\delta\bar\varphi(g_1,\ldots,g_4)}=0\,,
\een
which we can associate with the corresponding operator in the quantum theory,
\ben
\hat{\mathcal{C}}(g_I):=\int (\dd g')^4\;\mathcal{K}(g_1,\ldots,g_4,g'_1,\ldots,g'_4)\hat\varphi(g'_1,\ldots,g'_4)+\lambda\,\frac{\delta\hat{\mathcal{V}}[\hat\varphi,\hat\varphi^{\dagger}]}{\delta\hat\varphi^{\dagger}(g_1,\ldots,g_4)}\,.
\label{opdef}
\een
For a general classical potential term depending both on $\varphi$ and its complex conjugate, (\ref{opdef}) requires a choice of operator ordering, given that in general $[\hat\varphi(g_I),\hat\varphi^{\dagger}(g'_I)]\neq 0$. The usual procedure is to adopt a normal ordering prescription and we also adopt this standard choice\footnote{This does not suffice, of course, to make the equation well defined as an operator equation on the Fock space, from the rigorous functional analytic point of view. If the field operator is to be interpreted as an operator-valued distribution, in usual interacting quantum field theories one would not expect the operator $\hat{\mathcal{V}}$ or its functional derivative to be mathematically well-defined on the Fock space of the free theory without regularisation. We note however that relativistic QFT, where this would be the case, rests on Poincar\'e invariance and causality whose role in GFT is unclear, so that we cannot delve into a more detailed mathematical analysis here. Our discussions in this section are understood to implicitly assume that an appropriate regularisation has been chosen.}.

As we have mentioned in Sec.~\ref{gftintro}, the connection of operator equations of motion and the path integral is given by Schwinger--Dyson equations. These can be formally derived by using the ``fundamental theorem of functional calculus'' and assuming that there is no boundary term, so that
\ben
0 = \int\mathcal{D}\varphi\;\mathcal{D}\bar\varphi\;\frac{\delta}{\delta\bar\varphi(g_I)} \left(\mathcal{O}[\varphi,\bar\varphi]\;e^{- S[\varphi,\bar\varphi]}\right)=\left\langle \frac{\delta\mathcal{O}[\varphi,\bar\varphi]}{\delta\bar\varphi(g_I)} - \mathcal{O}[\varphi,\bar\varphi]\frac{\delta S[\varphi,\bar\varphi]}{\delta\bar\varphi(g_I)}\right\rangle
\een
for any functional of the field and its complex conjugate. The expectation value is to be interpreted as taken in the ``vacuum state'' specified by the boundary conditions of the path integral. Hence, the resulting equations are to be imposed on any state in the Fock space that is assumed to play the role of ``ground state'', not necessarily the Fock vacuum. In our setting, we will choose this state to be one of our condensate states, $|\sigma\rangle$ or $|\xi\rangle$.

The task will be to use the Schwinger--Dyson equations to extract an equation for the profile functions $\sigma$ or $\xi$ appearing in the definition of these states, which would encode the requirement that the corresponding states are approximate solutions of the full quantum dynamics. In a systematic treatment, one would have to prove that solutions to the simplest Schwinger--Dyson equations already approximate a fully dynamical solution to all of them. For our present purposes, this is a working assumption which can be justified to an extent from an analysis of the $n$-point functions of the theory, as we have outlined in Sec.~\ref{corrfun}. In the simplest case of the single-particle condensate, for example, we saw in (\ref{hart}) that all $n$-point functions are just products of the condensate wavefunction $\sigma$ and its complex conjugate $\bar\sigma$. This implies that the tower of Schwinger--Dyson equations involving all $n$-point functions just reduces to a set of (nonlinear) equations for $\sigma$. We are first looking for solutions to the simplest ones; all the higher-order equations would then be consistency conditions.

The simplest case occurs for $\mathcal{O}=1$ in which we obtain the requirement that
\ben
\langle\hat{\mathcal{C}}(g_I)\rangle_{\psi}:=\langle\psi|\hat{\mathcal{C}}(g_I)|\psi\rangle = 0\,,
\label{quanteom}
\een
where $|\psi\rangle$ is one of the condensate states we are considering.

For the single-particle condensate defined in (\ref{simple}), (\ref{quanteom}) takes a particularly simple form. As we have noted in Sec.~\ref{squeezedsec}, the states $|\sigma\rangle$ are eigenstates of the field operator $\hat\varphi$. Then, using the normal ordering prescription for $\hat{\mathcal{V}}$ in which all $\hat\varphi^{\dagger}$ are to the left of all $\hat\varphi$, the condition $\langle\sigma|\hat{\mathcal{C}}(g_I)|\sigma\rangle = 0$ reduces to (using that $\langle\sigma|\sigma\rangle>0$)
\ben
\int (\dd g')^4\;\mathcal{K}(g_1,\ldots,g_4,g'_1,\ldots,g'_4)\sigma(g'_1,\ldots,g'_4)+\lambda\,\frac{\delta\mathcal{V}[\varphi,\bar\varphi]}{\delta\bar\varphi(g_1,\ldots,g_4)}\Big|_{\varphi\rightarrow\sigma,\bar\varphi\rightarrow\bar\sigma}=0\,.
\label{simpleeq}
\een
Hence the expectation value of the quantum equation of motion reduces to the classical field equation, to be satisfied by the `condensate wavefunction' $\sigma$. This is the direct analogue in the group field theory context of the Gross--Pitaevskii equation for real Bose--Einstein condensates. For a general potential $\mathcal{V}$ (and specifically for the type of potentials typically considered in the GFT literature), this equation is nonlinear in $\sigma$, and nonlocal on the minisuperspace of homogeneous geometries (recall the interpretation of the domain of definition of $\sigma$ as implied by the reconstruction procedure of Sec.~\ref{approxgeo}). It bears close similarity to the equations studied in the nonlinear extension of loop quantum cosmology in \cite{nonlincosm} and in the simplified `group field theory' model of \cite{GFC}.

\

We interpret $\sigma$ as defining a probability distribution on the space of homogeneous spatial geometries, as anticipated. Again, this is analogous to Bose--Einstein condensates where the condensate wavefunction can directly be associated with particle density and momentum density as functions on space. Even though our equation is nonlinear in $\sigma$, this does not lead to any immediate issue with unitarity; $\hat{\mathcal{C}}$ has the interpretation of an initial-value constraint, not an evolution equation giving any notion of `time evolution' under which an inner product would have to be preserved. The nonlinearity will of course break the superposition principle of quantum mechanics that would be expected if $\sigma$, $\xi$, etc. are interpreted as wavefunctions. Linear combination of solutions of the Gross--Pitaevskii-like equations of motion will not be solutions themselves, in general. This is not an inconsistency, but it does prevent any straightforward interpretation of the equation as a standard quantum cosmology equation, as it would follow from the canonical quantisation of minisuperspace geometries. Rather, again in analogy with the theory of Bose--Einstein condensates, it suggests a re-interpretation of quantum cosmology itself as a form of hydrodynamics for quantum spacetime. 

\

The vanishing of the expectation value of $\hat{\mathcal{C}}$ is clearly just one condition to be satisfied by a genuine physical state. Any other condition of the form 
\ben
\langle\hat{\mathcal{O}}[\hat\varphi,\hat\varphi^{\dagger}]\hat{\mathcal{C}}(g_I)\rangle_\psi = \left\langle\frac{\delta\hat{\mathcal{O}}[\hat\varphi,\hat\varphi^{\dagger}]}{\delta\hat\varphi^{\dagger}(g_I)}\right\rangle_\psi\,,
\label{quantcond}
\een
for an arbitrary operator $\hat{\mathcal{O}}$, could be equivalently used to derive conditions on the profile functions $\sigma$ or $\xi$. Clearly, since there is an infinity of such conditions, one would have to show that not all of them are independent. Here we content ourselves with the approximation to the full quantum dynamics represented by the equation (\ref{simpleeq}) and with the estimate of the theoretical error obtained from the study of the $n$-point functions in the case of simple condensates. 

\

The philosophy followed for $|\sigma\rangle$ in deriving the analogue of the Gross--Pitaevskii equation can also be applied to the dipole condensate and its profile function $\xi$. Again, we start off by computing the expectation value (\ref{quanteom}), here in the state $|\xi\rangle$, obtaining
\ben
\int (\dd g')^4\;\mathcal{K}(g_1,\ldots,g_4,g'_1,\ldots,g'_4)\langle\hat\varphi(g'_1,\ldots,g'_4)\rangle_{\xi}+\lambda\,\left\langle\frac{\delta\hat{\mathcal{V}}[\hat\varphi,\hat\varphi^{\dagger}]}{\delta\hat\varphi^{\dagger}(g_1,\ldots,g_4)}\right\rangle_{\xi}=0\,.
\een
But the one-point function for $|\xi\rangle$ vanishes, leading us to conclude that
\ben
\left\langle\frac{\delta\hat{\mathcal{V}}[\hat\varphi,\hat\varphi^{\dagger}]}{\delta\hat\varphi^{\dagger}(g_1,\ldots,g_4)}\right\rangle_{\xi}=0\,.
\een
Similarly, we can compute an expectation value (\ref{quantcond}) with $\hat{\mathcal{O}}$ taken to be the field $\hat\varphi$, yielding (we use $\delta\varphi/\delta\bar\varphi=0$)
\ben
\int (\dd g')^4\;\mathcal{K}(g_I,g'_I)\langle\hat\varphi(g''_1,\ldots,g''_4)\hat\varphi(g'_1,\ldots,g'_4)\rangle_{\xi}+\lambda\,\left\langle\hat\varphi(g''_1,\ldots,g''_4)\frac{\delta\hat{\mathcal{V}}[\hat\varphi,\hat\varphi^{\dagger}]}{\delta\hat\varphi^{\dagger}(g_1,\ldots,g_4)}\right\rangle_{\xi}=0\,.
\label{cond2}
\een
As shown in Sec.~\ref{corrfun}, the two-point function in the state $|\xi\rangle$ satisfies
\ben
\langle\hat\varphi(g_I)\hat\varphi(h_I)\rangle_{\xi} = \xi(h_I^{-1} g_I)+\int (\dd g')^4  (\dd h')^4 \xi(g_I^{-1} g'_I)\xi(h_I^{-1} h'_I) \overline{\langle\hat\varphi(g'_I)\hat\varphi(h'_I)\rangle_{\xi}}\,,
\label{2point}
\een
so that (\ref{cond2}) becomes
\bena
0&=&\int (\dd g')^4\;\mathcal{K}(g_I,g'_I)\xi((g'_I)^{-1}g''_I)+\int (\dd g'\,\dd h\,\dd h')^4\;\xi(h_I^{-1} g''_I)\;\overline{\langle\hat\varphi(h_I)\hat\varphi(h'_I)\rangle_{\xi}}\;\mathcal{K}(g_I,g'_I)\xi((h'_I)^{-1} g'_I) \nonumber
\\&&+\lambda\,\left\langle\hat\varphi(g''_I)\frac{\delta\hat{\mathcal{V}}[\varphi,\bar\varphi]}{\delta\hat\varphi^{\dagger}(g_I)}\right\rangle_{\xi}\,.
\label{cond3}
\eena
Without explicit expressions for general $n$-point functions, this equation cannot directly be written as a condition on $\xi$. A simplification occurs if we assume that the interaction $\mathcal{V}$ is of odd order, as is indeed the case in many GFT models of 4$d$ quantum gravity. Then the second line vanishes since it only contains ($2n+1$)-point functions of $\xi$. If then, in addition, the kinetic operator $\mathcal{K}$ is invertible, no condensation of `dipoles' is possible; from (\ref{cond2}) and the invertibility of $\mathcal{K}$, the two-point function in the state $|\xi\rangle$ vanishes, but then (\ref{cond3}) states that $\xi$ must itself vanish. 

Let us assume that while the interaction is of odd order so that it does not contribute to (\ref{cond3}), the operator $\mathcal{K}$ has a nontrivial kernel. Then the equation to be satisfied is
\ben
\int (\dd g')^4\;\mathcal{K}(g_I,g'_I)\langle\hat\varphi(g'_1,\ldots,g'_4)\hat\varphi(g''_1,\ldots,g''_4)\rangle_{\xi} = 0\,.
\label{eqofmo}
\een
Using the relation (\ref{2point}), by recursion, one finds that the two-point function $\langle\hat\varphi(g_I)\hat\varphi(h_I)\rangle$ can be expressed in terms of a power series in $\xi$ and its complex conjugate:
\ben
\langle\hat\varphi(g_I)\hat\varphi(h_I)\rangle = \xi(g_I^{-1} h_I) + \int (\dd g')^4 \xi(g_I^{-1} g'_I) \;\Xi[\xi,\overline{\xi}](g'_I,h_I)\,.
\een
Following the same idea, one can rewrite (\ref{2point}) as 
\ben
\xi(g_I^{-1} h_I) = \langle\hat\varphi(g_I)\hat\varphi(h_I)\rangle_{\xi} - \int (\dd g')^4  (\dd h')^4 \xi(g_I^{-1} g'_I)\xi(h_I^{-1} h'_I) \overline{\langle\hat\varphi(g'_I)\hat\varphi(h'_I)\rangle_{\xi}}
\label{eq:recursionxi}
\een
and replace each of the $\xi$ in the right-hand side of \eqref{eq:recursionxi} with the expression given by \eqref{eq:recursionxi} to get a representation of $\xi$ as a power series in the two-point function and its complex conjugate,
\bena
\xi(g_I^{-1} h_I) &= &\langle\hat\varphi(g_I)\hat\varphi(h_I)\rangle_{\xi} + \int (\dd h')^4 \langle\hat\varphi(g_I)\hat\varphi(h'_I)\rangle_{\xi} \;\Gamma[\langle\hat\varphi\hat\varphi\rangle,\overline{\langle\hat\varphi\hat\varphi\rangle}](h'_I,h_I)\nonumber
\\&=:&\Phi[\langle\hat\varphi\hat\varphi\rangle,\overline{\langle\hat\varphi\hat\varphi\rangle}](g_I,h_I)\,.
\label{eq:firstbranch}
\eena
(\ref{2point}) is a quadratic equation for $\xi$ and there is a second branch of solutions for $\xi$ in terms of the two-point function. Namely, if there is an inverse $\mathfrak{K}$ for $\overline{\langle\hat\varphi(g_I)\hat\varphi(h_I)\rangle}$, in the sense that
\ben
\int (\dd g')^4 \mathfrak{K}(g_I,g'_I)\overline{\langle\hat\varphi(g'_I)\hat\varphi(h_I)\rangle}=\int \dd k\;\delta(g_Ikh_I^{-1})\,,
\een
with $\mathfrak{K}(g_I,h_I)=\mathfrak{K}(h_I,g_I)$ and $\mathfrak{K}(g_I,h_I)=\mathfrak{K}(g_I,h_I k)$, one can verify that
\ben
\xi(g_I^{-1} h_I) = - \mathfrak{K}(g_I,h_I) - \Phi[\langle\hat\varphi\hat\varphi\rangle,\overline{\langle\hat\varphi\hat\varphi\rangle}](g_I,h_I)
\label{eq:secondbranch}
\een
solves (\ref{eq:recursionxi}) if $\Phi$ does. Therefore, already without investigating issues of convergence of the power series (\ref{eq:firstbranch}), we see that the relation between the dipole wavefunction $\xi$ and the two-point functions is not one to one. The second branch \eqref{eq:secondbranch} describes non-perturbative condensate states that behave non-analytically when the condensate is diluted. A rescaling of the function $\xi$, $\xi\rightarrow\epsilon\,\xi$, leads to a different normalisation (\ref{normcons}), but this constant drops out of expectation values. It corresponds to a dilution of the condensate, as can be seen from looking at the expectation value of the total particle number
\ben
\hat{N}_{\tetrahedron}=\int (\dd g)^4\;\hat\varphi^{\dagger}(g_1,\ldots,g_4)\hat\varphi(g_1,\ldots,g_4)\,.
\een
When $\xi$ is rescaled in this way, the relative contribution of the nonlinear term in (\ref{2point}) becomes smaller and smaller, but only if we assume the two point-function not to grow faster than $\frac{1}{\epsilon}$ for small $\epsilon$. On the second branch of solutions (\ref{eq:secondbranch}), the relation between $\xi$ and $\langle\hat\varphi\hat\varphi\rangle$ is not analytic, and the two-point function blows up when $\xi$ goes to zero. This branch of solutions is not connected to the Fock vacuum, and we would be inclined to consider it as spurious: we expect that, if we deform the operator $\hat\xi$ to include different powers of $\hphid$, the structure of the equation relating the kernels to $\langle\hat\varphi\hat\varphi\rangle$ would change, leading to the disappearance of \eqref{eq:secondbranch}.

Focussing on the first branch of solutions (\ref{eq:firstbranch}) we see that 
\bena
\int (\dd g')^4\;\mathcal{K}(g_I,g'_I)\xi((g''_I)^{-1}g'_I) & = & \int (\dd g')^4\left(\mathcal{K}(g_I,g'_I)\langle\hat\varphi(g'_I)\hat\varphi(g''_I)\rangle_{\xi}\right.
\\&& + \left.\int (\dd h')^4 \mathcal{K}(g_I,g'_I)\langle\hat\varphi(g'_I)\hat\varphi(h'_I)\rangle_{\xi} \;\Gamma[\langle\hat\varphi\hat\varphi\rangle,\overline{\langle\hat\varphi\hat\varphi\rangle}](h'_I,g''_I)\right)\nonumber\,,
\eena
and so it follows that for the quantum equation of motion (\ref{eqofmo}) to hold, $\xi$ must satisfy the linear differential equation
\ben
\int (\dd g')^4\;\mathcal{K}(g_I,g'_I)\xi((g''_I)^{-1}g'_I) = 0\,.
\label{wdwtype}
\een
For GFT models with a $\mathcal{K}$ that has a nontrivial kernel, (\ref{wdwtype}) becomes a Wheeler--DeWitt-type equation for a function $\xi$ which can then be interpreted as a quantum cosmology wavefunction, encoding some part of the full GFT quantum dynamics. 

This is our key dynamical equation, under the approximations and assumptions made, for a quantum universe described by our dipole condensate.

For the second branch (\ref{eq:secondbranch}) this is no longer true, as we would in general have
\ben
\int (\dd g')^4\;\mathcal{K}(g_I,g'_I)\xi((g''_I)^{-1}g'_I) = -\int (\dd g')^4\;\mathcal{K}(g_I,g'_I)\mathfrak{K}(g'_I,g''_I)\neq 0\,.
\een
In the following, for the reasons explained, we focus on (\ref{wdwtype}).

\subsection{Simplicity constraints}

So far we have examined only the general structure of the equations of motion, and the related
problems, when restricted to the case of GFT condensates.
However, before being able to do the calculation for concrete models, we need to discuss the last important ingredient in the construction of spin foam and GFT models for quantum gravity, the geometricity or simplicity constraints implemented in the definition of current spin foam and GFT models for 4d quantum gravity.

\

The construction of spin foam models follows the interpretation of general relativity as a topological
BF theory with constraints. This is motivated by the fact that the quantisation of BF theories is under control, and they can
be easily discretised. The GFT actions proposed by Boulatov \cite{boulatov} and Ooguri \cite{Ooguri} are indeed designed to
provide a quantisation of BF theories.

In order to turn these models into candidate theories for quantum gravity, one has to find a way to impose the simplicity
constraints at the quantum level. The qualitative picture is that the classical simplicity constraints,
suitably discretised, will be translated into restrictions on the labels of the GFT Feynman amplitudes (or spin foams). The sum over Feynman amplitudes will then be converted to the sum of the terms that admit a geometric interpretation in terms of simplicial geometric variables, and only them.

The precise translation of the simplicity constraints of the continuum theory into the language 
of discrete models like spin foams is a delicate issue. A recent review, containing
a detailed discussion of the construction of some models can be found in \cite{SF}; other constructions are in \cite{listofallspinfoammodels}. 

In the language of GFT, the imposition of the simplicity constraints can be achieved with a
modification of the action such that the Feynman expansion of the partition function
involves the summation over geometric configuration only. 
There are several ways to do this. One way is to start with the GFT for BF theory for $\Sp(4)$ (or
$\SL(2,\bC)$ for Lorentzian models) and modify the action with a suitable \emph{constraint operator} $\mathbb{S}$
acting on the field, $\mathbb{S}\hat{\varphi}$. The choice of the precise form of the constraint operator (which can be given in terms of so-called fusion
coefficients) distinguishes the specific features encoded by the model at hand
\cite{listofallspinfoammodels}. Furthermore, 
for fixed form of the operator $\mathbb{S}$, 
we still have the freedom to decide where to apply it in the action: only in the kinetic term,
only in the interaction term, or in both. These three choices are not completely equivalent,
since $\mathbb{S}$, in general, is not a projector.
In addition, one might still have to worry about the contributions to the sum over amplitudes
of the configurations annihilated by $\mathbb{S}$.

\

As a result, different models will result also in different effective equations for cosmology.
Given the very simple correspondence between the microscopic equations of motion
of GFT and the macroscopic cosmological dynamics that we have described,
it is possible however to give a specific correspondence between the choice of constraint
operator $\mathbb{S}$ and the modifications to the effective dynamics.

\

Let us consider first the Riemannian case. The (linear) operator $\mathbb{S}$ restricts the
summation over representations by turning the field into a field over 
$\SU(2)^{4}/\SU(2)_{\mathrm{diag}}$,
\begin{equation}
\mathbb{S} : L^{2}(\Sp(4)^4/\Sp(4)_{\mathrm{diag}}) \rightarrow 
 L^{2}(\SU(2)^4/\SU(2)_{\mathrm{diag}})\, ,
\label{genmap}
\end{equation}
in the case of finite Immirzi parameter (the case of infinite Immirzi parameter has a different structure, but can be easily obtained via a limiting procedure \cite{bcaristidedaniele, listofallspinfoammodels}). 
Since we are using two different groups, to make the notation more transparent, in this section we will
write elements of $\SU(2)$ in lower case, and elements of $\Sp(4)$ or $\SL(2,\bC)$
in upper case.

The general form of $\mathbb{S}$ can be described by its action on a basis of functions. Generic
functions in the domain of $\mathbb{S}$ can be written in terms of a Peter--Weyl decomposition as
\begin{equation}
\varphi(G_{I}) = \sum_{\{(j_I^{+}, j_I^{-})\}} \sum_{i^{+} i^{-}} \sum_{\genfrac{}{}{0pt}{}{\{(p_I^+, p_I^-)\}}{\{(q_I^+, q_I^-)\}}}
\varphi^{i^{+}i^{-}(j_1^{+}, j_{1}^{-}),\ldots,(j_4^{+}, j_{4}^{-})}
_{(q_1^{+},q_{1}^{-}),\ldots,(q_4^{+}, q_{4}^{-})}
 I^{{i}^{+}i^{-}(j_{1}^{+}j_{1}^{-})\ldots
(j_{4}^{+}j_{4}^{-})}
_{(p_1^{+},p_{1}^{-}),\ldots,(p_4^{+}, p_{4}^{-})}
\prod_{I=1}^{4}  d_{j_{I}^{+}}d_{j_{I}^{-}}  
\mathcal{D}^{(j_{I}^{+},j_{I}^{-})}_{q_{I}^{+}q_{I}^{-}p_{I}^{+} p_{I}^{-}} (G_I)
\end{equation}
where we are using the splitting of the representation matrices $\mathcal{D}$
into representations of $\SU(2)$ using $\Sp(4)=\SU(2)\times\SU(2)$, $I$ is a four-valent $\Sp(4)$ intertwiner
and $i^{\pm}$ are additional angular
momenta labelling it. Similarly, functions in $ L^{2}(\SU(2)^4/\SU(2)_{\mathrm{diag}})$
can be decomposed as
\begin{equation}
\psi(g_{I}) = \sum_{\{J_I\}} \sum_{J,\{M_{I}, N_{I}\}}
\psi^{J J_{1}\ldots J_{4}}_{M_{1}\ldots M_{4}}
\iota^{J J_{1}\ldots J_{4}}_{N_{1}\ldots N_{4}}
\prod_{I=1}^{4}  d_{J_{I}}  
D^{J_{I}}_{M_{I}N_{I}}(g_I)\,.
\end{equation}
The operator $\mathbb{S}$ can be then specified in terms of its action on a basis, \ie by
the coefficients
\begin{equation}
\mathbb{S}^{J i^{+} i^{-} J_{1} \ldots J_{4}  (j_1^{+}, j_{1}^{-}),\ldots,(j_4^{+}, j_{4}^{-})} 
_{M_{1}\ldots M_{4}(q_1^{+},q_{1}^{-}),\ldots,(q_4^{+}, q_{4}^{-})}
\end{equation}
required to map between the coefficients of the expansions of the gauge invariant field functions in
representations.
The simplicity constraints, translated in the language of representation theory,
determine the coefficients $\mathbb{S}$. They are, in principle, four independent constraints on the different arguments of the field, and hence can be defined in terms of a map $S$ from the space of square integrable functions on 
a single copy of $\Sp(4)$ to the space of functions on $\SU(2)$. However, we must also take into account gauge invariance, \ie the fact that the field lives on a quotient space in (\ref{genmap}), by contracting indices of the representation matrices with four-valent intertwiners.

This means that the coefficients $\mathbb{S}$ imposing simplicity constraints can be written as \begin{equation}
\mathbb{S}^{J i^{+} i^{-} J_{1} \ldots J_{4}  (j_1^{+}, j_{1}^{-}),\ldots,(j_4^{+}, j_{4}^{-})} 
_{M_{1}\ldots M_{4}(q_1^{+},q_{1}^{-}),\ldots,(q_4^{+}, q_{4}^{-})}
= \sum_{\{N_I\},\{(p_I^+,p_I^-)\}}
\iota^{J J_{1}\ldots J_{4}}_{N_{1}\ldots N_{4}} I^{({i}^{+}i^{-})(j_{1}^{+}j_{1}^{-})\ldots
(j_{4}^{+}j_{4}^{-})}
_{(p_1^{+},p_{1}^{-}),\ldots,(p_4^{+}, p_{4}^{-})}\prod_{I=1}^{4} S^{J_{I}(j_{I}^{+}j_{I}^{-})}_{M_{I}N_{I} q_{I}^{+}q_{I}^{-}
p_{I}^{+}p_{I}^{-}
}\,,
\label{coeffconstraint}
\end{equation}
where $\iota$ is a four-valent $\SU(2)$ intertwiner and $I$ a four-valent $\Sp(4)$ intertwiner, and the coefficients of $S$ determine how simplicity is imposed.

For instance, with the EPRL prescription of embedding $SU(2)$ representations into
$\Sp(4)$ ones we get
\begin{equation}
S^{J(j^{+}j^{-})}_{MN q^{+}q^{-}
p^{+}p^{-}
} = 
\delta_{j^{+}, \frac{(1+\gamma)}{2}J}
\delta_{j^{-}, \frac{(1-\gamma)}{2}J}
C^{Jj^{+}j^{-}}_{M q^{+}q^{-}}
C^{Jj^{+}j^{-}}_{N p^{+}p^{-}} \, ,
\end{equation}
with $C$ being Clebsch--Gordan coefficients and $\gamma$ the Immirzi parameter.
Plugging these coefficients into \eqref{coeffconstraint} one obtains
\begin{equation}
\mathbb{S}^{J i^{+} i^{-} J_{1} \ldots J_{4}  (j_1^{+}, j_{1}^{-}),\ldots,(j_4^{+}, j_{4}^{-})} 
_{M_{1}\ldots M_{4}(q_1^{+},q_{1}^{-}),\ldots,(q_4^{+}, q_{4}^{-})}
= 
\left(\prod_{I=1}^{4}
C^{J_{I}j^{+}_{I}j^{-}_{I}}_{M q^{+}_{I}q^{-}_{I}}\right)
f^{J(i^{+},i^{-})}_{J_1\ldots J_{4}(j_{1}^{+}j_{1}^{-})\ldots
(j_{4}^{+}j_{4}^{-})}\,,
\label{coeffconstrgauge}
\end{equation}
where the coefficients $f$ are known as the fusion coefficients. Their expression is
\begin{equation}
f^{J(i^{+},i^{-})}_{J_1\ldots J_{4}(j_{1}^{+}j_{1}^{-})\ldots
(j_{4}^{+}j_{4}^{-})} = 
\sum_{\{N_I\},\{(p_I^+,p_I^-)\}}
\iota^{J J_{1}\ldots J_{4}}_{N_{1}\ldots N_{4}} I^{({i}^{+}i^{-})(j_{1}^{+}j_{1}^{-})\ldots
(j_{4}^{+}j_{4}^{-})}
_{(p_1^{+},p_{1}^{-}),\ldots,(p_4^{+}, p_{4}^{-})}\prod_{I=1}^{4} 
\delta_{j_{I}^{+}, \frac{(1+\gamma)}{2}J_{I}}
\delta_{j_{I}^{-}, \frac{(1-\gamma)}{2}J_{I}}
C^{J_{I}j^{+}_{I}j^{-}_{I}}_{N_{I} p^{+}_{I}p^{-}_{I}}
\,.
\end{equation}
If we impose the simplicity constraints in a different way \cite{listofallspinfoammodels} to define other models, we
obtain for the coefficients of $\mathbb{S}$ a similar expression, with different weights 
for the representations. It is then straightforward to obtain the corresponding explicit expression of the effective equations for other models.
A similar formula holds also in the case in which $\Sp(4)$ is
replaced with $\SL(2,\bC)$, provided that one deals with the regularisation issues
appearing due to the noncompactness of the group (which we discuss in Appendix~\ref{diverge}).

The case of infinite Immirzi parameter, associated to the Barrett--Crane model, is again slightly different, and one has to replace $\SU(2)$ with $\SL(2,\bC)/\SU(2)\simeq{\rm H}^3$. Apart from this, this case can be treated in exactly the same way.

\

Looking at the construction for condensate states describing homogeneous cosmologies, there are now several possible choices in the construction: one can insert
the constraint operators (in the appropriate form) in the states, in the kinetic term of the action, 
in the interaction term, or in combinations. Each choice will lead to slightly different
theories, with different Feynman rules.
Therefore, even with the same choices for $\mathcal{K}$ and $\mathcal{V}$, different ways to implement
the simplicity constraints (not only $\hat{\mathbb{S}}$, but also where it is inserted) will in general
change the effective dynamics.
 
To make the analysis more concrete and easy to follow, we will consider
only the case of simple condensates \eqref{simple}.
A first possibility is to insert the constraint operator only in the interaction term of the GFT action,
\ben
S[\varphi,\bar\varphi]=\mathcal{K}
[\varphi,\overline{\varphi}]+
\lambda\,\mathcal{V}
[\hat{\mathbb{S}}\varphi,\overline{\hat{\mathbb{S}}\varphi}]\,.
\een
The insertion of the constraint operator ensures that the constraints are imposed
at the level of the dynamics\footnote{Notice that, for consistency with the definition
of $\mathbb{S}$, we are using interactions for what would be an $\SU(2)$ GFT.}.
Then, the arguments of the function $\sigma$ cannot be
interpreted immediately in terms of geometric variables, since the
simplicity constraints have not yet been implemented. In other words, $\sigma$ cannot
be interpreted immediately as a distribution over minisuperspace. Only after the implementation of the (approximate) dynamics, and thus of the simplicity constraints, this interpretation will be allowed.

The quantum equation of motion is
\ben
\left(\int (\dd G')^4\;\mathcal{K}(G_1,\ldots,G_4,G'_1,\ldots,G'_4)\hat{\varphi}(G'_1,\ldots,G'_4)+
\lambda\,
\frac{\delta\hat{\mathcal{V}}[\hat{\mathbb{S}}\varphi,\overline{\hat{\mathbb{S}}\varphi}]}{\delta\hat\varphi^{\dagger}(G_1,\ldots,G_4)}\right)|\psi\rangle=0\,;
\label{quanteomconst}
\een
choosing $|\psi\rangle=|\sigma\rangle$ and multiplying with $\langle\sigma |$ as before, this means that in terms of $\sigma$,
\begin{equation}
\int (\dd G')^4\; 
\mathcal{K}(G_1,\ldots,G_4,G'_1,\ldots,G'_4)
\sigma(G'_1,\ldots,G'_4)+\lambda\,
\frac{\delta{\mathcal{V}}[\hat{\mathbb{S}}\sigma,\overline{\hat{\mathbb{S}}\sigma}]}{\delta\overline{\sigma}(G_1,\ldots,G_4)}=0\,.
\label{eqwithconstraints}
\end{equation}

It is important to stress that \eqref{eqwithconstraints} is a general result: it does not depend on
the particular form of the simplicity constraint operator $\hat{\mathbb{S}}$.
Consequently, it is applicable to
all the models for (Riemannian and Lorentzian) quantum gravity defined so far, with the only restriction being
the special form of the quantum state. It is the {\it general effective cosmological dynamics} extracted from the fundamental quantum gravity dynamics (at least in the approximate sense clarified above) for a generic 4$d$ GFT (thus spin foam) model.

As an example, consider GFT models with a trivial kinetic term, $\mathcal{K}(g_I,g'_I)=\delta(g_I^{-1}g_I')$, plus a general potential. The effective equation (\ref{eqwithconstraints}) simplifies to
\ben
\sigma(G_1,\ldots,G_4)+\lambda\,\frac{\delta{\mathcal{V}}[\hat{\mathbb{S}}\sigma,\overline{\hat{\mathbb{S}}\sigma}]}{\delta\overline{\sigma}(G_1,\ldots,G_4)}=0\,.
\label{effoog}
\een
Non-geometric tetrahedra are not allowed dynamically. Splitting the function $\sigma$ on (\ref{effoog}) as $\sigma=\sigma_{\mathfrak{K}}+\tilde\sigma$ where $\sigma_{\mathfrak{K}}$ is the component of $\sigma$ in the kernel of $\mathbb{S}$, we find that
\ben
\sigma_{\mathfrak{K}}(G_1,\ldots,G_4)=0
\een
as the potential in (\ref{effoog}) only depends on $\tilde\sigma$: geometric and non-geometric configurations decouple. 

For instance, we can take a simplicial interaction term. For generic 
fusion coefficients, the equation in components gives:
\begin{align}
0= & \sig{1}{1}{2}{3}{4} + \lambda  \sum_{\{ J, q^{+},q^{-}, i^{+},i^{-},j^{+},j^{-} \}} \{15j\}_{SU(2)} 
\prod_{\tau} f(\tau) \times \nonumber \\
& \sig{2}{5}{6}{7}{8} 
 \sig{3}{9}{10}{11}{12} \times \nonumber \\
& \sig{4}{13}{14}{15}{16}
\sig{5}{17}{18}{19}{20} \times \nonumber\\
&
\inter{1}{1}{20}
\inter{2}{2}{15}
\inter{3}{3}{10}
\inter{4}{4}{5}
\inter{5}{6}{19}
\times \nonumber \\
&
\inter{6}{7}{14}
\inter{7}{8}{9}
\inter{8}{11}{18}
\inter{9}{12}{13}
\inter{10}{16}{17} \, ,
\end{align}
where $\prod_{\tau} f(\tau)$ denotes the product of the fusion coefficients associated
to the labels of the five tetrahedra of the single four-simplex associated to the GFT vertex.
Once the specific values of the fusion coefficients has been given, and thus a specific model chosen, one can then
try to solve this equation, obtaining the effective dynamics for geometric configurations.

There is a second way to impose the constraints, which makes the correspondence with the
spin foam models treated in the literature more clear, and the correspondence 
with geometric data more direct. In a certain sense, this might be seen as
the imposition of the constraints directly in the kinetic term, and, automatically, in the states
used.
Instead of working with a map like the one
described above, which has the net effect of reducing the $\Sp(4)$ (or $\SL(2,\bC)$)
theory to an $\SU(2)$ theory, one can start with a group field theory
defined over four copies of $\SU(2)$ and embed it into a covariant theory for $\Sp(4)$ (or $\SL(2,\bC)$)\,.

This can be done using a suitable map\footnote{In the case of Barrett--Crane models,
one needs to replace $\SU(2)$ with the homogeneous space $\SL(2,\bC)/\SU(2)$, see the discussion in Sec.~\ref{efffried2}.}
\begin{equation}
\varpi : 
L^{2}(\SU(2)^4/\SU(2)_{\mathrm{diag}})
 \rightarrow 
L^{2}(G^4/G_{\mathrm{diag}})
 \, ,
\end{equation} 
where $G$ is either $\Sp(4)$ or $\SL(2,\bC)$, according to the case that one wants to consider.
This map can be constructed in the same way in which $\mathbb{S}$ has been constructed. 
In fact, $\varpi$
 can be seen as the transpose of $\mathbb{S}$, embedding directly the simple components,
 identified with representations of $\SU(2)$, into the full representations of the four-dimensional
 gauge group. 

There are two advantages in this second approach.
First, the theory is defined directly in terms of
the geometric variables (\ie simple bivectors), with no room for additional
degrees of freedom, whether they are decoupled from the geometric ones or not. Second, it 
makes straightforward the comparison with the
amplitudes in terms of which spin foam models are defined, \ie in terms of amplitudes
of $\Sp(4)$ ($\SL(2,\bC)$) BF theories modified by the insertion of the simplicity constraints.

The model then can be constructed in terms of a GFT over $\SU(2)^4$ (or $(\SL(2,\bC)/\SU(2))^4$
for Barrett--Crane-like models), involving some field 
$\varphi: \SU(2)^4/\SU(2)_{\mathrm{diag}} \rightarrow \mathbb{C}$, with the interactions now being the interaction terms
for a GFT for the full four-dimensional gauge group, evaluated on the embedded field
$(\varpi \varphi)(G_I)$, where $G_{I}$ are $\Sp(4)$ or $\SL(2,\bC)$ group elements:
\ben
S[\varphi,\bar\varphi]=\mathcal{K}_{\SU(2)}
[\varphi,\overline{\varphi}]+
\lambda\,\mathcal{V}_{H}
[\varpi\varphi,\overline{\varpi\varphi}] \, ,
\een
where $H$ denotes $\Sp(4)$ or $\SL(2,\bC)$ according to the signature chosen.

The operator equation of motion is analogous to what we had before,
\ben
\left(\int (\dd g')^4\;\mathcal{K}(g_1,\ldots,g_4,g'_1,\ldots,g'_4)\hat{\varphi}(g'_1,\ldots,g'_4)+
\lambda\,
\frac{\delta\hat{\mathcal{V}}[\varpi\varphi,\overline{\varpi\varphi}]}{\delta\hat\varphi^{\dagger}(g_1,\ldots,g_4)}\right)|\psi\rangle=0\,.
\label{quanteomemb}
\een
Considering again only the case of the expectation value of the operator appearing in this
equation in the state $|\sigma\rangle$, we obtain
the equation (in Riemannian signature)
\bena
0&=&\sigma^{i_{1} J_{1}\ldots J_{4}}_{M_{1}\ldots M_{4}}
+
\lambda
\sum_{J_{5}\ldots J_{10}, i_{1},\ldots i_{5}}
\left(\prod_{\tau=1}^{5} f(\tau)\right) 
\left(\prod_{f=1}^{10} A_{f}\right)  \nonumber\\
&&\times\sigma^{i_{2} J_{4}J_{5}J_{6} J_{7}}_{M_{4}M_{5} M_{6} M_{7}}
\sigma^{i_{3} J_{7}J_{3}J_{8} J_{9}}_{M_{7}M_{3} M_{8} M_{9}}
\sigma^{i_{4} J_{9}J_{6}J_{2} J_{10}}_{M_{9}M_{6} M_{2} M_{10}}
\sigma^{i_{5} J_{10}J_{8}J_{5} J_{1}}_{M_{10}M_{8} M_{5} M_{1}}
\{15j\}_{\Sp(4)}
\label{eqcondensateconstr}
\eena
where we are using a shortened notation to make the equation look a bit
more compact. With $\{15 j\}_{\Sp(4)}$ we denote
the $\Sp(4)$ invariant with the combinatorics of a four-simplex, analogous to
the $\{15j\}$ symbol for $\SU(2)$, labelled by the representations of $\Sp(4)$
related to the ones of $\SU(2)$ by the fusion coefficients.
The label $\tau$ denotes the five tetrahedra associated to the four-simplex, and 
$f(\tau)$ is the fusion coefficient associated to the given tetrahedron with the
given decoration in terms of $\SU(2)$ and $\Sp(4)$ representations.
The summation over $\Sp(4)$ representations entering the $\{15j\}_{\Sp(4)}$ symbol, with
the appropriate measure,
is left implicit.
Finally, the Clebsch--Gordan coefficients in \eqref{coeffconstrgauge}
determine the additional face amplitudes $A_{f}$ associated to the
faces of each tetrahedron,
\begin{equation}
A_{f} = \sum_{M,q^{+},q^{-}}\left(C^{J_{f}j_{f}^{+}j_{f}^{-}}
_{M_{f}q_{f}^{+}q_{f}^{-}}\right)^2 \, .
\end{equation}

Once more, the equation 
\eqref{eqcondensateconstr}
describes the dynamics of condensates as determined by the chosen 
spin foam/GFT model, encoded in the face amplitudes and fusion coefficients. 
It is clear that further simplifications are needed, as the equation itself
is rather complicated. Nonetheless it is useful to display it even without proposing an explicit solution, since it highlights
the generality of the procedure that we are proposing. The various spin foam
models proposed so far are easily included with no essential modification in
the equations. The Lorentzian models lead as well to the same equations.

\

This detailed discussion shows that the true challenges are then the proof
that these condensate states can be used as reliable approximations of the physical
states on one hand, and the development of methods of approximation for
the solution of \eqref{eqcondensateconstr} (and similar equations obtained for other condensates), for specific models. All this is indeed work in progress.

\subsection{Effective modified Friedmann equation: a concrete example} 
\label{efffried}

Let us now look at the resulting effective dynamics for GFT condensates in a specific simple example. We have seen that, assuming that the kinetic operator used in the definition of the GFT has a nontrivial kernel and that one focusses on states that are well-behaved in a regime of low particle density, the `dipole' function $\xi$ has to satisfy the linear equation (\ref{wdwtype}) which is of Wheeler--DeWitt type. Similarly, the same linear equation would result from considering the simple condensate given by the function $\sigma$ and taking a weak-coupling limit of the non-linear equation (\ref{simpleeq}). 

In this section, to make the previous rather general discussions more concrete in a specific model, we show that the equation (\ref{wdwtype}) can, in a semiclassical (WKB) limit, reproduce an effective cosmological dynamics that resembles what one would expect from a classical gravitational theory; in the isotropic case, the dynamics reduces to precisely the classical (vacuum) Friedmann equation with quantum corrections depending on the choice of state. We will also highlight the limitations of this simple example, which in fact should not be taken too seriously: for one thing, most of the dynamics in spin foam models and group field theories is supposed to be encoded in the vertex term, not only in the kinetic term. The example has to be taken only as a template of how one should go about extracting an approximate classical geometric equation in our general scheme: take the fundamental GFT/SF dynamics, and extract an effective cosmological dynamics for some simple condensate state. 

For generic GFT models, we have seen how to construct GFT condensate states from the elementary GFT field $\varphi$, defined on four copies of $\Sp(4)$ and satisfying the gauge invariance property
\ben
\varphi(g_1,\ldots,g_4) = \varphi(g_1 h,\ldots,g_4 h)\quad\forall h\in\Sp(4)\,.
\label{gauge}
\een
In order to obtain models that can include gravitational degrees of freedom, and can thus be relevant for cosmology, we now need to consider simplicity constraints as well, to ensure that one describes only geometric configurations (\ie those for which $B_i^{AB}={\epsilon_i}^{jk}e_j^A e_k^B$ for some triad $e_i^A$).

We have discussed the most general constructions introduced in the literature, leading to different models, at length. For concreteness, for the purposes of this section we impose simplicity as in \cite{bcgft}, by requiring that
\ben
\varphi(g_1,\ldots,g_4) = \varphi(g_1 h_1,\ldots,g_4 h_4)\quad\forall h_I\in\SU(2)_{X_0}\subset \Sp(4)\,,
\label{simpl}
\een
where $\SU(2)_{X_0}$ denotes the subgroup of $\Sp(4)$ (acting transitively on ${\rm S}^3$) which stabilises a fixed $X_0\in{\rm S}^3$. Hence, $\varphi$ is really a function on four copies of the coset space $\Sp(4)/\SU(2)_{X_0}\sim{\rm S}^3$, or equivalently four copies of $\SU(2)$ (since $\Sp(4)$ itself is simply $\SU(2)\times\SU(2)$); in this section we work with a field on $\SU(2)^4$. This is the type of prescription corresponding to the Barrett--Crane model \cite{bcaristidedaniele}.

There is an issue with imposing both (\ref{gauge}) and (\ref{simpl}), which can be understood most simply by noticing that there is no natural way to define a right action of $\Sp(4)$ on the coset space $\Sp(4)/\SU(2)_{X_0}$. When one imposes (\ref{gauge}) and (\ref{simpl}) by group averaging, one finds that the two operations do not commute. The issue was resolved in \cite{bcaristidedaniele}, where a normal $X\in{\rm S}^3$ is added as an argument of the GFT field, thus basing the whole formalism explicitly on projected spin networks \cite{projected}. The simplicity constraints are then imposed by
\ben
\varphi(g_1,\ldots,g_4;X) = \varphi(g_1 h_1,\ldots,g_4 h_4;X)\quad\forall h_I\in\SU(2)_X\subset \Sp(4)\,,
\label{simpl2}
\een
where $X$ is now the argument of the field and no longer fixed; the gauge invariance property is then
\ben
\varphi(g_1,\ldots,g_4;X) = \varphi(g_1 h^{-1},\ldots,g_4 h^{-1};h\cdot X)\quad\forall h\in\Sp(4)\,.
\label{gauge2}
\een
This is now consistent with the reduction to the coset space. After imposing simplicity, the arguments on the left-hand side of (\ref{gauge2}) are elements of $\Sp(4)/\SU(2)_X$ while those on the right-hand side live in $\Sp(4)/\SU(2)_{h\cdot X}$. The map
\ben
R_{h^{-1}}:\Sp(4)/\SU(2)_X\rightarrow \Sp(4)/\SU(2)_{h\cdot X}\,,\quad [g]\mapsto[g\, h^{-1}]
\een
between coset spaces is well-defined, as one can easily verify.

One can use (\ref{gauge2}) to gauge-fix the normal $X$ to a fixed $X_0$, reducing the $\Sp(4)$ invariance to the subgroup $\SU(2)_{X_0}$ that leaves this $X_0$ invariant. In the GFT model proposed in \cite{bcaristidedaniele} there is no explicit
coupling of normal vectors in the GFT interaction term, and in this sense they have no dynamics. One can then reduce again to a formulation where the GFT field depends only on four copies of $\SU(2)$, together with invariance under an action of $\SU(2)_{X_0}$. In the case considered here, this action would be trivial, given by the right action of $\SU(2)_{X_0}$ on $\Sp(4)/\SU(2)_{X_0}$.  

Let us now assume the normals have been gauge-fixed and hence can be removed from the formalism, and define the condensate (\ref{dipolestate}) by
\bena
|\xi\rangle &:=& \mathcal{N}(\xi) \exp\left(\hat\xi\right)|0\rangle\,, \nonumber\\  \quad \hat\xi &:=& 
\half\int (\dd g)^4 (\dd h)^4\; \xi(g_1^{-1} h_1,\ldots,g_4^{-1} h_4)\hat\varphi^{\dagger}(g_1,\ldots,g_4)\hat\varphi^{\dagger}(h_1,\ldots,h_4)\nonumber
\eena
as previously. Because of the simplicity condition (\ref{simpl}), the function $\xi$ satisfies $\xi(g_I)=\xi(k_I g_I k'_I)$ for all $k_I,\,k'_I\in\SU(2)_{X_0}$. It is hence a function on four copies of the space $\SU(2)_{X_0}\backslash\Sp(4)/\SU(2)_{X_0}$ (which is simply a compact interval), or alternatively a function on four copies of ${\rm S}^3\sim\Sp(4)/\SU(2)_{X_0}$ which is invariant under separate left actions of four copies of $\SU(2)_{X_0}$. This $\SU(2)_{X_0}$ acts on ${\rm S}^3$ by rotating around a fixed axis given by $X_0$. We will adopt the second interpretation.

Once simplicity constraints are imposed, the configuration space of the dipole function $\xi$ is no longer simply the gauge-invariant configuration space of a single tetrahedron, but a quotient of that by the left actions of $\SU(2)$: $\xi$ is just a function of the `absolute values' of four parallel transports. They do not admit a direct interpretation as `Hubble parameters' since they are still subject to a closure condition, and their geometric significance at the level of simplicial geometry is not obvious at this stage. One can always work at the level of four ${\rm S}^3$ elements and take care of the additional $\SU(2)^4$ symmetry, as we do in the following.

To proceed, we assume that there is a closure condition meaning that only three of the four arguments of $\xi$ are independent. This condition is the one we would get automatically if the field $\varphi$ satisfied an $\SU(2)$ closure condition of the form (\ref{gauge}),
\ben
\varphi(g_1,\ldots,g_4)=\varphi(g_1\,h,\ldots,g_4\,h)\quad\forall h\in \SU(2)\,,
\een
where the arguments $g_I$ are now in $\SU(2)$. Concretely, we impose $\xi(g_I)=\xi(k g_I k')$ for any $k,k'\in\SU(2)$. Note that this is not a special case of the previous invariance property; the action on $\SU(2)$ on itself is transitive, while the action of $\SU(2)_{X_0}$ on ${\rm S}^3$ stabilises $X_0$, for instance. With all this being done, the arguments of the collective wave function admit now the required geometric interpretation as minisuperspace variables.

\

After all these preliminaries, we can now proceed to derive the effective cosmological dynamics from GFT models with a particular kinetic term. We focus on models whose kinetic operator is the Laplace--Beltrami operator on $\SU(2)^4$, together with a `mass term'. A motivation for this choice is that the presence of the 
Laplacian seems to be required by GFT renormalisation 
\cite{GFTrenorm}. The equation (\ref{wdwtype}) for the function $\xi$ then becomes (setting the $g''_I$ which are arbitrary equal to the identity)
\ben
\left(\sum_{I} \Delta_{g'_I}+\mu\right)\xi(g'_I) = 0\,,
\label{wdw2}
\een
where we recall that $\xi$ is a function on $\SU(2)^4\sim (\Sp(4)/\SU(2)_{X_0})^4$ with additional symmetries.
Using the parametrisation for $\SU(2)$ given by
\ben
g = \sqrt{1-\vec{\pi}^2}\,{\bf 1} - \im\vec{\sigma}\cdot\vec\pi\,,\quad|\vec{\pi}|\le 1\,,
\een
where $\sigma^i$ are the Pauli matrices, the Laplace-Beltrami operator on $\SU(2)$ is 
\ben
\Delta_g f(\pi[g]) = \left(\delta^{\alpha\beta}-\pi^\alpha\pi^\beta\right)\partial_\alpha\partial_\beta f(\pi) - 3\pi^{\alpha}\partial_{\alpha}f(\pi)\,.
\een
(Note that the second term, which will drop out of the WKB analysis, was missing in \cite{prl}.) This parametrisation of $\SU(2)$ associates a Lie algebra element $\pi$ to every group element $g$. We can associate this Lie algebra element with a gravitational connection whose parallel transport is $g$: if we assume that the gravitational connection is constant over the dual link we are considering, the path-ordered exponential reduces to the usual exponential, $g=\mathcal{P}\exp(\int\omega)=\exp(\omega_1)$, in coordinates in which the corresponding link has unit coordinate length in the $x_1$ direction. Expanding this in the basis of Pauli matrices, we have $\omega_1 = \im \vec\sigma\cdot\vec\omega_1$ and
\ben
g = \cos(|\vec\omega_1|)\,{\bf 1} + \im\,\vec\sigma\cdot\vec\omega_1\,\frac{\sin(|\vec\omega_1|)}{|\vec\omega_1|}\,.
\een
In this sense, the fundamental dynamical variable, the Lie algebra element $\pi$, corresponds to the ``sine of the connection'', rather than the connection itself, which is very much reminiscent of what happens in loop quantum cosmology (LQC). Since we assume that for our configurations all gauge-invariant combinations of these parallel transports are close to the identity, we can approximate $\sin(|\vec\omega_1|)\approx |\vec\omega_1|$, and the corresponding Lie algebra elements can be interpreted directly as a gravitational connection at leading order.

In order to take the semiclassical (WKB) limit, we can then substitute the coordinate expression of the Laplace--Beltrami operator into (\ref{wdw2}), rewrite $\xi(\pi_I[g_I])=A[\pi_I]\exp(\im S[\pi_I]/\kappa)$ in terms of slowly varying amplitude and rapidly varying phase, and take the (formal) eikonal limit $\kappa \rightarrow 0$. The equation we obtain is
\ben
\sum_I\left(B_I\cdot B_I - (\pi_I\cdot B_I)^2\right) = O(\kappa)\,,
\label{friedmann}
\een
where $\cdot$ is the Killing form on $\mathfrak{su}(2)$ and $B_I:=\partial S/\partial \pi_I$  is the momentum conjugate to $\pi_I$, a bivector associated to one of the faces of a tetrahedron. For this scheme to be self-consistent, the phase of the function $\xi$ has to vary rapidly compared to the modulus, which is itself peaked near the identity in $\SU(2)^4$. (\ref{friedmann}) contains only the leading term in the WKB expansion, and the term in $\mu$, being of higher order ($\kappa^2$), does not appear. In the WKB approximation, (\ref{friedmann}) becomes the Hamilton--Jacobi equation for the classical action $S$.

Because of the symmetries of the function $\xi$, and hence the function $S$, the variables appearing in (\ref{friedmann}) are not all independent. Let us consider this in more detail. First, there is an invariance of $\xi$ under separate left actions of $\SU(2)_{X_0}$ as rotations of the three-sphere (using the map $\SU(2)\rightarrow\SO(3)\simeq\SU(2)/\mathbb{Z}_2$). Identifying $X_0\in{\rm S}^3$ with the identity in $\SU(2)$, this action corresponds to rotating the coordinate vector $\vec\pi$, $\vec\pi\rightarrow O\,\vec\pi$, or infinitesimally $\vec\pi\rightarrow\vec\pi+\vec\tau\times\vec\pi$ where $\times$ is the standard cross product in $\bR^3\simeq\mathfrak{su}(2)$. Invariance of $S$ under these transformations tells us that
\ben
S[\vec\pi_I] = S\left[\vec\pi_I+\vec\tau_I\times\vec\pi_I\right]\simeq S[\vec\pi_I]+\sum_I\vec{\tau}_I\cdot\left(\vec\pi_I\times\vec{\nabla}_I S[\pi_I]\right)\,,
\een
\ie the WKB ``angular momenta'' $\vec\pi_I\times B_I = [\pi_I,B_I]$ vanish; $\pi_I$ and $B_I$ are proportional as elements of $\mathfrak{su}(2)$.\footnote{In class A Bianchi models, this condition can be satisfied for the canonical pairs of dynamical variables by choosing them as diagonal in a given fiducial basis, \eg for Ashtekar--Barbero variables in a Bianchi IX model in \cite{lqcbianchi}.} Second, we have an invariance under the simultaneous action of $\SU(2)$ on all arguments, $S[\pi(g_I)]=S[\pi(kg_Ik')], \, \forall k,k' \in \SU(2)$. The transformation property of the $\SU(2)$ coordinates $\pi$ under a left multiplication is
\ben
\pi(kg_I)=\vec{\kappa}\sqrt{1-\vec{\pi_I}^2}+\vec{\pi_I}\sqrt{1-\vec{\kappa}^2}+\vec{\kappa}\times\vec{\pi_I}
\een
where $\pi(k)=:\vec{\kappa}$. Similarly, under right multiplication we have
\ben
\pi(g_Ik')=\vec{\kappa}'\sqrt{1-\vec{\pi_I}^2}+\vec{\pi_I}\sqrt{1-\vec{\kappa}'^2}-\vec{\kappa}'\times\vec{\pi_I}\,.
\een
Infinitesimally, the six independent possible transformations (acting on all four arguments of $S$) can be parametrised by
\ben
\vec{\pi_I} \mapsto \vec{\pi_I}+\delta\vec{\pi_I}= \vec{\pi_I} + \vec{\epsilon}\sqrt{1-\vec{\pi_I}^2} + \vec{\eta}\times\vec{\pi_I}
\een
corresponding to translations and rotations acting on the three-sphere. Invariance under rotations has already been used previously. The invariance of the function $S$ under a simultaneous translations of the group elements $g_I$ by an $\mathfrak{su}(2)$ element $\vec{\epsilon}$ means that
\ben
S[\pi_I] = S\left[\vec{\epsilon}\sqrt{1-\vec{\pi}_I^2}+\vec{\pi}_I\right]\simeq S[\pi_I]+\vec{\epsilon}\cdot\sum_I \left(\sqrt{1-\vec{\pi}_I^2}\,\vec{\nabla}_I S[\pi_I]\right)\,,
\een
which implies the conservation law $\sum_I \sqrt{1-\vec{\pi}_I^2}\,B_I =0$. This can be solved to express $B_4$ in terms of the other momenta $B_i$. Furthermore, we can use the invariance of $S$ to set $\vec{\pi}_4=0$ by a suitable left multiplication. The equation (\ref{friedmann}) then becomes (indices $i$ and $j$ run from 1 to 3)
\ben
\sum_i\left(B_i\cdot B_i - (\pi_i\cdot B_i)^2\right)+\sum_{i,j}\sqrt{1-\vec{\pi}_i^2}\sqrt{1-\vec{\pi}_j^2}\,B_i\cdot B_j = O(\kappa)\,.
\een

In order to identify the $B_i$ and conjugate $\pi_i$ with cosmological variables, we follow their geometric interpretation as bivectors and conjugate infinitesimal parallel transports, and write $B_i=A_i\,T_i$ and $\pi_i = p_i V_i$, where each $T_i$ and $V_i$ is a dimensionless normalised Lie algebra element. $A_i$ is an area element which we identify with the usual scale factors as $A_1=a_2 a_3$ and cyclically. Then (\ref{friedmann}) becomes
\ben
\sum_i 2A_i^2\left(1 - p_i^2\right)+\sum_{i\neq j}A_iA_j\gamma_{ij}\sqrt{1-p_i^2}\sqrt{1-p_j^2} = O(\kappa)
\label{friedmann2}
\een
where the dimensionless quantities $\gamma_{ij}=T_i\cdot T_j$ depend on the state, and we used $T_i\cdot V_i=\pm1$. 

At the level of the WKB approximation that we have employed, (\ref{friedmann2}) is the effective `Hamiltonian constraint' satisfied by the WKB phase space variables $A_i$ and $p_i$, which can be interpreted in terms of classical gravitational dynamics. Before discussing (\ref{friedmann2}) in full generality, let us first specialise to an isotropic geometry. Here we can set $A_i=\mu_i A,\,p_i=\nu_i p$ for constants $\mu_i$ and $\nu_i$, and we further assume that the state is such that the anisotropic contributions to (\ref{friedmann2}) vanish: $\gamma_{ij}=0$ (meaning that all $T_i$ are pairwise orthogonal in $\mathfrak{su}(2)$). We then obtain 
\ben
p^2 - k = O\left(\frac{\kappa}{a^2}\right)\,,
\label{friedmann4}
\een
where $k=\left(\sum_i \mu_i^2\right)/\left(\sum_i \mu_i^2 \nu_i^2\right)$. At leading order in $\kappa$, this is the classical Friedmann equation for an empty universe with spatial curvature $k$, with the modification of replacing the connection by its sine (here represented by the variable $p$), just as in loop quantum cosmology (LQC). Since $k>0$, this interpretation is consistent when $\GG=\SU(2)$, where $\GG$ is the group of isometries of spatial hypersurfaces that has to be chosen to interpret the condensate states. 

\

We now recall that our reconstruction procedure, providing a geometric interpretation to the variables appearing in our quantum states, required connections that were flat on the scale of the tetrahedra. We have used the gauge freedom to set $\vec{\pi}_4=0$ and therefore we should assume that $p_i\ll 1$ for our setting to be consistent, which also allows us to interpret $p$ directly as the gravitational connection (which, if one assumes it to be the Levi-Civita connection of an FRW metric, is $p\propto\frac{\dot{a}}{N}$). The assumption $p_i\ll 1$ was not needed for the derivation of the equation (\ref{friedmann4}), nor for the definition of the quantum states, but it is imposed for the geometric interpretation we want to give to our variables, \ie for relating discrete to continuum variables, using the procedure described in Sec.~\ref{approxgeo}.

The constant $k$ appearing in (\ref{friedmann4}) is of order one, so that the condition (\ref{friedmann4}) at lowest order in the WKB approximation, while identical to the classical Friedmann equation, has no consistent solutions. ($p=\const$ would correspond to flat $\bR^4$ foliated by round spheres of varying radius.) One could of course stick to the equation so obtained, and assume its validity and geometric interpretation beyond the domain in which our classical reconstruction procedure would allow it. As said, in fact, the $p \ll 1$ condition does not come from the dynamics of the theory, nor is it needed for deriving the above effective equation. Moreover, we have already pointed out how the same reconstruction procedure needs to be further developed, in particular for what concerns the role of various distance scales (as in loop quantum cosmology). However, at the moment we take this as an indication that our choices of GFT dynamics (here totally neglecting the interaction term), quantum ensembles, condensate states, and approximation scheme have to be improved to fully reproduce General Relativity in a cosmological setting.

\

Let us now return to the general anisotropic case given by (\ref{friedmann2}). Since we assume that $p_i\ll 1$, in order to interpret the effective classical dynamics satisfied by the GFT condensate, we can expand (\ref{friedmann2}) in powers of $p_i$,
\ben
2\sum_i A_i^2 +\sum_{i\neq j}A_i A_j\,\gamma_{ij} - 2\sum_i p_i^2 A_i^2  -\half \sum_{i\neq j}A_i A_j \gamma_{ij}(p_i^2+p_j^2) = O(\kappa,p_i^3)
\label{friedmann3}
\een
where we are only looking at terms up to quadratic order in $p_i$. The terms neglected can be viewed as higher derivative corrections to the effective gravitational dynamics which become negligible in the low-curvature regime we are looking at. The GFT dynamics allows also to compute explicitly the 
corrections to (\ref{friedmann3}), including both the subdominant terms in the WKB approximation 
of the above equation
and the corrections coming from the higher order terms in the effective cosmological dynamics. 

In the general anisotropic case, the dynamics of homogeneous universes in General Relativity is given by one of the Bianchi models, depending on the group of isometries of spatial hypersurfaces. Consistency with the isotropic case seems to suggest that this should be the Bianchi IX model, with $\GG=\SU(2)$, which we derive for the convenience of the reader in Appendix \ref{bianchi}. In metric variables, one finds a kinetic term quadratic in the momenta $p_i$ plus a potential representing spatial curvature which is independent of $p$. For instance, for the choice of variables $A_1=a_2 a_3$ and cyclically, where $a_i$ are the scale factors appearing in the 3-metric $\sum_k a_k^2({}^0 e^{k}\otimes {}^0 e^{k})$ for a given basis $\{{}^0 e^{i}\}$ of left-invariant one forms on $\SU(2)$, the (Riemannian) Hamiltonian constraint is
\ben
\mathcal{H}=A_1A_2p_1p_2+A_1A_3p_1p_3+A_2A_3p_2p_3+\left(\frac{A_2 A_3}{2A_1}\right)^2+\left(\frac{A_1 A_3}{2A_2}\right)^2+\left(\frac{A_1 A_2}{2A_3}\right)^2-\half\sum_i A_i^2\,.
\een
The effective `Hamiltonian constraint' (\ref{friedmann3}) giving the dynamics of our GFT condensate states has the general form of a quadratic kinetic term and an `anisotropy potential', independent of $p$, that should be interpreted as a non-zero spatial curvature. 

While there is a freedom in the choice of state that amounts to a tuning of the coefficients $\gamma_{ij}$ in (\ref{friedmann3}), none of the possible choices exactly reproduce the classical Bianchi IX dynamics in the truncation to quadratic order in momenta. We have made certain choices in obtaining this result: a choice of condensate state which neglects higher order multi-particle correlations and leads to an effective dynamics to which only the kinetic term in the GFT action contributes, and a choice of GFT model. In particular, from the LQG perspective an approximation in which the GFT potential $\mathcal{V}[\varphi,\bar\varphi]$ does not contribute to the effective equations cannot capture an essential aspect of GFT dynamics, which is in the potential and its prescription for determining the ways tetrahedra are glued to simplices or other building blocks. A better approximation would certainly have to involve the GFT potential in some way.

Again, the semiclassical analysis of anisotropic condensates shows that these choices (reduction to kinetic term only, quantum states, etc.) have to be improved in order to be able to reproduce General Relativity in the effective dynamics. Nevertheless, the example demonstrates the general applicability of our procedure for the derivation of an effective semiclassical dynamics from the quantum dynamics of condensate states in GFT.

\subsection{Effective modified Friedmann equation: Lorentzian case} \label{efffried2}
We can now try to repeat the constructions of Sec.~\ref{efffried} for GFT models corresponding to Lorentzian signature. As we discuss in Appendix~\ref{diverge}, the only additional difficulty is that symmetries of the GFT field can lead to divergences under integration. Repeating our definition of the dipole condensate (\ref{dipolestate}) for the case of gauge group $\Sp(4)$,
\bena
|\xi\rangle &:=& \mathcal{N}(\xi) \exp\left(\hat\xi\right)|0\rangle \quad \text{with}\nonumber \\  \quad \hat\xi &:=& 
\half\int (\dd g)^4 (\dd h)^4\; \xi(g_1^{-1} h_1,\ldots,g_4^{-1} h_4)\hat\varphi^{\dagger}(g_1,\ldots,g_4)\hat\varphi^{\dagger}(h_1,\ldots,h_4)\, ,\nonumber
\eena
we can replace $\varphi(g_1,\ldots,g_4)=\psi(g_1 g_4^{-1},g_2 g_4^{-1}, g_3 g_4^{-1})$ and change variables to go to a formulation without redundant integrations. One obtains
\ben
\hat\xi  = \half\int (\dd g)^4 (\dd h)^4\; \xi(g_1^{-1} h_1,g_2^{-1} h_2,g_3^{-1} h_3,e)\hat\psi^{\dagger}(g_1,g_2,g_3)\hat\psi^{\dagger}(h_1,h_2,h_3)
\een
using the invariance of $\xi$ under simultaneous left or right multiplication of its arguments. It is now explicit that the integration over $g_4$ and $h_4$ is redundant, and so, using the normalisation of the Haar measure that $\vol(\Sp(4))=1$,
\ben
\hat\xi = \half\int (\dd g)^3 (\dd h)^3\; \xi(g_1^{-1} h_1,g_2^{-1} h_2,g_3^{-1} h_3,e)\hat\psi^{\dagger}(g_1,g_2,g_3)\hat\psi^{\dagger}(h_1,h_2,h_3)\,.
\label{xigaugefix}
\een
For the compact group $\Sp(4)$ this is just a rewriting of the same definition. We could now try to use (\ref{xigaugefix}) for non-compact gauge groups such as $\SL(2,\bC)$ where the previous definition would be ill-defined due to the infinite volume of this group.

However, a closer look at (\ref{xigaugefix}) reveals that, due to the invariance of the function $\xi$ under conjugation,
\ben
\xi(g_1^{-1} h_1,g_2^{-1} h_2,g_3^{-1} h_3,e) = \xi(kg_1^{-1} h_1k^{-1},kg_2^{-1} h_2k^{-1},kg_3^{-1} h_3k^{-1},e)\quad\forall k\in \Sp(4)\,,
\een
the integrals appearing in (\ref{norm}), such as $\int (\dd g)^3 (\dd h)^3 |\xi(g_i^{-1} h_i)|^2$, are still infinite for gauge group $\SL(2,\bC)$; the resulting state $|\xi\rangle$ is not normalisable.
As we show in Appendix~\ref{diverge}, this state of affairs is due only
to the fact that, for convenience, we do not work with functions defined over the correct
space, \ie $\SL(2,\bC)^4/\SL(2,\bC)_{{\rm diag}} $. However,
this is just a technical, not a conceptual problem, which is easily circumvented with
a suitable choice of gauge fixing. 

This discussion ignores the imposition of the simplicity constraints, which, as we saw in Sec.~\ref{efffried}, affect the gauge invariance property imposed on the field $\varphi$. In Lorentzian signature, imposing them as we did for the Barrett--Crane prescription used in Sec.~\ref{efffried} means that we now require
\ben
\varphi(g_1,g_2,g_3,g_4) = \varphi(g_1 h_1,g_2 h_2, g_3 h_3, g_4 h_4) \quad\forall h_I\in\SU(2)_{X_0}\subset\SL(2,\bC)\,,
\een
where $X_0\in{\rm H}^3$. The GFT field then becomes a function on four copies of the homogeneous space $\SL(2,\bC)/\SU(2)$, which is 3-dimensional hyperbolic space, or ${\rm Hom(2)}$ as a group manifold (see Appendix \ref{hom2} for a discussion of this group and its geometry). As before, in order to impose gauge invariance properly, one should go to an extended formalism including a normal that is now an element of ${\rm H}^3$. For the purposes of extending the example of Sec.~\ref{efffried} to Lorentzian signature, we shall assume the normals have been gauge-fixed to $X_0$, and there is no further gauge invariance property of the field $\varphi$. This means we can proceed as before, using the definition (\ref{dipolestate}). The function $\xi$ is now interpreted as a function on four copies of $\SL(2,\bC)/\SU(2)$, separately invariant under left actions of $\SU(2)$ on the four arguments.

Again, we can assume that the kinetic term of the GFT model consists of a Laplacian and a mass term. For the coordinates on ${\rm H}^3$ that are the analogue of the coordinates on ${\rm S}^3$ chosen above, the Laplacian is
\ben
\Delta_{{\rm H}^3} f(\pi) = \left(\delta^{\alpha\beta}+\pi^\alpha \pi^\beta\right)\partial_\alpha\partial_\beta f(\pi)\,,
\een
so that (\ref{wdwtype}) reduces to
\ben
\left(\sum_{I} \Delta_{g'_I}+\mu\right)\xi(g'_I) = 0\,,
\een
the WKB approximation will give
\ben
\sum_I\left(B_I\cdot B_I + (\pi_I\cdot B_I)^2\right) = O(\kappa)\,.
\label{friedmannlor}
\een
This corresponds to the ``analytic continuation'' $\pi\rightarrow\im\pi$ when compared with (\ref{friedmann}) which is precisely the transformation between Riemannian and Lorentzian signature gravity. 

Again, the symmetries of $\xi$ and $S$ give constraints on the variables appearing in (\ref{friedmannlor}). The invariance of $S$ under separate left actions of $\SU(2)_{X_0}$ on hyperbolic space, $\vec\pi\rightarrow O\,\vec\pi$ or infinitesimally $\vec\pi\rightarrow\vec\pi+\vec\tau\times\vec\pi$, means that
\ben
S[\vec\pi_I] = S\left[\vec\pi_I+\vec\tau_I\times\vec\pi_I\right]\simeq S[\vec\pi_I]+\sum_I\vec{\tau}_I\cdot\left(\vec\pi_I\times\vec{\nabla}_I S[\pi_I]\right)\,,
\een
and again $\vec\pi_I\times B_I = 0$. To identify the quanta that make up the GFT condensate with geometric tetrahedra, we then also need to impose a closure condition on $\xi$. The most natural choice is to use the condition analogous to what is used in Sec.~\ref{efffried}, namely invariance under the simultaneous action of $\SL(2,\bC)$ on all four arguments of $\xi$,
\ben
\vec{\pi_I} \mapsto \vec{\pi_I}+\delta\vec{\pi_I}= \vec{\pi_I} + \vec{\epsilon}\sqrt{1+\vec{\pi_I}^2} + \vec{\eta}\times\vec{\pi_I}
\een
which are the translations and rotations of hyperbolic space. This is again a non-compact symmetry which will make $|\xi\rangle$ non-normalisable. However, the linear effective equation we have derived for the dipole condensate does not depend the normalisation of the state explicitly. One can hence define a regularised normalisable condensate state where the redundant integration over $\SL(2,\bC)$ is cut off at some `maximal boost'. This regulator can be taken to infinity without affecting the effective equations. For the purposes of this calculation, we can continue without considering this regularisation explicitly. We then find that
\ben
S[\pi_I] = S\left[\vec{\epsilon}\sqrt{1+\vec{\pi}_I^2}+\vec{\pi}_I\right]\simeq S[\pi_I]+\vec{\epsilon}\cdot\sum_I \left(\sqrt{1+\vec{\pi}_I^2}\,\vec{\nabla}_I S[\pi_I]\right)\,,
\een
and so, as before, we can use $\sum_I \sqrt{1+\vec{\pi}_I^2}\,B_I=0$ to express $B_4$ in terms of the other momenta $B_i$. We can then also set $\vec{\pi}_4=0$ by a suitable translation. From (\ref{friedmannlor}) we thus obtain
\ben
\sum_i\left(B_i\cdot B_i + (\pi_i\cdot B_i)^2\right)+\sum_{i,j}\sqrt{1+\vec{\pi}_i^2}\sqrt{1+\vec{\pi}_j^2}\,B_i\cdot B_j = O(\kappa)\,.
\een
Again, we can introduce cosmological variables $B_i=A_i\,T_i$ and $\pi_i = p_i V_i$, and (\ref{friedmannlor}) becomes
\ben
\sum_i 2A_i^2\left(1 + p_i^2\right)+\sum_{i\neq j}A_iA_j\gamma_{ij}\sqrt{1+p_i^2}\sqrt{1+p_j^2} = O(\kappa)\,.
\een

Since this is simply the analytically continued version of (\ref{friedmann2}), which we were not able to match to any Bianchi model in Riemannian signature, it does not correspond precisely to any Bianchi model in Lorentzian signature. Nevertheless, the structure of the equation shows that changing the GFT gauge group corresponds precisely to the change of signature in the metric formulation that one would expect, and so everything seems consistent as we argued. In particular, going to the isotropic limit we obtain
\ben
p^2 + k = O\left(\frac{\kappa}{a^2}\right)\,,
\label{friedmann5}
\een
where the spatial curvature is still positive. This is an interesting result which suggests that the models we investigate generically describe spatially closed universes, and that the spatial curvature we obtained before was not simply a result of the positive curvature of the gauge group $\SU(2)$ used as configuration space for the GFT field. In fact, here the configuration space is ${\rm H}^3$ which is negatively curved.

\section{Beyond vacuum and homogeneity: matter and perturbations}
\label{beyond}

The GFT models we have discussed so far are candidates for a theory of pure quantum geometry. We have seen that a certain class of states in these models captures the degrees of freedom of spatially homogeneous cosmologies, and we have obtained indications that their dynamics can reproduce, under a few assumptions, that of a classical theory of gravity. However, in order to connect to any realistic model of cosmology, it is essential to be able to describe two more ingredients: matter fields and perturbations in the gravitational field corresponding to inhomogeneities.

\subsection{Adding matter: a scalar field}

The most direct way of including matter degrees of freedom is to do what one would do in Wheeler--DeWitt geometrodynamics, that is to enlarge superspace to include not only the gravitational but also matter fields. This approach is also followed in loop quantum gravity \cite{LQG} and can be adapted easily to the GFT setting (for matter fields in this context, see \cite{danielejimmy}). An alternative would be to look for matter degrees of freedom in the effective theory of perturbations over background solutions of the GFT dynamics, \ie look for {\it emergent matter}. This last route has been tentatively explored in \cite{emergentmatter}. 

Let us consider a (real) scalar field, the most natural matter in cosmology. A scalar under diffeomorphisms is naturally associated to the vertex of a spin network, or in the dual simplicial geometry picture to an elementary tetrahedron. 

In continuum loop quantum gravity, in order to be able to apply the mathematical framework of compact groups to a scalar field, the physical variable used in the quantum theory is a `point holonomy'
\ben
U_{\zeta,x}(\phi):=\exp(\im\zeta\phi(x))\,,
\een
an element of ${\rm U}(1)$ associated to the vertex. The parameter $\zeta$ can be seen as a regulator to be taken to zero after the quantum theory has been defined. 

The restriction to compact groups does not seem strictly necessary in our context. We have seen that Lorentzian models can be defined, although one needs to be careful in order to avoid divergences. It is then not clear whether we need to compactify the scalar degree of freedom at the vertex by introducing a point holonomy. For dimensional reasons, we must in either case introduce a parameter $\zeta$ with dimensions of length, so that the new variable for the GFT field is $\tilde\phi:=\zeta\phi$. Note that the only dimensionful parameter that has appeared so far is $\kappa$ (which has dimensions of area) used in the WKB approximation. It is not clear at this stage how $\zeta$ and $\kappa$ are related. We will see that in the effective Friedmann equation $\zeta$ determines the coupling of matter to gravity, and so just like $\kappa$ should be related to Newton's constant.

The extended GFT model including a scalar field is then defined by a kinetic term
\ben
S_{{\rm k}}[\varphi,\bar\varphi]=\int (\dd g)^4 (\dd g')^4 \dd\tilde\phi\,\dd\tilde\phi'\;\bar\varphi(g_1,\ldots,g_4,\tilde\phi)\mathcal{K}(g_1,\ldots,g_4,g'_1,\ldots,g'_4,\tilde\phi,\tilde\phi')\varphi(g'_1,\ldots,g'_4,\tilde\phi')
\een
plus a potential. At this abstract level, all we have done is to extend the domain space of the GFT from four copies of $\Sp(4)$ to $\Sp(4)^4\times\bR$ or $\Sp(4)^4\times{\rm U}(1)$, in the case of Riemannian signature that we now restrict to for technical simplicity. There is no additional gauge invariance property for $\varphi$ since the scalar is left invariant by local $\Sp(4)$ transformations. The basic commutation relations for the quantum field are then
\ben
\left[ \hat{\varphi}(g_I,\tilde\phi),\hat{\varphi}^\dagger(g'_I,\tilde\phi') \right]= \mathbb{I}_G(g_I, g_I')\delta(\tilde\phi-\tilde\phi')\,,\quad \left[ \hat{\varphi}(g_I,\tilde\phi),\hat{\varphi}(g'_I,\tilde\phi') \right] = \left[ \hat{\varphi}^\dagger(g_I,\tilde\phi),\hat{\varphi}^\dagger(g'_I,\tilde\phi') \right] = 0\,,
\een
where the coordinate $\tilde\phi$ takes values in $[0,2\pi)$ for a point holonomy and in all of $\bR$ otherwise. It should then be clear that the constructions of Sec.~\ref{effcosm} go through just as before. Again, if we assume an interaction of odd order, for the `dipole' condensate defined by
\bena
|\xi'\rangle &:=& \mathcal{N}(\xi') \exp\left(\hat\xi'\right)|0\rangle \quad \text{with} \\  \quad \hat\xi' &:=& 
\half\int (\dd g)^4 (\dd h)^4 \;\dd\tilde\phi\;\xi'(g_1^{-1} h_1,\ldots,g_4^{-1} h_4,\tilde\phi)\hat\varphi^{\dagger}(g_1,\ldots,g_4,\tilde\phi)\hat\varphi^{\dagger}(h_1,\ldots,h_4,\tilde\phi) \, , 
\eena
the effective dynamics splits into two separate equations corresponding to the kinetic and potential terms, the former being
\ben
\int (\dd g')^4\;\dd\tilde\phi'\;\mathcal{K}(g_I,g'_I;\tilde\phi,\tilde\phi')\langle\hat\varphi(g'_1,\ldots,g'_4,\tilde\phi')\hat\varphi(g''_1,\ldots,g''_4,\tilde\phi'')\rangle_{\xi'} = 0\,.
\label{eqofmomat}
\een
The two-point function can be computed to be
\ben
\langle\varphi(g_I,\tilde\phi)\varphi(h_I,\tilde\Phi)\rangle=\xi(g_I^{-1} h_I ,\tilde\phi)\delta(\tilde\phi-\tilde\Phi)+\int(\dd g')^4 (\dd h')^4\;\xi(g_I^{-1} g_I',\tilde\phi)\xi(h_I^{-1} h_I',\tilde\Phi)\overline{\langle\varphi(g'_I,\tilde\phi)\varphi(h'_I,\tilde\Phi)\rangle}\,,
\een
and following the discussion in Sec.~\ref{effcosm}, we can focus on the class of solutions to (\ref{eqofmomat}) that satisfy
\ben
\int (\dd g')^4\;\mathcal{K}(g_I,g'_I;\tilde\phi,\tilde\phi')\xi((g''_I)^{-1} g'_I,\tilde\phi') = 0\,,
\een
which is the same equation that we would obtain from considering the weak-coupling limit of the effective equation for the single condensate $\sigma'$, defined in analogy to (\ref{simple}).

If we now focus on models with a Laplacian kinetic term on the extended domain space,
\ben
\mathcal{K}(g_I,g'_I;\tilde\phi,\tilde\phi')=\delta(g_I^{-1}g_I')\delta(\tilde\phi-\tilde\phi')\left(\sum_I\Delta_{g_I}+\tau\,\Delta_{\tilde\phi}+\mu\right)
\een
where we allow for a nontrivial relative weight $\tau$ between the gravitational and matter parts of the kinetic term, we can redo the analysis of Sec.~\ref{efffried} to obtain in the WKB limit (using that $f(x)\delta(x)=0$ implies that $f(0)=0$)
\ben
\sum_I\left(B_I\cdot B_I - (\pi_I\cdot B_I)^2\right)+\tau\, p_{\tilde\phi}^2 = O(\kappa)\,,
\een
where $p_{\tilde\phi}:=\partial S/\partial\tilde\phi$ is the momentum conjugate to $\tilde\phi$. For isotropic states, we saw before that the gravitational part of this can be written as $a^4(k-p^2)$ for some state-dependent constant $k>0$, so that we finally get
\ben
\frac{k-p^2}{a^2}=-\frac{\tau\,p_{\tilde\phi}^2}{a^6} + O(\kappa)\,.
\een
This is just the classical Friedmann equation with a massless free scalar field if the sign of $\tau$ is chosen appropriately ($\tau>0$ for Riemannian signature, it would be $\tau<0$ in the Lorentzian case), the dimensionful parameter $\zeta$ is chosen appropriately and physical units are restored. Just as in usual quantum cosmology, the dynamics of gravity coupled to such a scalar field is described by a positive definite Laplacian (for Riemannian signature gravity) and a wave operator of Lorentzian signature (for Lorentzian gravity) on superspace.\footnote{In quantum cosmology, just like in our discussion in GFT, the choice of kinetic operator can be viewed as a choice of metric on superspace. For a geometric discussion of this metric and its Lorentzian signature, see \cite{hawkingpage}.} Since the Laplacian on ${\rm U}(1)$ is the same as on $\bR$, we obtain the same Friedmann-type equation from the WKB approximation, independent of whether the scalar field is compactified using point holonomies in the definition of the GFT model.

\

These calculations show that a very natural extension of the GFT kinetic term to the matter sector leads to the correct coupling to gravity in the Friedmann equation obtained in the WKB limit. Clearly, they are just an example of how matter can be included into the GFT models. More work is needed in the fundamental definition of GFT models for quantum gravity models with matter fields, or in understanding whether matter degrees of freedom are already present in the existing models, without adding them by hand.

\subsection{Perturbations/inhomogeneities}
An important point remains to be addressed: the dynamics of the perturbations above a spatially homogeneous universe.
 This subject requires considerable extension of the analysis of the solutions to the GFT
equations of motion presented earlier in the paper. Nonetheless, we can provide a rough and tentative 
picture of the various points that need to be addressed.

Since condensates are candidates for quantum states that will be able to describe the
cosmological sector of GFT, it is natural to assume that inhomogeneities will be encoded
in small deviations from condensate states. In other words, it is natural to look for the physics of cosmological perturbations in the regime of fluctuations above the GFT condensates, i.e. GFT {\it phonons}.

The deviation from homogeneity complicates significantly the treatment, not least because everything has to be expressed in a coordinate-free language. 
In order to encode in a GFT state perturbations of an otherwise homogeneous metric,
one has to find first a complete set of invariants that can be used to store the shape of the perturbations. For instance, one could give the spectrum
of the perturbations in terms of the Fourier modes on a spatial slice.
Second, one has to translate this information into a state that stores it in terms of correlations
among GFT quanta. 

This is a subproblem
of the general sampling and reconstruction problem that is found in the construction of
semiclassical states in quantum gravity, and that we have discussed in Sec.~\ref{approxgeo}. Thus, the reconstruction procedure has to be extended to this case. This will be a first task ahead.

\

A second task will have to do with solving the quantum dynamics of the theory, which we are not able to solve exactly even for the simple condensate states. 
If $\ket{\Psi}$ is a state solving the equations of motion in the homogeneous sector,
then the equations of motion for any small deviation from it (not necessarily homogeneous) can be understood in terms of
perturbation theory,
\begin{equation}
\ket{\Psi} \rightarrow \ket{\Psi} + \epsilon\ket{\delta \Psi}\,.
\end{equation}
The perturbation has to be such that, if $\ket{\Psi}$ is a solution of the equations of motion, so is $\ket{\Psi} + \epsilon\ket{\delta \Psi}$, at least in the approximate sense that we consider only limited information about the quantum states and ask that this is compatible with them solving the equations of motion.

As discussed in Sec.~\ref{effcosm}, the quantum equations of motion can be turned into an infinite set of equations
for all the correlation functions,
\begin{equation}
\bra{\Psi} \hat{\mathcal{O}}[\hat\varphi,\hat\varphi^{\dagger}] \hat{\mathcal{C}}(g_I) \ket{\Psi} = 0\,,
\end{equation}
and in general we will be interested in a certain set of observables $\{\hat{\mathcal{O}}_{\mu}\}$. We will adapt 
the state $|\sigma\rangle$ in such a way that we reproduce the results of the exact solution
$\ket{\Psi}$ with a given accuracy $\eta$,
\begin{equation}
\left|\frac{ \langle \hat{\mathcal{O}}_{\mu} \rangle_{\Psi} - \langle \hat{\mathcal{O}}_{\mu} \rangle_{\sigma}  }{\langle \hat{\mathcal{O}}_{\mu} \rangle_{\sigma}}\right|
< \eta\,.
\end{equation}
In general it will be possible to choose observables for which this inequality is violated, and they
will set the theoretical error of the effective theory that we are going to develop. Therefore, the preliminary step before using our machinery to discuss inhomogeneous cosmologies
is the enumeration of the observables of the GFT that we want to keep under control. 

At this point, for the restricted set of observables we can replace $\ket{\Psi}$ with $\ket{\sigma}$,
making a relative error at most of magnitude $\eta$. In order to be consistent with our replacement of the exact solution $\ket{\Psi}$ with the state $\ket{\sigma}$, for example, the split of the state into homogeneous part and 
inhomogeneities  requires that $\eta \ll \epsilon$. If this is the case, for the observables $\mathcal{O}_{\mu}$ we can use $\ket{\sigma}$ instead of $\ket{\Psi}$, and one can obtain a set of equations involving the condensate wavefunction $\sigma$ as well as the various quantities used in the parametrisation of inhomogeneities. 

The assumption that we are dealing with a small amount of inhomogeneities has
to be translated into the appearance of a dimensionless expansion parameter, $\epsilon \ll1$,
that controls the deviation from the perfect condensate state,
\begin{equation}
|\sigma \rangle \rightarrow | \sigma \rangle + \epsilon |\delta \Psi \rangle\,.
\end{equation}
We stress that the meaning of the parameter $\epsilon$ is given by the structure of the state $|\delta\Psi\rangle$, and
can be understood only once the geometric content of the state is specified. 

These considerations are so far very general. To turn them into concrete calculations, one would need to identify the elementary excitations above the homogeneous backgrounds in terms of operators. In analogy with the case of quantum fluids, it is reasonable to conjecture that
the elementary excitations might be described by effective ``phononic'' fields whose relationship with
the fundamental field operators in terms of which GFT is formulated might not be a simple
linear transformation (\eg a Bogoliubov transformation). 

An approximate way to study this dynamics of perturbations would be to simply work at the classical GFT level and study perturbations around (condensate) solutions of the classical GFT equations, obtaining their effective action. Indeed, this analysis has been already carried out in simple cases (simple solutions of the equations of motion, special types of perturbations around them) and shown to give rise to effective field theories for scalar fields over non-commutative flat spacetimes \cite{emergentmatter}. This line of work goes then in the direction of identifying {\it emergent matter} from collective excitations of the very same degrees of freedom constituting spacetime itself. These results should now be improved, generalised, and reanalysed in light of the results presented in this paper.

On the basis of these considerations, then, we can expect to obtain an analogue of the Bogoliubov--de Gennes formalism for Bose--Einstein condensates. It is also worth stressing that this method, going
beyond a mean-field approach, might be useful, in addition to the analysis of deviation from
homogeneity, in the description of phase transitions, as it can provide
a first estimate of the breakdown of the regime that can be described in terms of a semiclassical
background geometry.

\

Last comes the issue of re-interpreting and rewriting the effective dynamics for the GFT phonons as an effective field theory on the continuum (spatially homogeneous) spacetime, defined by the background GFT condensate, as opposed to a field theory on minisuperspace.

To achieve this, it could be crucial to use the directions in the GFT configuration space that we have so far neglected: our analysis of the condensate states describing homogeneous cosmologies was restricted to {\em gauge-invariant} configurations, invariant under local actions of $\Sp(4)$ or $\SL(2,\bC)$. While this restriction is motivated by the interpretation of the fundamental degrees of freedom in discrete geometry, where indeed geometric observables can only depend on gauge-invariant quantities, it means we are throwing away information about a local {\em reference frame} generally encoded in a GFT quantum state. When going beyond the homogeneous condensate, the information in such a reference frame could be used for the construction of a coordinate system in which the perturbations can be localised. One would then generically look for gauge-variant perturbations over the gauge-invariant GFT condensate. All the geometric information needed to define the localisation and the state of motion of the GFT perturbations on the spacetime defined by the background GFT condensate state (assumed to be also a semiclassical state) should be extracted from the latter.

\section{Conclusions} 

In this paper, we have addressed one fundamental issue faced by all approaches to quantum gravity: to extract an effective macroscopic continuum dynamics directly from the fundamental microscopic quantum dynamics of the theory. 

We have constructed a class of condensate states that can be interpreted as macroscopic, spatially homogeneous geometries of the type usually considered in cosmology. These states are non-perturbative in that they contain contributions from arbitrary numbers of excitations of the (no-space) Fock vacuum of the theory. Our construction is thus a concrete realisation of the picture of spacetime as a quantum fluid or condensate, advocated previously in the context of group field theory (GFT) \cite{GFTfluid} and more generally in \cite{hu}. This picture can then be investigated dynamically in GFT: because these condensate states are analogous to coherent or squeezed states, their $n$-point functions can be computed and used to derive effective equations for the `condensate wavefunctions' used to define the states. These effective equations take the form of generalised nonlinear and nonlocal quantum cosmology equations. We have shown all of this in full generality, for a general choice of GFT model and for the different types of condensate states we are considering. It is easy to specialise the effective equations to specific GFT/spin foam models of 4$d$ quantum gravity. We have investigated a simple example where a particular choice of kinetic term leads to an effective equation that reduces, in a WKB approximation and in the isotropic case, to the classical Friedmann equation in vacuum. This example can be extended to Lorentzian signature and to include a massless scalar field, as we have shown. 

\

There are many directions for future work. Perhaps the most pressing one is to understand more carefully the nature and regime of the different approximations involved. The picture of spacetime as a GFT condensate involves a hydrodynamic approximation of the fundamental quantum dynamics. It required us to assume that the basic geometric building blocks describe near-flat configurations at the scale of the same building blocks, and specific approximations to the full GFT dynamics. In terms of cosmological variables, we are working in a regime outside of high curvature. In the homogeneous and isotropic case, where in our example the lowest order in momenta (\ie Hubble rates) corresponded to what one would expect from GR, we also employed a WKB approximation enforcing semiclassicality on our quantum states. It is not clear at this stage which of the approximations would have to be considered first in computing the leading order corrections.

More fundamentally, as one extrapolates to higher curvatures and follows the evolution of the universe backwards in time towards the Big Bang (or forwards approaching a Big Crunch), one expects the hydrodynamic approximation to break down. As we have argued, this would be signalled by large quantum fluctuations over the mean field whose effective dynamics is described by our quantum cosmological equations. Just like in the physics of Bose--Einstein condensates, this means that the ansatz one has made for the quantum state is no longer a good approximation and has to be replaced by something else; the Gross--Pitaevskii equation for the mean field no longer captures the relevant quantum dynamics. If we take this analogy seriously for the case of quantum gravity and quantum cosmology, it would mean that a high-curvature (presumably Planckian) regime cannot be described by a state consisting of near-flat, weakly interacting building blocks of geometry. Instead, one might expect a quantum phase transition, presumably a transition from a pre-geometric phase to a phase of an approximately smooth metric geometry -- a scenario that often goes under the name of {\em geometrogenesis} \cite{GFTfluid, geomgen}. Understanding this deep quantum-gravity regime will require methods that go beyond the ones used in this paper, and that will be explored in future work. Similarly, an important direction of future research, as we tried to discuss, is the study of fluctuations over the condensate states we have considered. The physics of such fluctuations should encode, we conjecture, the physics of cosmological perturbations (inhomogeneities) and one should try to recast the effective dynamics of such GFT perturbations in the form of an effective field theory over the background homogeneous geometries defined by our condensate states. The ultimate goal, of course, is to use such effective dynamics, directly extracted from the fundamental quantum gravity dynamics, to obtain predictions of testable quantum gravity effects in cosmology.

\begin{acknowledgments}
We thank Emanuele Alesci, Abhay Ashtekar, Frank Hellmann, Jo\~{a}o Magueijo, Roberto Percacci, Carlo Rovelli, Ed Wilson-Ewing, and all the members of the quantum gravity group at the AEI for useful discussions.
Research at Perimeter Institute is supported by the Government of Canada through Industry Canada and by the Province of Ontario through the Ministry of Research \& Innovation. DO acknowledges financial support from the A. von Humboldt Stiftung with a Sofja Kovalevskaja Award.
\end{acknowledgments}

\begin{appendix}

\section{Regularisation of Lorentzian models}
\label{diverge}

The generalisation of the results obtained in the Riemannian case to Lorentzian signature
requires some care in the definition and manipulation of the various quantities. In this Appendix
we consider briefly the key points that need to be addressed.
No significant modifications in the essence of
the procedure introduced in the paper arise, and there are only some small technical adjustments. 

In the models we consider, the signature
of the metric tensor that has to be recovered is encoded in the choice of local gauge group.
Therefore, models for Lorentzian spacetimes have to be based, in four dimensions, on
$\SL(2,\bC)$. 
The noncompactness of the group leads to a number of technical difficulties when working with a GFT field defined on four copies of 
$\SL(2,\bC)$, \eg when defining integrals and structures as a non-commutative Fourier transform on the group. 
 
A primary concern is that already the classical GFT action in Lorentzian signature is ill-defined, since the  imposition of the closure constraint as an invariance property of the GFT field leads to spurious integrations over one or more copies of $\SL(2,\bC)$. For the kinetic term, assuming that the kinetic operator $\mathcal{K}$ has the same symmetries as the GFT field $\varphi$, this is straightforward to see, as
\bena
&&\int (\dd g)^4 (\dd g')^4\;\bar\varphi(g_1,\ldots,g_4)\mathcal{K}(g_1,\ldots,g_4,g'_1,\ldots,g'_4)\varphi(g'_1,\ldots,g'_4)\nonumber
\\&=&\int (\dd g)^4 (\dd g')^4\;\bar\varphi(g_1g_4^{-1},g_2g_4^{-1},g_3g_4^{-1},e)\mathcal{K}(g_1,\ldots,g_4,g'_1,\ldots,g'_4)\varphi(g'_1(g'_4)^{-1},g'_2(g'_4)^{-1},g'_3(g'_4)^{-1},e)\nonumber
\\&=&\int (\dd h)^3 (\dd h')^3\,\dd g_4\,\dd h_4\;\bar\varphi(h_1,h_2,h_3,e)\mathcal{K}(h_1,h_2,h_3,h'_1,h'_2,h'_3)\varphi(h'_1,h'_2,h'_3,e)
\eena
where we have defined $h_i\equiv g_i g_4^{-1},\;h'_i\equiv g'_i (g'_4)^{-1}$. Integration over $g_4$ and $h_4$ now leads to factors proportional to the volume of the group. In the case of compact groups this is just a finite constant which can be set to one by a normalisation of the Haar measure, but in the Lorentzian case it means that the action can only be zero or infinite, and so does not define a variational principle.

One way out is the observation that the closure constraint is equivalent to a restriction
of the domain of the field to the homogeneous space $\SL(2,\bC)^4/\SL(2,\bC)_{{\rm diag}} \sim
\SL(2,\bC)^3$. One can rewrite the theory in terms of a field on this second group manifold without divergences arising from redundant integrations. In order to define the models, one can introduce a gauge-fixed field\footnote{Notice that in terms of the original four copies
of $\SL(2,\bC)$ there will be several different parametrisations, which will be equivalent
representations of the same model.}
\ben
\psi:\SL(2,\bC)^3 \rightarrow \mathbb{C}\,,\quad \psi(g_1,g_2,g_3)=\varphi(g_1,g_2,g_3,e)=\varphi(g_1g_4,g_2g_4,g_3g_4,g_4)\,,
\een
and rewrite the theory in terms of this, removing redundant integrations, {\em viz.},
\ben
S_{{\rm k}}^{{\rm reg}}[\psi,\bar\psi]=\int (\dd h)^3 (\dd h')^3\;\bar\psi(h_1,h_2,h_3)\mathcal{K}(h_1,h_2,h_3,h'_1,h'_2,h'_3)\psi(h'_1,h'_2,h'_3)
\een
for the kinetic term we looked at. There is of course no conceptual
difficulty in rewriting any action in this way, once we keep track in each term of the reduced
dependence of the fields on the arguments. By its definition, the field $\psi$ has no gauge invariance property corresponding to closure and so the action defined this way can be finite. The quantum field $\hat\psi$ satisfies standard commutation relations,
\ben
[\hat\psi(g_1,g_2,g_3),\hat\psi^{\dagger}(g'_1,g'_2,g'_3)] = \delta_{\SL(2,\bC)^3}(g_1g_1'^{-1},g_2 g_2'^{-1}, g_3g_3'^{-1})
\een
consistent with the commutation relations of $\hat\varphi$,
\ben
[\hat\varphi(g_1,\ldots,g_4),\hat\varphi^{\dagger}(g'_1,\ldots,g'_4)]= \delta_{\SL(2,\bC)^3}(g_1g_1'^{-1},
\ldots, g_4g_4'^{-1})\,,
\een
where $\delta_{\SL(2,\bC)^3}$ denotes the Dirac delta over the homogeneous space obtained from
$\SL(2,\bC)^4$ after imposing gauge invariance.

Let us work out explicitly the regularised version for the Ooguri model for BF theory in four dimensions. Here the kinetic term is
\ben
S_{{\rm k}}=\int (\dd g)^4 \bar\varphi(g_1,g_2,g_3,g_4)\varphi(g_1,g_2,g_3,g_4)=\left(\int \dd g\right)\int(\dd h)^3 \bar\psi(h_1,h_2,h_3)\psi(h_1,h_2,h_3)
\een
with a single redundant integration. There is also an interaction term, given by
\ben
\mathcal{V}_{{\rm Oo}} = \int (\dd g)^{10}
\varphi(g_1,g_2,g_3,g_4) 
\varphi(g_4,g_5,g_6,g_7) 
\varphi(g_7,g_3,g_8,g_9) 
\varphi(g_9,g_6,g_2,g_{10}) 
\varphi(g_{10},g_8,g_5,g_1) \,.
\een
Replacing the field $\varphi$ by the gauge-fixed field $\psi$, this becomes
\bena
\mathcal{V}_{{\rm Oo}} &=& \int (\dd g)^{10}
\psi(g_1g_4^{-1},g_2g_4^{-1},g_3g_4^{-1}) 
\psi(g_4g_7^{-1},g_5g_7^{-1},g_6g_7^{-1}) 
\psi(g_7g_9^{-1},g_3g_9^{-1},g_8g_9^{-1}) \times\nonumber
\\&&\times\psi(g_9g_{10}^{-1},g_6g_{10}^{-1},g_2g_{10}^{-1}) 
\psi(g_{10}g_1^{-1},g_8g_1^{-1},g_5g_1^{-1}) \,.
\label{oogpot}
\eena
We can now change variables to
\bena
&&h_1=g_1g_4^{-1}\,,\quad h_2=g_2g_4^{-1}\,,\quad h_3=g_3g_4^{-1}\,,\quad h_4=g_4g_7^{-1}\,,\quad h_5=g_5g_7^{-1}\,,\nonumber
\\&&h_6=g_6g_7^{-1}\,,\quad h_7=g_7g_9^{-1}\,,\quad h_8=g_8g_9^{-1}\,,\quad h_9=g_9g_{10}^{-1}\,,
\eena
so that (\ref{oogpot}) becomes
\bena
\mathcal{V}_{{\rm Oo}}& = &\left(\int \dd g\right)\int (\dd h)^9 
\psi(h_1,h_2,h_3) \psi(h_4,h_5,h_6) \psi(h_7,h_3 h_4 h_7,h_8)\times\nonumber
\\&&\times\psi(h_9,h_6h_7h_9,h_2h_4h_7h_9) 
\psi(h_9^{-1}h_7^{-1}h_4^{-1}h_1^{-1},h_8h_7^{-1} h_4^{-1}h_1^{-1},h_5h_4^{-1}h_1^{-1}) \,.
\eena
Notice that this interaction term and the kinetic term have the same redundant integration, which
can be then factorised and removed when the model is regularised.

A regularised version of these models, with potentially finite but non-zero action, can be given either in terms of the gauge-fixed field $\psi$ or in terms of the original field $\varphi$, since all we have done is change variables to make the redundant integration explicit. The regularised Ooguri action $S_{{\rm Oo}}^{{\rm reg}}=S_{{\rm k}}^{{\rm reg}}+
\mathcal{V}_{{\rm Oo}}^{{\rm reg}}$ is equal to
\bena
S_{{\rm Oo}}^{{\rm reg}}&=&\int (\dd g)^3 \;\bar\varphi(g_1,g_2,g_3,g_0)\varphi(g_1,g_2,g_3,g_0)
\\&&+\int (\dd g)^{9} \varphi(g_1,g_2,g_3,g_4)\varphi(g_4,g_5,g_6,g_7)\varphi(g_7,g_3,g_8,g_9)\varphi(g_9,g_6,g_2,g_0)\varphi(g_0,g_8,g_5,g_1) \,.\nonumber
\eena
This now appears to be a function of $g_0$ which is not integrated over, but is in fact independent of $g_0$ because of the gauge invariance property of $\varphi$, and one can set $g_0=e$, for instance.

The various quantities (actions, convolutions of operators, etc.) appearing in GFTs for noncompact
groups then have to be understood as defined in terms of the homogeneous space obtained
after imposing the closure constraint, thus eliminating the redundant integrations. 
However, for convenience in the notation and in the presentation of the various structures,
we are not going to write them explicitly in this form.

\section{Dynamics of the Bianchi IX model}
\label{bianchi}

For completeness and to clarify our choice of variables, we derive the dynamics of the anisotropic but homogeneous Bianchi IX universe in general relativity from scratch. Similar derivations can be found in textbooks such as \cite{cosmobook,martinbook}. 

Although this is in general not a consistent procedure, it turns out that one can substitute the ansatz of a spatially homogeneous geometry, with spatial slices given by 3-spheres with the round metric, into the Einstein--Hilbert action. This ansatz corresponds to a tetrad given by
\ben
e^i = a^{(i)}\,{}^0 e^{(i)}\;(i=1,2,3)\,,\quad e^0 = N\,dt\,,
\een
where $a^i$ and $N$ are functions of time only, there is no summation over the index $i$, and ${}^0 e^{i}$ define a (fiducial) basis of left-invariant one forms on $S^3\simeq\SU(2)$ satisfying the Maurer--Cartan relations
\ben
d{}^0 e^{i} = -\frac{1}{2}{\epsilon^i}_{jk} {}^0 e^{j} \wedge {}^0 e^{k}\,.
\een
Solving Cartan's equation of structure $de^I = -{\omega^I}_J \wedge e^J$ for the Levi-Civita connection $\omega$ gives
\ben
{\omega^i}_0 = \frac{\dot{a}^{(i)}}{a^{(i)}\,N}\, e^{(i)}\,,\quad {\omega^1}_2=\half\left(-\frac{a^1}{a^2 a^3}-\frac{a^2}{a^1 a^3}+\frac{a^3}{a^1 a^2}\right)e^3\,,\quad{\rm cyclically\;for}\;{\omega^3}_1\,,\;{\omega^2}_3\,.
\een
When computing the Riemann tensor $R^{IJ}=d\omega^{IJ}+{\omega^I}_J\wedge{\omega^{JK}}$ one has to keep in mind that only the components ${R^{IJ}}_{IJ}$ contribute to the Ricci scalar. The relevant contributions are
\bena
R^{i0}&\supset&\left(\frac{\ddot{a}^{(i)}}{a^{(i)}N^2}-\frac{\dot{a}^{(i)}\dot{N}}{a^{(i)} N^3}\right)e^i\wedge e^0\,,
\\R^{12}&\supset&\left(\frac{\dot{a}^1\dot{a}^2}{a^1a^2N^2}+\frac{1}{4}\left(\left(\frac{a^1}{a^2a^3}\right)^2+\left(\frac{a^2}{a^1a^3}\right)^2-3\left(\frac{a^3}{a^1a^2}\right)^2+\frac{2}{(a^1)^2}+\frac{2}{(a^2)^2}-\frac{2}{(a^3)^2}\right)\right)e^1\wedge e^2\nonumber
\eena
and cyclically for $R^{31}$ and $R^{23}$. The Ricci scalar is 
\ben
R = {R^{IJ}}_{IJ} = \sum_i\left(\frac{\ddot{a}^{i}}{a^{i}N^2}-\frac{\dot{a}^{i}\dot{N}}{a^i N^3}+\frac{1}{2(a^{i})^2}\right) + \frac{\dot{a}^1\dot{a}^2}{a^1a^2N^2}+{\rm 2\; more}-\frac{1}{4}\left(\frac{a^1}{a^2a^3}\right)^2+{\rm 2\; more}
\een
where the ``2 more'' terms denote cyclic permutations. Up to a constant coming from integration over the three-sphere that we set to one, the Lagrangian $\mathcal{L}=\int \dd^3x\;|e|\,R$ is
\ben
\mathcal{L}=\frac{\ddot{a}^1 a^2 a^3}{N}-\frac{\dot{a}^1 \dot{N}a^2 a^3}{N^2}+\frac{\dot{a}^1\dot{a}^2 a^3}{N}+\frac{N}{4}\left(-\frac{(a^1)^3}{a^2 a^3}+\frac{2a^2 a^3}{a^1}\right)+{\rm cyclic\; perm.}
\een
or after integration by parts, discarding the boundary term,
\ben
\mathcal{L}= -\frac{\dot{a}^1\dot{a}^2 a^3}{N} +\frac{N}{4}\left(-\frac{(a^1)^3}{a^2 a^3}+\frac{2a^2 a^3}{a^1}\right)+{\rm cyclic\; perm.}
\een
Various choices for the canonical variables can be found in the literature. If one sticks with the $a^i$ and their conjugate momenta $p_i$, $p_1=-\frac{1}{N}(a^2\dot{a}^3+\dot{a}^2 a^3)$ etc., the Hamiltonian $\mathcal{H}=\dot{a}^i p_i - \mathcal{L}$ is
\ben
\mathcal{H}=\frac{N}{4a^1a^2a^3}\left(- 2a_1a_2p_1p_2 + {\rm 2\;more}+\sum_i a_i^2 p_i^2 \right) +\frac{N}{4}\left(\frac{(a^1)^3}{a^2 a^3}-\frac{2a^2 a^3}{a^1}+{\rm cyclic\; perm.}\right)\,,
\een
which takes the form of kinetic term plus anisotropy potential (this would vanish for Bianchi I). The conventional choice for the lapse function is $N=a^1 a^2 a^3 $ which simplifies the form of $\mathcal{H}$ greatly:
\bena
\mathcal{H}&=&\frac{1}{4}\sum_i a_i^2 p_i^2- \half\left(a_1a_2p_1p_2+a_1a_3p_1p_3+a_2a_3p_2p_3\right) \nonumber
\\&&+ \frac{(a^1)^4+(a^2)^4+(a^3)^4}{4}-\frac{(a^1)^2 (a^2)^2+(a^1)^2 (a^3)^2+(a^2)^2 (a^3)^2}{2}\,.
\eena
For the geometric variables appearing in the GFT Fock space, one might choose different variables which are quadratic in the scale factors $a^i$, such as $A^1=a^2 a^3$ and cyclically. In terms of the $A^i$ the Lagrangian becomes
\ben
\mathcal{L}= \frac{\sqrt{A^1 A^2 A^3}}{4N}\left(\left(\frac{\dot{A}^1}{A^1}\right)^2-\frac{2\dot{A}^1\dot{A}^2}{A^1 A^2}\right)+\frac{N}{4}\left(-\frac{(A^2 A^3)^{3/2}}{(A^1)^{5/2}}+\frac{2(A^1)^{3/2}}{\sqrt{A^2 A^3}}\right)+{\rm cyclic\; perm.}
\een
and the Hamiltonian is, with the same choice of lapse $N=\sqrt{A^1 A^2 A^3}$,
\ben
\mathcal{H}=-(A^1A^2p_1p_2+A^1A^3p_1p_3+A^2A^3p_2p_3)+\left(\frac{A^2 A^3}{2A^1}\right)^2+\left(\frac{A^1 A^3}{2A^2}\right)^2+\left(\frac{A^1 A^2}{2A^3}\right)^2-\half\sum_i(A^i)^2
\een
where $p_i$ are now the conjugate momenta to $A^i$.

A different common choice of variables is given by $h^i=(a^i)^2$ so that the spatial metric is given by $\sum_i h^i({}^0 e^i\otimes {}^0 e^i)$. In these variables, also used in \cite{martinbook}, the Lagrangian is
\ben
\mathcal{L}=-\frac{1}{4N}\dot{h}^1\dot{h}^2\sqrt{\frac{h^3}{h^1 h^2}} + {\rm 2\;more}+\frac{N}{4}\left(-\frac{(h^1)^{3/2}}{\sqrt{h^2 h^3}}+2 \sqrt{\frac{h^1 h^2}{h^3}}+ {\rm cyclic\;perm.}\right)
\een
and again choosing $N=\sqrt{h^1 h^2 h^3}$ we obtain the Hamiltonian (compare Sec. 4.1.2 in \cite{martinbook})
\ben
\mathcal{H} = \sum_i (h^i)^2 p_i^2 - 2\left(h^1h^2p_1p_2 +h^1h^3p_1p_3+h^2h^3p_2p_3\right) + \frac{1}{4}\sum_i (h^i)^2 - \half \left(h^1 h^2 + h^1 h^3 + h^2 h^3\right)\,.
\een
with $p_i$ now conjugate to $h^i$.

Choosing a different Bianchi model would correspond to a different anisotropy potential (essentially the same terms with different coefficients) but the same kinetic term. Going to Riemannian signature corresponds to changing the overall sign of the kinetic term.

\section{The homogeneous space $\SL(2,\bC)/\SU(2)$}
\label{hom2}
Here we discuss the geometry of this homogeneous space which appears as the space of unit timelike normal vectors in 4$d$ Lorentzian geometry.

First, we note that the Lie algebra $\mathfrak{sl}(2,\bC)$ consists of all $2\times 2$ complex traceless matrices. A convenient basis for this (real) Lie algebra is given by
\bena
&&T_1=\begin{pmatrix}0 & 1 \\ 0 & 0\end{pmatrix}\,,\quad T_2=\begin{pmatrix}0 & \im \\ 0 & 0 \end{pmatrix}\,,\quad T_3 = \begin{pmatrix}1 & 0 \\ 0 & -1 \end{pmatrix}\,,\nonumber
\\&& T_4=\im\sigma_1=\begin{pmatrix} 0 & \im \\ \im & 0 \end{pmatrix}\,,\quad T_5=\im\sigma_2=\begin{pmatrix} 0 & 1 \\ -1 & 0 \end{pmatrix}\,,\quad T_6=\im\sigma_3=\begin{pmatrix}\im & 0 \\ 0 & -\im \end{pmatrix}\,,
\eena
where $\sigma^i$ are the Pauli matrices. Clearly $T_4, T_5$ and $T_6$ are generators of an $\SU(2)$ subgroup; there is also a Bianchi V subalgebra generated by $T_1, T_2$ and $T_3$:
\ben
[T_1,T_2]=0\,,\quad[T_1,T_3]=-2T_1\,,\quad [T_2,T_3]=-2T_2\,.
\een
This corresponds to the Lie algebra of the group ${\rm Hom}(2)$ of homotheties of the plane, which here appears as the Borel subgroup of $\SL(2,\bC)$. A general $\SL(2,\bC)$ element can then be written as
\ben
g=\begin{pmatrix}e^{\lambda} & e^{\lambda}\,w \\ 0 & e^{-\lambda}\end{pmatrix}\begin{pmatrix}x & y \\ -\bar{y} & \bar{x}\end{pmatrix}=\begin{pmatrix}e^\lambda(x-w\,\bar{y}) & e^\lambda(y+w\,\bar{x}) \\ -e^{-\lambda}\bar{y} & e^{-\lambda}\bar{x}\end{pmatrix}\,,\lambda\in\bR\,,\,x,y,w\in\bC\,,\,|x|^2+|y|^2=1\,.
\label{decomp}
\een
In this decomposition, $\lambda$ and the real and imaginary parts of $w=\alpha+\im\beta$ can be seen as coordinates on the homogeneous space $\SL(2,\bC)/\SU(2)$. The coordinates $\vec{\pi}$ on $\SU(2)$ that we introduced in Sec.~\ref{efffried} (and which cover only half of $\SU(2)$) correspond to $y=\pi_2+\im\pi_1$ and $x=\sqrt{1-\vec{\pi}^2}+\im\pi_3$.

One can construct a left-invariant metric on the homogeneous space from left-invariant one-forms on $\SL(2,\bC)$, obtained from expanding
\ben
g^{-1}\,\dd g = \hat{g}^{-1}\,\dd \hat{g} + \hat{g}^{-1}\,h^{-1}\,\dd h\,\hat{g}
\label{expand}
\een
in the basis of $\mathfrak{sl}(2,\bC)$, and we are using the decomposition $g=h\,\hat{g}$ as in (\ref{decomp}). The first term alone gives the left-invariant forms on $\SU(2)$,
\ben
\omega_i=\epsilon_{ijk}\,\pi_j\,\dd\pi_k+\sqrt{1-\vec{\pi}^2}\,\dd\pi_i+\frac{\pi_i}{\sqrt{1-\vec{\pi}^2}}\vec{\pi}\cdot\dd\vec{\pi}\,;
\een
contracting these with a multiple of the $\mathfrak{su}(2)$ Killing form gives the bi-invariant metric on $\SU(2)$,
\ben
\delta^{ij}\,\omega_i\otimes\omega_j = \left[\delta_{ij}+\pi_i\pi_j\frac{1}{1-\vec{\pi}^2}\right]\dd\pi^i\,\dd\pi^j\,,
\een
which is the round metric on the three-sphere. From the second term in (\ref{expand}) one can compute the remaining contributions to the $\SL(2,\bC)$ left-invariant forms to get
\bena
\tau_1&=&\nonumber 4\dd\lambda(\pi_1\pi_3+\pi_2\sqrt{1-\vec{\pi}^2})+(2\alpha\,\dd\lambda+\dd\alpha)(1-2(\pi_2^2+\pi_3^2))+2(2\beta\,\dd\lambda+\dd\beta)(\pi_3\sqrt{1-\vec{\pi}^2}-\pi_1\pi_2)
\\\tau_2&=&4\dd\lambda(\pi_1\sqrt{1-\vec{\pi}^2}-\pi_3\pi_2)-2(2\alpha\,\dd\lambda+\dd\alpha)(\pi_3\sqrt{1-\vec{\pi}^2}+\pi_1\pi_2)+(2\beta\,\dd\lambda+\dd\beta)(1-2(\pi_1^2+\pi_3^2))\nonumber
\\\tau_3&=&\dd\lambda(1-2(\pi_1^2+\pi_2^2))+(2\alpha\,\dd\lambda+\dd\alpha)(\pi_1\pi_3-\pi_2\sqrt{1-\vec{\pi}^2})-(2\beta\,\dd\lambda+\dd\beta)(\pi_2\pi_3+\pi_1\sqrt{1-\vec{\pi}^2})\nonumber
\\\tau_4&=&2\dd\lambda(\pi_3\pi_2-\pi_1\sqrt{1-\vec{\pi}^2})+2(2\alpha\,\dd\lambda+\dd\alpha)\pi_1\pi_2+(2\beta\,\dd\lambda+\dd\beta)(\pi_1^2-\pi_2^2)+\omega_1\nonumber
\\\tau_5&=&-2\dd\lambda(\pi_1\pi_3+\pi_2\sqrt{1-\vec{\pi}^2})+(2\alpha\,\dd\lambda+\dd\alpha)(\pi_2^2-\pi_1^2)+2(2\beta\,\dd\lambda+\dd\beta)\pi_1\pi_2+\omega_2\nonumber
\\\tau_6&=&(2\beta\,\dd\lambda+\dd\beta)(\pi_1\pi_3-\pi_2\sqrt{1-\vec{\pi}^2})+(2\alpha\,\dd\lambda+\dd\alpha)(\pi_2\pi_3+\pi_1\sqrt{1-\vec{\pi}^2})+\omega_3\,.
\eena

In the basis we have chosen, the Killing form on $\mathfrak{sl}(2,\bC)$ is non-diagonal; its (normalised) non-zero elements are
\ben
K_{15}=K_{24}=-1\,,\quad K_{33}=2\,,\quad K_{44}=K_{55}=K_{66}=-2\,.
\een
It has three positive and three negative eigenvalues, corresponding to the compact and noncompact directions. The bi-invariant metric on $\SL(2,\bC)$ is hence
\ben
{\bf g}_{\SL(2,\bC)}=-(\tau_1\otimes\tau_5+\tau_5\otimes\tau_1)-(\tau_2\otimes\tau_4+\tau_4\otimes\tau_2)+2\tau_3\otimes\tau_3-2\left(\tau_4\otimes\tau_4+\tau_5\otimes\tau_5+\tau_6\otimes\tau_6\right)\,.
\een

A natural left-invariant metric on the homogeneous space is now obtained by orthogonally projecting ${\bf g}_{\SL(2,\bC)}$ to the orbits of the action of $\SU(2)$, {\em \`a la} Kaluza-Klein:
\ben
{\bf g}_{\SL(2,\bC)}=-2\left(\tau_4+\frac{1}{2}\tau_2\right)^2-2\left(\tau_5+\frac{1}{2}\tau_1\right)^2-2\tau_6^2+2\tau_3^2+\half\tau_1^2+\half\tau_2^2\,.
\een
The last three terms then give a metric on the quotient space $\SL(2,\bC)/\SU(2)$ which is simply
\ben
{\bf g}_{\SL(2,\bC)/\SU(2)}=2\,\dd\lambda^2+\half(2\alpha\,\dd\lambda+\dd\alpha)^2+\half(2\beta\,\dd\lambda+\dd\beta)^2\,.
\label{hyper}
\een
As expected from the construction, any dependence on $\pi$ drops out. The metric (\ref{hyper}) has constant negative curvature, and is explicitly given in terms of the left-invariant forms on ${\rm Hom}(2)$. The left action of ${\rm Hom}(2)$ on itself is a subgroup of the group of isometries $\SL(2,\bC)$.

(\ref{hyper}) viewed as a metric on the group ${\rm Hom}(2)$ is {\em not} right-invariant (with respect to the right action of ${\rm Hom}(2)$ on itself), as can be seen from computing the right-invariant forms on ${\rm Hom}(2)$,
\ben
\upsilon_1 = \dd\lambda\,,\quad\upsilon_2 = e^{2\lambda}\,\dd\alpha\,,\quad\upsilon_3=e^{2\lambda}\,\dd\beta\,.
\een
The right-invariant metric on ${\rm Hom}(2)$ given by $\delta_{ij}\upsilon^i\otimes\upsilon^j$ gives another metric of constant negative curvature with isometry group $\SL(2,\bC)$. Note that already the left-invariant and right-invariant volume elements on ${\rm Hom}(2)$ differ; the group is not unimodular, and its Killing form is degenerate.

The relation of the coordinates $(\lambda,\alpha,\beta)$ to more familiar coordinate systems on hyperbolic space is the following: consider the embedding of hyperbolic space into $\bR^{3,1}$ by
\ben
-t^2 +x^2 +y^2 +z^2 = -1
\label{eq:hyperboloid}
\een
and choose null coordinates
\ben
u = \frac{t+x}{2}\,, \qquad v= \frac{t-x}{2}\,.
\een
One can now solve (\ref{eq:hyperboloid}) for $u$, describing hyperbolic space as the submanifold of $\bR^{3,1}$ given by
\ben
(t,x,y,z)=\left(\frac{1+y^2+z^2}{4v}+v,\frac{1+y^2+z^2}{4v}-v,y,z\right)\,.
\een
The induced metric is found to be
\ben
{\bf g}_{{\rm H}^3} = \frac{1+y^2+z^2}{v^2}\,\dd v^2 - \frac{2y}{v}\,\dd v\,\dd y - \frac{2z}{v}\,\dd v\,\dd z + \dd y^2 + \dd z^2\,.
\een
Changing coordinates to $v=e^\omega$ (notice that $v>0$ if one considers the hyperboloid of future timelike vectors in Minkowski space), this reduces to
\ben
{\bf g}_{{\rm H}^3} = \left(1+y^2+z^2\right)\dd \omega^2 - 2y\,\dd \omega\,\dd y - 2z\,\dd \omega\,\dd z + \dd y^2 + \dd z^2\,.
\een
The identification with the coordinates used in (\ref{hyper}) is then $\omega=\sqrt{2}\lambda$, $y=-\alpha/\sqrt{2}$, $z=-\beta/\sqrt{2}$.

\end{appendix}


\end{document}